\documentstyle[12pt]{report}
\input epsf
\def\thefootnote{\fnsymbol{footnote}}
\def\ref#1{$^{#1)}$}
\def\Tr{{\rm Tr}}
\newcommand{\dilaton}{{\ell}}
\newcommand{\Lag}{{\cal L}}
\newcommand{\superint}{\int \diff^{4}\theta}
\newcommand{\lowest}{|_{\theta =\bar{\theta}=0}}
\newcommand{\diff}{\mbox{d}}
\newcommand{\Diff}{{\cal D}}
\newcommand{\WaWa}{\Tr({\cal W}^{\alpha}{\cal W}_{\alpha})}
\newcommand{\WbWb}{\Tr({\cal W}_{\dot{\alpha}}{\cal W}^{\dot{\alpha}})}
\newcommand{\DaDa}{{\cal D}^{\alpha}{\cal D}_{\alpha}}
\newcommand{\DbDb}{{\cal D}_{\dot{\alpha}}{\cal D}^{\dot{\alpha}}}

\newcommand{\hs}{\hspace{0.2cm}}
\newcommand{\dg}{g_{_{\dilaton}}}
\newcommand{\dgg}{g_{_{\dilaton\dilaton}}}
\newcommand{\df}{f_{_{\dilaton}}}
\newcommand{\dff}{f_{_{\dilaton\dilaton}}}
\newcommand{\ldgl}{\dilaton g_{_{\dilaton}}}
\newcommand{\zdgz}{z g_{_{z}}}
\newcommand{\xdgx}{x g_{_{x}}}
\newcommand{\ldfl}{\dilaton f_{_{\dilaton}}}
\newcommand{\zdfz}{z f_{_{z}}}
\newcommand{\xdfx}{x f_{_{x}}}

\newcommand{\ldhl}{\dilaton h_{_{\dilaton}}}
\newcommand{\zdhz}{z h_{_{z}}}
\newcommand{\xdhx}{x h_{_{x}}}
\newcommand{\axion}{a}
\newcommand{\ep}{\varepsilon}
\def\[{\left [}
\def\]{\right ]}
\def\({\left (}
\def\){\right )}
\def\lbr{\left\{}
\def\rbr{\right\}}
\def\pp{\partial}
\def\hB{\hat{B}}
\def\D{{\cal D}}
\def\G{{\cal G}}
\def\tB{{\tilde B}}
\def\tb{{\tilde b}}

\def\m{\bar{m}}
\def\CO{{\cal O}}

\begin{document}

\begin{titlepage}
\pagestyle{empty}
\begin{center}
 \hfill          \begin{tabular}{r}
                      LBNL-40273 \\
                      UCB-PTH-97/18 \\
                 \end{tabular}

\vskip .5in

{\large \bf Supersymmetry Breaking in Superstring Theory \\ by Gaugino 
            Condensation and its Phenomenology}\footnote{This work was 
supported by the Director, Office of Energy Research, Office of High Energy 
and Nuclear Physics, Division of High Energy Physics of the U.S. Department 
of Energy under Contract DE-AC03-76SF00098.}

\vskip .25in
(Ph.D. Dissertation, University of California at Berkeley)

\vskip .25in

\vskip .2in
Yi-Yen Wu
\footnote{Electronic address:  {\tt YYWU@THEORM.LBL.GOV}}

{\em Theoretical Physics Group\\
     Lawrence Berkeley National Laboratory\\
     and\\
     Department of Physics\\
     University of California\\
     Berkeley, California  94720}
\end{center}

\vskip .5in

\newpage
\begin{abstract}
The weakly-coupled heterotic string is known to have problems of dilaton/moduli
stabilization, supersymmetry breaking (by hidden-sector gaugino condensation), 
gauge coupling unification (or the Newton's constant), QCD axion, as well as 
cosmological problems involving dilaton/moduli and axion. We study these 
problems by adopting the point of view that they arise mostly due to our 
limited calculational power, little knowledge of the full vacuum structure, 
and an inappropriate treatment of gaugino condensation. It turns out that 
these problems can be solved or are much less severe after a more consistent 
and complete treatment. 

There are two kinds of non-perturbative effects in our construction of string
effective field theory: the field-theoretical non-perturbative effects of
gaugino condensation (with a constraint ignored in the past) and the
stringy non-perturbative effects conjectured by S.H. Shenker, which are best 
described using the linear multiplet formalism. Stringy non-perturbative 
corrections to the K\"ahler potential are invoked to stabilize the dilaton at 
a value compatible with a weak coupling regime. Modular invariance is ensured 
through the Green-Schwarz counterterm and string threshold corrections which, 
together with hidden matter condensation, lead to moduli stabilization at the 
self-dual point where the {\em vev}'s of moduli's $F$ components vanish. In 
the vacuum, supersymmetry is broken at a realistic scale with vanishing 
cosmological constant. As for soft supersymmetry breaking, our model always 
leads to a dilaton-dominated scenario. For the strong CP problem, the 
model-independent axion has the right properties to be the QCD axion. 
Furthermore, there is a natural hierarchy between the dilaton/moduli mass and 
the gravitino mass, which could solve both the cosmological moduli problem and 
the cosmological problem of the model-independent axion.
\end{abstract}
\end{titlepage}

\pagestyle{plain}
\renewcommand{\thepage}{\roman{page}}
\setcounter{page}{2}
\mbox{ }

\vskip 1in

\begin{center}
{\bf Disclaimer}
\end{center}

\vskip .2in

\begin{scriptsize}
\begin{quotation}

This document was prepared as an account of work sponsored by the United
States Government. While this document is believed to contain correct 
information, neither the United States Government nor any agency
thereof, nor The Regents of the University of California, nor any of their
employees, makes any warranty, express or implied, or assumes any legal
liability or responsibility for the accuracy, completeness, or usefulness
of any information, apparatus, product, or process disclosed, or represents
that its use would not infringe privately owned rights.  Reference herein
to any specific commercial products process, or service by its trade name,
trademark, manufacturer, or otherwise, does not necessarily constitute or
imply its endorsement, recommendation, or favoring by the United States
Government or any agency thereof, or The Regents of the University of
California.  The views and opinions of authors expressed herein do not
necessarily state or reflect those of the United States Government or any
agency thereof, or The Regents of the University of California.

\end{quotation}
\end{scriptsize}

\vskip 2in

\begin{center}
\begin{small}
{\it Lawrence Berkeley Laboratory is an equal opportunity employer.}
\end{small}
\end{center}

\newpage
\section*{Acknowledgements}
\hspace{\parindent}
I would like to express my most sincere gratitude to my adviser,
Professor Mary K. Gaillard, for guiding me into the fascinating area 
of string phenomenology and theoretical physics, for her very inspiring 
advice and informative remarks, and for her kindness and support. Her 
perceptiveness in and dedication to physics have always inspired me. Special 
thanks go to Professors Mary K. Gaillard and Pierre Bin\'etruy with whom I 
collaborated. I have benefitted a lot from their pioneering 
works and the invaluable collaboration.

I would like to thank Professors Bruno Zumino, Orlando Alvarez and Stanley 
Mandelstam for teaching me a lot of physics. I thank Professors Ian Hinchliffe,
Mahiko Suzuki and Majorie Shapiro for warm help and encouragement, and 
Professors Jim Gates, Hitoshi Murayama and Hirosi Ooguri for inspiring 
comments and discussions. I must also thank the postdocs as well as my fellow 
graduate students, especially Nima Arkani-Hamed, Jonathan Feng, 
Hsin-Chia Cheng, Chong-Sun Chu, Pei-Ming Ho and Takeo Moroi for much useful 
discussion and help.

This work was supported by the Director, Office of Energy
Research, Office of High Energy and Nuclear Physics, Division of High
Energy Physics of the U.S. Department of Energy under Contract
DE-AC03-76SF00098.

\newpage
\tableofcontents
\listoffigures

\newpage
\renewcommand{\thepage}{\arabic{page}}
\renewcommand{\theequation}{\arabic{chapter}.\arabic{equation}}
\renewcommand{\thefigure}{\arabic{chapter}.\arabic{figure}}
\def\thefootnote{\arabic{footnote}}
\setcounter{page}{1}
\chapter{Preamble}
\setcounter{equation}{0} 
\setcounter{footnote}{0}
\newpage
\hspace{0.8cm}
How the electroweak symmetry is broken is one of the fundamental questions 
of particle physics. In the standard model, the scalar Higgs doublet 
acquires a non-vanishing vacuum expectation value ({\em vev}), and therefore 
breaks the electroweak symmetry. However, the field-theoretical loop 
corrections to the masses of scalar particles are quadratically divergent. 
Therefore, the scale of electroweak symmetry breaking is in fact unstable 
against radiative corrections, and how the very large hierarchy between the 
Planck scale and the scale of electroweak symmetry breaking is generated 
remains a mystery. Currently, weak scale supersymmetry \cite{Haber85} is the 
most promising solution this hierarchy problem. Supersymmetric theories are 
free from quadratic divergences due to delicate cancellations between boson 
and and fermion loop corrections, and therefore can stabilize the hierarchy 
between the Planck scale and the electroweak scale. However, supersymmetry
itself does not explain the origin of the electroweak scale. Furthermore, 
supersymmetry introduces new particles ({\it i.e.}, supersymmetric partners 
of the standard model particles.) Therefore, as a requirement of particle 
phenomenology, supersymmetry must be broken and the resulting theory is a 
supersymmetric extension of the standard model with supersymmetry {\em softly} 
broken at the electroweak scale. The experimental search for supersymmetric 
partners is very important to our understanding of electroweak symmetry 
breaking. It will also shed light on the mechanism of supersymmetry breaking 
as well as the physics at (and possibly above) the scale where supersymmetry 
is broken. On the other hand, constructing a realistic scheme of supersymmetry 
breaking remains one of the big challenges to supersymmetry phenomenology. 
Although it is possible, without knowing the details of the supersymmetry 
breaking 
mechanism, to parametrize the effects of softly broken supersymmetry in an
effective description, yet it involves a huge numbers of unknown parameters
and thus makes phenomenological analyses highly intractable. It is therefore 
desirable to have a realistic supersymmetry breaking scheme which predicts all 
the soft supersymmetry breaking parameters in terms of only a few parameters. 

   It is well known that superstring theory offers, according to the above
consideration, the most powerful scheme of supersymmetry phenomenology. More
precisely, all the {\em parameters} appearing in the effective description 
of the superstring are in principle determined by the dynamics of superstring 
alone, {\it i.e.}, by the {\em vev}'s of certain fields ({\it e.g.}, the string 
dilaton and moduli.) Besides, the most compelling reason to study superstring 
theory is the fact that it is the only known candidate theory of quantum 
gravity. However, at the perturbative level the superstring has many vacua 
parametrized by flat directions ({\it e.g.}, the string dilaton and moduli) 
which will be lifted only after non-perturbative effects are 
included\footnote{It is very possible that the same non-perturbative effects 
are also responsible for 
supersymmetry breaking.}. Even with the recent progress of string duality, 
there is still little knowledge of these non-perturbative effects and hence 
how the above powerful feature of superstring theory is realized. Earlier 
attempts to 
study the phenomenology of superstrings \cite{Brignole} have either ignored 
the non-perturbative effects responsible for stabilizing the string 
dilaton/moduli or relied on the racetrack model\footnote{As will be discussed 
later, the racetrack model suffers from a negative cosmological constant 
problem as well as an un-naturalness problem.} \cite{racetrack}, and therefore 
their results may not be reliable. It is believed and will be shown in the 
following chapters that it is possible to draw reliable predictions from 
superstrings only after the relevant non-perturbative effects are fully taken 
into account.

   Our study of superstring phenomenology contains two kinds of 
non-perturbative effects: the stringy non-perturbative effects generated above
the string scale, and the field-theoretical non-perturbative effects of gaugino
condensation generated by strongly-interacting gauge groups below the string 
scale. As for stringy non-perturbative effects, they have always been ignored
in the past. The existence of significant stringy non-perturbative effects was 
first conjectured by S.H. Shenker \cite{Shenker90}. The recent development of 
string duality has provided further evidence \cite{Shenker96,eva} for 
Shenker's conjecture. It was first noticed by T. Banks and M. Dine that 
significant stringy non-perturbative effects could have interesting
implications \cite{Banks94}. Here we will study in detail the phenomenological 
implications of stringy non-perturbative effects using the linear multiplet 
formalism of superstring effective theory. It was first pointed out in 
\cite{Gates} that the field-theoretical limit of the weakly-coupled heterotic 
string theory should be described using the linear multiplet formalism rather 
than the chiral multiplet formalism.\footnote{A subtlety associated with the
chiral multiplet formalism of gaugino condensation in the past will be 
discussed later.} On the other hand, there exists a
chiral-linear duality between these two formalisms \cite{Burgess95}, and 
therefore in principle these two formalisms should be equivalent. However, the
chiral-linear duality is apt to be very complicated, especially when full
quantum corrections are included. Therefore, there should exist a formalism
where the physics allows a simpler description. It has been argued in
\cite{Adamietz93,Derendinger94} that, according to the above 
consideration, the linear multiplet formalism should be the more appropriate
formalism.\footnote{A more detailed discussion can be found in Section 2.1.} 
Furthermore, our study represents a concrete realization of this 
point of view. As we shall see in Chapter 2, in the linear multiplet formalism 
the string coupling 
is the linear multiplet $\,L\,$ which is the natural parametrization of 
stringy physics. On the other hand, the coupling of string effective field 
theory is $\,L/\(1+f(L)\)\,$ which is the natural parametrization of 
field-theoretical effects; it is modified in the presence of stringy effects 
$f(L)$. Therefore, the linear multiplet formalism naturally distinguishes 
stringy effects from field-theoretical effects, and it is this feature that 
enables one to keep track of both effects within the effective field theory.
This advantage of the linear multiplet formalism is very crucial to our study 
where both stringy and field-theoretical non-perturbative effects are 
considered. As we will see, stringy non-perturbative effects do
play an important role in stabilizing the string dilaton/moduli and in 
breaking supersymmetry via the field-theoretical non-perturbative effects of 
gaugino condensation \cite{dilaton,yy,multiple}.

   As for the field-theoretical non-perturbative effects, gaugino condensation 
has always played a unique role: at low energy, the strong 
dilaton-Yang-Mills interaction leads to gaugino condensation which not only 
breaks supersymmetry spontaneously but also generates a non-perturbative 
potential which may eventually stabilize the dilaton\footnote{In general there 
is also matter condensation which generates a non-perturbative potential for
string moduli.}. In the scheme of gaugino condensation the stabilization of 
string dilaton/moduli and the breaking of supersymmetry are therefore unified 
in the sense that they are two aspects of a single non-perturbative 
phenomenon. Furthermore, gaugino condensation has its own important 
phenomenological motivations: gaugino condensation occurs in the hidden sector 
of a generic string model \cite{Nilles82,Dine85}; it can break supersymmetry 
at a sufficiently small scale and may induce viable soft supersymmetry 
breaking effects in the observable sector through gravity and/or an anomalous 
U(1) gauge interaction \cite{messenger}. However, although gaugino 
condensation has been studied since 1982, it still has several long-standing 
problems in the context of superstrings. Firstly, superstring phenomenology 
based on the scheme of gaugino condensation has been long plagued by the 
infamous dilaton runaway problem \cite{Banks94,Dine85}. That is, (assuming 
that the tree-level K\"ahler potential of the dilaton is a good approximation) 
one generally finds that the supersymmetric vacuum with vanishing coupling 
constant and no gaugino condensation is the only stable minimum in the 
weak-coupling regime. Secondly, modular invariance is a very important property
of the heterotic string, and it should have unique phenomenological 
predictions. However, most of the studies of gaugino condensation had
neither complete nor correct treatments of modular invariance. As we shall see, 
a fully modular invariant treatment of gaugino condensation has non-trivial
phenomenological implications.\footnote{The unique phenomenological predictions
associated with modular invariance will be discussed in Chapters 4 and 5. As
will be explained in Section 4.7, we emphasize that these unique predictions 
do not necessarily follow from any framework with modular invariance; in the
context of weakly-coupled heterotic string these predictions are the 
consequences of both modular invariance and an appropriate treatment of gaugino 
condensation. This may explain why these predictions are absent in those works
\cite{ppp} where modular invariance is correctly incorporated but a constraint 
on gaugino condensation (to be discussed below) has not been included.}
Thirdly, in the past the gaugino condensate has 
always been described by an {\em unconstrained} chiral superfield $U$ which 
corresponds to the bound state of $\,{\cal W}^{\alpha}{\cal W}_{\alpha}\,$ in 
the underlying theory. It was pointed out recently that $U$ should be a 
{\em constrained} chiral superfield \cite{Binetruy95,sduality,Burgess95,Pillon} 
due to the constrained superspace geometry of the underlying Yang-Mills theory:
\begin{eqnarray}
U\,&=&\,-(\DbDb-8R)V, \nonumber \\ 
\bar{U}\,&=&\,-(\DaDa-8R^{\dagger})V,
\end{eqnarray}
where $V$ is an unconstrained vector superfield. Fourthly, superstring 
phenomenology based on gaugino condensation suffers from several cosmological
problems such as the cosmological moduli problem \cite{Kaplan93} and the 
cosmological bound on the invisible axion \cite{Abbott83}. These cosmological 
problems either destroy the successful nucleosynthesis or overclose the
universe. 

   These formidable problems might make one think that the weakly-coupled 
heterotic string theory is in grave danger. On the other hand, these problems 
are not unrelated to one another because the superstring has a highly 
constrained and 
predictive framework. As we shall see, in fact these problems arise from our 
poor understanding of non-perturbative string dynamics as well as 
inappropriate/incomplete treatments of superstring phenomenology in the past. 
Once we know how to proceed in the right direction, these problems turn out 
to be solved or much less serious. For the first problem, we emphasize the
advantage of using the linear multiplet formalism and show that stringy 
non-perturbative effects may stabilize the dilaton at a value compatible 
with a weak coupling regime \cite{dilaton,yy}. For the second and the third 
problems, full modular invariance is ensured through the Green-Schwarz term and 
string threshold corrections, and the constraint on the gaugino condensate $U$ 
is explicitly solved using the linear multiplet formalism 
\cite{dilaton,yy,multiple}. They do lead to unique predictions of 
superstring theory about supersymmetry breaking, the compactification 
scale, and axion physics\footnote{These unique predictions were unknown in the
past due to the aforementioned first three problems.}. For example, string
moduli are stabilized at the self-dual point, and therefore they do not 
participate in supersymmetry breaking because the {\em vev}'s of moduli's 
$F$ terms vanish \cite{multiple}. This is certainly a desirable feature in 
consideration of flavor changing neutral current (FCNC) because non-vanishing 
{\em vev}'s of moduli's $F$ terms generically lead to non-universal 
contributions to the soft supersymmetry breaking parameters.\footnote{In
contrast, the FCNC could be a serious problem for the strong coupling limit of 
heterotic string theory ({\it i.e.}, M-theory compactified on 
R$^{10}$$\times$S$^{1}$/Z$_{2}$). As pointed out in \cite{aq97} recently,
in that case supersymmetry is broken in the 5D bulk; both the supergravity
multiplet and the bulk moduli play the roles of messengers for carrying the 
effects of supersymmetry breaking to the observable sector at one of the 
two boundaries. Therefore, the contribution to the soft supersymmetry breaking 
parameters from the bulk moduli is non-universal generally.} In other words,
simply as a consequence of modular invariance and an appropriate treatment of
gaugino condensation, we have a dilaton-dominated scenario of supersymmetry
breaking. Therefore, the weakly-coupled heterotic string offers a rationale 
for the well-known dilaton-dominated scenario of supersymmetry breaking in a
very elegant way. For the fourth problem, let's recall the standard lore of 
superstring phenomenology which tells us that, based on a very naive 
order-of-magnitude estimate, string dilaton and moduli gain from supersymmetry 
breaking masses of order of the gravitino mass. Since the gravitino mass is of 
order of the electroweak scale, these small masses 
of the dilaton and moduli lead to the cosmological moduli problem. On the other 
hand, our model is realistic enough for us to discuss these issues based on 
actual computations rather than educated guesses: it turns out that the string 
dilaton and moduli are in fact much heavier than the gravitino, which may be 
sufficient to solve the cosmological moduli problem \cite{review}. Furthermore,
the large entropy produced by the decays of the heavy moduli in our model will 
dilute the axion density and therefore raise the cosmological bound on the 
axion decay constant. As we shall see, this could solve the cosmological 
problem of the invisible axion. 

   Finally, let's make a brief comment on how the recent development of string 
duality might affect the status of weakly-coupled heterotic string theory. 
There have been claims in the literature in favor of the strongly-coupled 
heterotic string theory by arguing that it is unlikely that the weakly-coupled 
heterotic string theory can solve the dilaton runaway problem. However, the 
recent observation of string dualities actually implies that the strong 
coupling limit of heterotic string theory, which can be described by another 
weakly-coupled theory ({\it i.e.}, M-theory compactified on 
R$^{10}$$\times$S$^{1}$/Z$_{2}$ \cite{hw}), is plagued by a similar runaway 
problem\footnote{For example, one has to worry about the
runaway problem associated with the interval, $\rho_{11}$, along the 11th 
dimension. In particular, $\rho_{11}$ controls supersymmetry breaking, and
supersymmetry is unbroken in the limit $\rho_{11}\rightarrow\infty$.} 
\cite{stringduality}. Therefore, there seem to be only two logical 
options for solving the runaway problem: either a truly non-perturbative 
heterotic string theory which does not allow a weakly-coupled description, or 
a weakly-coupled theory ({\it i.e.}, the weakly-coupled heterotic string 
theory or the strong coupling limit of heterotic string theory). So far the 
first option remains a very interesting yet remote possibility.\footnote{Some 
recent attempts at a non-perturbative formulation of heterotic string theory 
can be found in \cite{Rey}.} On the other hand, as for the second option both 
the weakly-coupled heterotic string theory and the strong coupling limit of 
heterotic string theory certainly deserve further study\footnote{Although 
recently there is an argument of coupling unification preferring the strong 
coupling limit of heterotic string theory to the weakly-coupled heterotic 
string theory \cite{Witten96}, it involves assumptions that are not true 
generically. For example, it is assumed in \cite{Witten96} that the 
compactification scale, $V_{comp}^{-1/6}$, is of order $M_{GUT}^{(MSSM)}$, 
where $M_{GUT}^{(MSSM)}$ is the grand unification scale of the MSSM. However, 
in our model the moduli associated with compactification are stabilized at the 
self-dual point, and therefore the argument of \cite{Witten96} is not valid.
More details of this discussion can be found in Section 5.5}. It is first
proposed by T. Banks and M. Dine \cite{Banks94} that significant stringy 
non-perturbative effects could stabilize the dilaton. Based on this idea, it 
is our purpose here to show that the weakly-coupled heterotic string theory 
could solve the dilaton runaway problem as well as lead to a satisfactory 
phenomenology \cite{review}.\footnote{A similar point of view is advocated in
\cite{Casas96}. However, among the aforementioned problems of weakly-coupled
heterotic string, only the dilaton runaway problem was treated in 
\cite{Casas96}.}

   In Chapter 2, a simple string orbifold model with a hidden E$_{8}$ gauge 
group and no hidden matter is used to illustrate the studies of the linear
multiplet formalism, the incorporation of stringy non-perturbative effects, 
static gaugino condensation, and the dilaton runaway problem. In Chapter 3, 
we give the motivations for studying dynamical gaugino condensation, and then
show that static gaugino condensation is indeed the appropriate low-energy
effective description of dynamical gaugino condensation. In Chapter 4, we 
extend our previous studies to a generic string orbifold model. The resulting
model is generic and realistic enough, and we are therefore in a position to 
address several important phenomenological issues. In Chapter 5, we discuss
phenomenological issues such as the dilaton and moduli masses, axion physics,
soft supersymmetry breaking parameters, gauge coupling unification, as well 
as cosmological issues.
\newpage
\setcounter{chapter}{1}
\chapter{The Stringy Story of Gaugino Condensation}
\hspace{0.8cm}
\setcounter{equation}{0}
\setcounter{figure}{0}
\setcounter{footnote}{0}
\newpage

\section{Introduction}
\hspace{0.8cm}
Constructing a realistic scheme of supersymmetry breaking is one of the big
challenges to supersymmetry phenomenology. However, in the context of
superstring phenomenology, there are actually more challenges. As is well 
known, a very powerful feature of superstring phenomenology is that all 
the {\em parameters} of the model are in principle dynamically determined by 
the {\em vev}'s of certain fields. One of these important fields is the 
string dilaton whose {\em vev} determines the gauge coupling constants. On the
other hand, how the dilaton is stabilized is outside the reach of perturbation
theory since the dilaton's potential remains flat to all order in perturbation
theory according to the non-renormalization theorem. Therefore, understanding 
how the dilaton is stabilized ({\it i.e.}, how the gauge coupling constants are 
determined) is of no less significance than understanding how supersymmetry is 
broken. Gaugino condensation has been playing a unique role in these issues: 
Gaugino condensation not only breaks supersymmetry but also generates a
non-perturbative dilaton potential which may eventually stabilize the dilaton. 
Furthermore, gaugino condensation has its own important phenomenological 
motivations \cite{Nilles82,Dine85,messenger}. Unfortunately, this beautiful 
scheme of gaugino condensation has been long plagued by the infamous dilaton 
runaway problem \cite{Banks94,Dine85}. (The recent observation of string 
dualities further implies that the strong-coupling regime is plagued by a 
similar runaway problem \cite{stringduality}.) Only a few solutions to the 
dilaton runaway problem have been proposed. Assuming the scenario of two or 
more gaugino condensates, the racetrack model stabilizes the dilaton and 
breaks supersymmetry with a more complicated dilaton superpotential generated 
by multiple gaugino condensation \cite{racetrack}. However, stabilization of 
the dilaton in the racetrack model requires a delicate cancellation between 
the contributions from different gaugino condensates, which is not very 
natural. Furthermore, it has a large and negative cosmological constant when 
supersymmetry is broken. The other solutions generically require the 
presence of an additional source of supersymmetry breaking ({\it e.g.}, 
a constant term in the superpotential) \cite{Dine85,constant}. It is therefore
fair to say that there is no satisfactory solution so far.

Recently, there have been several new developments and insights in superstring 
phenomenology. It is our purpose to show that these new ingredients play 
important roles in the above issues and can eventually lead to a promising 
solution. One of these new ingredients is the linear multiplet formalism of
superstring effective theories \cite{Gates,Adamietz93,Derendinger94}: 
Among the massless string modes, a real scalar (dilaton), an antisymmetric 
tensor field (the Kalb-Ramond field) and their supersymmetric partners can be 
described either by a chiral superfield $S$ or by a linear multiplet $L$, 
which is known as the chiral-linear duality \cite{Linear}. By definition, the 
linear multiplet $L$ is a vector superfield that satisfies the following 
constraints \cite{Linear}:
\begin{eqnarray}
-(\DbDb-8R)L\,&=&\,0, \nonumber\\
-(\DaDa-8R^{\dagger})L\,&=&\,0.
\end{eqnarray}
The lowest component of $L$ is the dilaton field $\dilaton$, and its {\em vev} 
is related to the gauge coupling constant as follows\footnote{However, as we 
shall see in Section 2.2.2, this identification of gauge coupling constant in 
terms of $\langle\,\dilaton\,\rangle$ will be modified in the presence of stringy 
non-perturbative effects \cite{Shenker90}.}: 
$g^{2}(M_{s})\,=\,2\langle\,\dilaton\,\rangle$, where $M_{s}$ is the string scale 
\cite{Gaillard92,KL}. Although the chiral-linear duality is obvious at tree 
level, it becomes obscure when quantum effects are included. Although 
scalar-2-form field strength duality, which is contained in chiral-linear 
duality, has been shown to be preserved in perturbation theory~\cite{frad}, 
the situation is less clear in the presence of non-perturbative effects, which 
are important in the study of gaugino condensation. It has recently been 
shown~\cite{Binetruy95,Burgess95} that gaugino condensation can be formulated 
directly using a linear multiplet for the dilaton. Although a formal 
equivalence between the chiral and linear multiplet formalisms has been 
shown~\cite{Burgess95}, the content of the resulting chiral-linear duality 
transformation is in general very complicated. If there is an elegant 
description of gaugino condensates in the context of superstring effective 
theories, it may be simple in only one of these formalisms, but not 
in both. Therefore, a pertinent issue is: which formalism is better?
Here we will construct the effective theory of gaugino condensation directly 
in the linear multiplet formalism without referring to the chiral multiplet 
formalism. There is reason to believe that the linear multiplet formalism is 
in fact more appropriate. The stringy reason for choosing the linear multiplet 
formalism is that the precise field content of the linear multiplet appears 
in the massless string spectrum, and $\langle L\rangle$ plays the role of 
string loop expansion parameter. Therefore, string information is more 
naturally encoded in the linear multiplet formalism of string effective 
theory. Furthermore, as we will see in Chapter 2, stringy effects are believed 
to be important in the stabilization of the dilaton and supersymmetry breaking 
by gaugino condensation; therefore, it is more appropriate to study these 
issues in the linear multiplet formalism. 

The other new ingredient concerns the effective description of gaugino 
condensation. In the known models of gaugino condensation using the chiral 
superfield representation for the dilaton, the gaugino condensate has always 
been described by an {\em unconstrained} chiral superfield $U$ which 
corresponds to the bound state of $\,{\cal W}^{\alpha}{\cal W}_{\alpha}\,$ in 
the underlying theory. It was pointed out recently that $U$ should be a 
{\em constrained} chiral superfield \cite{Binetruy95,sduality,Burgess95,Pillon} 
due to the constrained superspace geometry of the underlying Yang-Mills theory:
\begin{eqnarray}
U\,&=&\,-(\DbDb-8R)V, \nonumber \\ 
\bar{U}\,&=&\,-(\DaDa-8R^{\dagger})V,
\end{eqnarray}
where $V$ is an unconstrained vector superfield. Furthermore, in the linear 
multiplet formalism the linear multiplet $L$ and the constrained $U$, $\bar{U}$
nicely merge into an unconstrained vector superfield $V$ \cite{Binetruy95}, 
and therefore the effective Lagrangian can elegantly be described by $V$ alone.

The third new ingredient is the stringy non-perturbative effect conjectured by
\mbox{S.H. Shenker} \cite{Shenker90}. It is further argued in \cite{Banks94} 
that the K\"ahler potential can in principle receive significant stringy
non-perturbative corrections although the superpotential cannot generically. 
Significant stringy non-perturbative corrections to the K\"ahler potential imply
that the usual dilaton runaway picture is valid only in the weak-coupling
regime; as pointed out in \cite{Banks94}, these corrections may naturally
stabilize the dilaton.\footnote{Choosing a specific form for possible 
non-perturbative corrections to the K\"ahler potential, \cite{Casas96} has
discussed the possibility of stabilizing the dilaton in a model of gaugino 
condensation using chiral superfield representation for the dilaton. 
However, neither the issue of modular anomaly cancellation nor the constraint
(2.2) was taken into account. As will be discussed in Section 4.7, these two 
issues are essential to string phenomenology.} 

In the next section we describe the linear multiplet formalism of string 
effective Yang-Mills theory, whose effective theory below the condensation 
scale is constructed and analyzed in Section 2.3. It is then shown in Section 
2.4 that supersymmetry is broken and the dilaton is stabilized in a large 
class of models of static gaugino condensation. Here we use the K\"{a}hler 
superspace formulation~\cite{Binetruy90} of supergravity, suitably extended to 
incorporate the linear multiplet~\cite{Binetruy91}.
\section{The Linear Multiplet Formalism}
\subsection{Effective Yang-Mills Theory from Superstring}
\hspace{0.8cm}
In the realm of superstring effective Yang-Mills theory, there are two 
important ingredients, namely, the symmetry group of modular transformations 
and the linear multiplet. In order to make the discussion as explicit as 
possible in this chapter, we consider here orbifolds with gauge 
group\footnote{As for phenomenological consideration, it is more desirable
to discuss a generic orbifold. Such a non-trivial generalization will be made
in Chapter 4.} $\mbox{E}_{8}\otimes\mbox{E}_{6}\otimes\mbox{U(1)}^{2}$, 
which have been studied most extensively in the context of modular symmetries 
\cite{Gaillard92,KL,Dixon90}. They contain three untwisted (1,1) moduli 
$T^{I}$, $I=1,\,2,\,3$, which transform under SL(2,Z) as follows:
\begin{equation}
T^{I}\;\rightarrow\;\frac{aT^{I}-ib}{icT^{I}+d},\;\;\;
ad-bc=1,\;\;\;a,b,c,d\;\in\mbox{Z}.
\end{equation}
The corresponding K\"{a}hler potential is
\begin{equation}
G\,=\,\sum_{I}g^{I}\,+\,
      \sum_{A}\exp(\sum_{I}q_{A}^{I}g^{I})|\Phi^{A}|^{2}\,+\,
      \CO(|\Phi|^{4}),
\end{equation}
where $g^{I}\,=\,-\ln(T^{I}+\bar{T}^{I})$, and the modular weights
$q_{A}^{I}$ depend on the particular matter field $\Phi^{A}$
as well as on the modulus $T^{I}$. However, it is well known 
that the effective theory obtained from the massless truncation 
of superstring is {\it not} invariant under the modular 
transformations (2.3) at one loop \cite{Derendinger92,Ovrut93}. 
Counterterms, that correspond to the result of integrating out massive modes, 
have to be added to the effective theory in order to restore modular invariance 
since string theory is known to be modular invariant to all orders of the loop 
expansion \cite{mod}. Two types of such counterterms have been discussed in 
the literature \cite{Gaillard92,Dixon90,Ovrut93}, the so-called $f$-type 
counterterms ({\it i.e.}, string threshold corrections) and the Green-Schwarz counterterm. 
The Green-Schwarz counterterm, which is analogous to the Green-Schwarz anomaly 
cancellation mechanism in D=10, is naturally implemented with the linear 
multiplet formalism~\cite{Linear}. In Chapters 2 and 3 we consider only those 
orbifolds in which the full modular anomaly is cancelled by the Green-Schwarz 
counterterm alone ({\it i.e.}, orbifolds with universal modular anomaly cancellation), 
and more generic orbifolds with both types of counterterms present will be 
considered in Chapter 4. Indeed, an orbifold has universal modular anomaly 
cancellation unless its modulus $T^{I}$ corresponds to an internal plane which 
is left invariant under some orbifold group transformations, which may happen 
only if an $N$=2 supersymmetric twisted sector is present~\cite{ant}. Therefore, 
a large class of orbifolds, including the $\mbox{Z}_{3}$ and $\mbox{Z}_{7}$ 
orbifolds, is under consideration in this chapter.

The antisymmetric tensor field of superstring theories undergoes Yang-Mills 
gauge transformations. In the effective theory, it can be incorporated into a 
gauge invariant vector superfield $L$, the so-called modified linear multiplet, 
coupled to the Yang-Mills degrees of freedom as follows:
\begin{eqnarray}
-(\DbDb-8R)L\,&=&\,(\DbDb-8R)\Omega\,=\,\sum_{a}\WaWa^{a}, 
\nonumber \\ 
-(\DaDa-8R^{\dagger})L\,&=&\,(\DaDa-8R^{\dagger})\Omega\,=\,
\sum_{a}\WbWb^{a},
\end{eqnarray}
where $\Omega$ is the Yang-Mills Chern-Simons superform. The summation extends 
over the indices $a$ numbering simple subgroups of the full gauge group. The 
modified linear multiplet $L$ contains the linear multiplet as well as the 
Chern-Simons superform, and its gauge invariance is ensured by imposing 
appropriate transformation properties for the linear multiplet. The generic 
lagrangian describing the linear multiplet coupled to supergravity and matter 
in the presence of Yang-Mills Chern-Simons superform is \cite{Gaillard92}:
\begin{eqnarray}
K\,&=&\,\ln L\,+\,g(L)\,+\,G, \nonumber \\
\Lag\,&=&\,\superint\,E\,\left\{\,-2\,+\,f(L)\,\right\}
\,+\,
\superint\,E\,\{\,bL\sum_{I}g^{I}\,\}, \\
b\,&=&\,\frac{\,C}{\,8\pi^{2}}\,=\,
        \frac{\,2\,}{\,3\,}b_{0},
\end{eqnarray}
where $L$ is the modified linear multiplet and $C\,=\,30$ is the Casimir 
operator in the adjoint representation of $\mbox{E}_{8}$. $b_{0}$ is the 
$\mbox{E}_{8}$ one-loop $\beta$-function coefficient. The first term of 
$\Lag$ is the superspace integral which yields the kinetic actions for the 
linear multiplet, supergravity, matter and Yang-Mills fields. The second term 
in (2.6) is the Green-Schwarz counterterm, which is ``minimal'' in the sense 
of \cite{Gaillard92}. Furthermore, arbitrariness in the two functions $g(L)$ 
and $f(L)$ is reduced by the requirement that the Einstein term in $\Lag$ be 
canonical. Under this constraint,  $g(L)$ and $f(L)$ are related to each other 
by the following first-order differential equation~\cite{Binetruy91}: 
\begin{equation}
L\frac{\diff g(L)}{\diff L}\,=\,
-L\frac{\diff f(L)}{\diff L}\,+\,f(L),
\end{equation}
The complete component lagrangian of (2.6) with the tree-level K\"{a}hler 
potential ({\it i.e.}, $g(L)=0$ and $f(L)=0$) has been presented in 
\cite{Adamietz93} based on the K\"{a}hler superspace formulation. Similar 
studies have also been performed in the superconformal formulation of 
supergravity \cite{Gates,Derendinger94}. In the following, we are interested 
in the effective lagrangian of (2.6) below the condensation scale.
\subsection{Stringy Effects versus Field-Theoretical Effects} 
\hspace{0.8cm}
In this section we would like to illustrate how stringy effects are naturally
incorporated with the superstring effective field theory using the linear 
multiplet formalism. Consider again the effective field theory defined at the 
string scale $M_{s}$. The quantum corrections, $g(L)$ and $f(L)$, to the 
tree-level K\"ahler potential of (2.6) are naturally interpreted as stringy 
effects. Indeed, in the context of superstring $L$ plays the role of string 
loop expansion parameter ({\it i.e.}, the string coupling), and therefore stringy 
effects are naturally parametrized by $L$. Although perturbative contributions 
to $g(L)$ and $f(L)$ are generically small, yet, as first pointed out by 
Shenker \cite{Shenker90}, there can be significant stringy non-perturbative 
contributions. It is then interesting to ask how the usual relation between the
dilaton $\dilaton$ and the gauge coupling constant of the effective field 
theory, $g^{2}(M_{s})\,=\,2\langle\,\dilaton\,\rangle$, might get modified in the 
presence of stringy effects? It is straightforward to compute the gauge 
coupling constant at the string scale, $g(M_{s})$, defined by the effective 
field theory (2.6) as follows:
\begin{equation}
g^{2}(M_{s})\,=\,
\left\langle\,\frac{2\dilaton}{\,1\,+\,f(\dilaton)\,}\,\right\rangle.
\end{equation}
Indeed, the presence of stringy effects do affect the usual interpretation of
the gauge coupling constant of the effective field theory in terms of the
string dilaton. More precisely, the linear multiplet formalism naturally 
distinguishes stringy effects from field-theoretical effects; that is, 
$\dilaton$ is the natural parametrization of stringy effects and 
$\left\langle\,2\dilaton/\left(1+f(\dilaton)\right)\,\right\rangle$ is on the
other hand the natural parametrization of field-theoretical effects. Therefore,
in the linear multiplet formalism of superstring effective field theory it is
much easier to keep track of both stringy and field-theoretical effects. 
As mentioned before, this unique feature of
linear multiplet formalism is crucial to our study here, since stringy
non-perturbative effects do play an important role in the stringy story of 
gaugino condensation. 

On the other hand, in the chiral multiplet formalism where the string dilaton is
described by a chiral superfield $S$ chiral superfield ($s=S\lowest$), $S$ has
to be re-defined order by order in perturbation, which is clear from the
perturbative chiral-linear duality. Furthermore, in the chiral multiplet
formalism there is no clear distinction between stringy effects and 
field-theoretical effects;
more precisely, we always have from the chiral multiplet formalism of the
superstring effective field theory
$\,g^{2}(M_{s})\,=\,\langle\,2/(s+\bar{s})\,\rangle$
even when stringy effects are included. One may also derive this result by a 
duality transformation from the linear multiplet formalism (2.6) to the
corresponding chiral multiplet formalism of (2.6). It has been shown 
\cite{Gaillard92} that $\,1/(S+\bar{S})\,$ corresponds to $\,L/(1+f)\,$ through
this duality transformation, and therefore the interpretations of 
$g^{2}(M_{s})$ in both formalisms are consistent with each other. In conclusion,
we emphasize the advantage of using the linear multiplet formalism over the
chiral multiplet formalism in telling the stringy story of gaugino 
condensation.\footnote{We emphasize that, in consideration of the
chiral-linear duality shown in \cite{Burgess95}, in principle the linear
multiplet formalism should be equivalent to the {\em constrained} chiral
multiplet formalism. However, as discussed in Section 2.1, the chiral-linear
duality is apt to be very complicated, especially when the full quantum
corrections are included; therefore, there should exist a formalism where the
physics allows a simpler description.} More evidence of this advantage will be 
discovered in the following sections.
\subsection{Low-Energy Effective Degrees of Freedom}
\hspace{0.8cm}
Below the condensation scale at which the gauge interaction becomes
strong, the effective lagrangian of the Yang-Mills sector can be 
described by a composite chiral superfield $U$, which corresponds 
to the chiral superfield $\WaWa$ of the underlying theory. (We 
consider here gaugino condensation of a simple gauge group.)
The scalar component of $U$ is naturally interpreted as the 
gaugino condensate. It was pointed out only recently that the composite 
field $U$ is actually a constrained chiral superfield 
\cite{sduality,Burgess95,Pillon}. The constraint on $U$ can be 
seen most clearly through the constrained superspace geometry of the 
underlying Yang-Mills theory. As a consequence of this constrained 
geometry, the chiral superfield $\WaWa$ and its hermitian conjugate 
$\WbWb$ satisfy the following constraint:
\begin{equation}
(\DaDa-24R^{\dagger})\WaWa\,-\,(\DbDb-24R)\WbWb
\,=\,\mbox{total derivative.}
\end{equation}
(2.10) has a natural interpretation in the context of a 3-form
supermultiplet, and indeed $\WaWa$ can be interpreted as the degrees 
of freedom of the 3-form field strength \cite{3form}. The explicit 
solution to the constraint (2.10) has been presented in \cite{Pillon},
and it allows us to identify the constrained chiral superfield $\WaWa$
with the chiral projection of an unconstrained vector superfield $L$:
\begin{eqnarray}
\WaWa\,&=&\,-(\DbDb-8R)L, \nonumber \\ 
\WbWb\,&=&\,-(\DaDa-8R^{\dagger})L.
\end{eqnarray}
Below the condensation scale, the constraint (2.10) is replaced by the 
following constraint on $U$ and $\bar{U}$:
\begin{equation}
(\DaDa-24R^{\dagger})U\,-\,(\DbDb-24R)\bar{U}
\,=\,\mbox{total derivative.} 
\end{equation}
Similarly, the solution to (2.12) allows us to identify the constrained
chiral superfield $U$ with the chiral projection of an unconstrained
vector superfield $V$:
\begin{eqnarray}
U\,&=&\,-(\DbDb-8R)V, \nonumber \\ 
\bar{U}\,&=&\,-(\DaDa-8R^{\dagger})V.
\end{eqnarray}
(2.13) is the explicit constraint on $U$ and $\bar{U}$.

In fact, the constraint on $U$ and $\bar{U}$ enters the linear multiplet 
formalism of gaugino condensation very naturally. As described in Section 
2.2.1, the linear multiplet formalism of supersymmetric Yang-Mills theory is 
described by a gauge-invariant vector superfield $L$ which satisfies
\begin{eqnarray}
-(\DbDb-8R)L\,&=&\,(\DbDb-8R)\Omega\,=\,\WaWa, \nonumber \\ 
-(\DaDa-8R^{\dagger})L\,&=&\,(\DaDa-8R^{\dagger})\Omega\,=\,
\WbWb.
\end{eqnarray}
For the linear multiplet formalism of the superstring effective lagrangian below
the condensation scale, (2.14) is replaced by
\begin{eqnarray}
-(\DbDb-8R)V\,&=&\,U,\nonumber \\ 
-(\DaDa-8R^{\dagger})V\,&=&\,\bar{U},
\end{eqnarray}
where $U$ is the gaugino condensate chiral superfield, and $V$ contains the 
linear multiplet as well as the ``fossil'' Chern-Simons superform. In view of 
(2.15), it is clear that the constraint on $U$ and $\bar{U}$ arises naturally 
in the linear multiplet formalism of gaugino condensation. Furthermore, the 
low-energy degrees of freedom ({\it i.e.}, the linear multiplet and the gaugino 
condensate) are nicely merged into a single vector superfield $V$, and 
therefore the linear multiplet formalism of gaugino condensation can elegantly 
be described by $V$ alone in the context of superstring. The detailed 
construction of the effective lagrangian for the vector superfield $V$ will be 
presented in the next section.
\section{Gaugino Condensation in Superstring Theory} 
\subsection{A Simple Model}
\hspace{0.8cm}
Constructing the linear multiplet formalism of gaugino condensation requires 
the specification of two functions of the vector superfield $V$, namely, the 
superpotential and the K\"{a}hler potential. In the linear multiplet formalism, 
there is no classical superpotential~\cite{sduality}, and the quantum 
superpotential originates from the non-perturbative effects of gaugino 
condensation. This non-perturbative superpotential, whose form was dictated by 
the anomaly structure of the underlying theory, was first obtained by 
Veneziano and Yankielowicz \cite{vy,tom,bg89,fmtv}. The details of its 
generalization to the case of matter coupled to $N$=1 supergravity in the 
K\"{a}hler superspace formulation has been presented in \cite{chiral91}, and 
the superpotential term in the Lagrangian reads: 
\begin{eqnarray}
\superint\,\frac{E}{R}\,e^{K/2}W_{VY}\,&=&\,
\superint\,\frac{E}{R}\,\frac{1}{8}bU\ln(e^{-K/2}U/\mu^{3}),
\nonumber\\
\superint\,\frac{E}{R^{\dagger}}\,e^{K/2}\bar{W}_{VY}\,&=&\,
\superint\,\frac{E}{R^{\dagger}}
\,\frac{1}{8}b\bar{U}\ln(e^{-K/2}\bar{U}/\mu^{3}),
\end{eqnarray}
where $\,U\,=\,-(\DbDb-8R)V\,$ is the 
constrained gaugino condensate chiral superfield with K\"{a}hler weight 2, and 
$\mu$ is a constant with dimension of mass that is left undetermined by the 
method of anomaly matching.

As for the K\"{a}hler potential for $V$, there is little knowledge beyond tree 
level. The best we can do at present is to treat all physically reasonable 
K\"{a}hler potentials on the same footing and to look for possible general 
features and/or interesting special cases. In particular, we are interested in 
a specific class of K\"ahler potentials where there are significant stringy 
non-perturbative corrections as pointed out in \cite{Shenker90,Banks94}.
Before discussing this general analysis, it is instructive to examine a simple 
yet un-realistic linear multiplet model for gaugino condensation defined as 
follows \cite{sduality}:
\begin{eqnarray}
K\,&=&\,\ln V\,+\,G, \nonumber \\
\Lag_{eff}\,&=&\,\superint\,E\,\{\,-2\,+\,bVG\,\} \,+\, 
\superint\,\frac{E}{R}\,e^{K/2}W_{VY} \,+\,
\superint\,\frac{E}{R^{\dagger}}\,e^{K/2}\bar{W}_{VY},
\nonumber \\
G\,&=&\,-\sum_{I}\ln(T^{I}+\bar{T}^{I}).
\end{eqnarray}
This simple model describes the effective theory for (2.6) below the 
condensation scale, where the K\"{a}hler potential of $V$ assumes its 
tree-level form. It is a ``static'' model of gaugino condensation in the sense 
that no kinetic term for $U$ is included. From the viewpoint of the anomaly 
structure, static as well as dynamical models of gaugino condensation are 
interesting in their own right. However, as will be discussed in Chapter 3, 
dynamical models rather than static models generically occur in the context of 
superstrings. Dynamical models of gaugino condensation in the linear multiplet 
formalism \cite{Binetruy95,Burgess95} have been studied less extensively.  On 
the other hand, as will also be shown in Chapter 3, after integrating out the 
heavy modes the static model of gaugino condensation is proven to be the 
appropriate effective description for the dynamical model\footnote{Unlike 
studies using the chiral multiplet formalism in the past, proving such a 
connection between static and dynamical gaugino condensation is much more 
non-trivial in the linear multiplet formalism with the constraint on $U$ 
incorporated consistently.}. Therefore, in Chapter 2 we will concentrate on 
static models of gaugino condensation, and there will be no loss of generality. 

With $\,U\,=\,-(\DbDb-8R)V\,$ and 
$\,\bar{U}\,=\,-(\DaDa-8R^{\dagger})V\,$, we can rewrite the superpotential 
terms of $\Lag_{eff}$ as a single $D$ term by superspace partial integration.
For example, for any chiral superfield $X$ of zero K\"ahler weight:
\begin{equation}
{1\over8}\superint\,{E\over R}U_a\ln X + {\rm h.c.}=
\superint\,E V_a\ln(X\bar{X}) $$ $$
- \pp_m\(\superint\,{E\ln X\over8R}\D_{\dot\alpha}
V_aE^{\dot\alpha m}  + {\rm h.c.}\) , 
\end{equation}
where $E^{\dot\alpha m}$ is an element of the supervielbein, and the 
total derivative on the right hand side contains the chiral anomaly 
($\propto \pp_mB^m \simeq F^a_{mn}{\tilde F}^{mn}_a $) of the $F$ term on the 
left hand side. Therefore, up to a total derivative, the simple model (2.17) 
can be rewritten as follows:
\begin{eqnarray}
K\,&=&\,\ln V\,+\,G, \nonumber \\
\Lag_{eff}\,&=&\,\superint\,E\,\{\,-2 \,+\, bVG \,+\,
bV\ln(e^{-K}\bar{U}U/\mu^{6})\,\}.
\end{eqnarray}
In (2.19), the modular anomaly cancellation by the Green-Schwarz 
counterterm is transparent~\cite{sduality}. The Green-Schwarz counterterm 
$\,bVG\,$ 
and the superpotential $D$ term $\,bV\ln(e^{-K}\bar{U}U/\mu^{6})\,$ are {\em not} 
modular invariant separately, but their sum is modular invariant, which 
ensures the modular invariance of the full theory. In fact, the Green-Schwarz 
counterterm cancels the $T^{I}$ moduli-dependence of the superpotential 
completely. This is a unique feature of the linear multiplet formalism, and, 
as we will see later, has interesting implications for the moduli-dependence 
of physical quantities. 

Throughout this paper only the bosonic and gravitino parts of the component 
lagrangian are presented, since we are interested in the vacuum configuration
and the gravitino mass. In the following, we enumerate the definitions of 
bosonic component fields of the vector superfield $V$.
\begin{eqnarray}
\dilaton\,&=&\,V\lowest,\nonumber\\
\sigma^{m}_{\alpha\dot{\alpha}}B_{m}\,&=&\,
\frac{1}{2}[\,\Diff_{\alpha},\Diff_{\dot{\alpha}}\,]V\lowest\,+\,
\frac{2}{3}\dilaton\sigma^{a}_{\alpha\dot{\alpha}} b_{a},\nonumber\\
u\,&=&\,U\lowest\,=\,-(\bar{\Diff}^{2}-8R)V\lowest,\nonumber\\
\bar{u}\,&=&\,\bar{U}\lowest\,=\,-(\Diff^{2}-8R^{\dagger})V\lowest,
\nonumber \\
D\,&=&\,\frac{1}{8}\Diff^{\beta}(\bar{\Diff}^{2}-8R)
      \Diff_{\beta}V\lowest\nonumber\\
   &=&\,\frac{1}{8}\Diff_{\dot{\beta}}(\Diff^{2}-8R^{\dagger})
      \Diff^{\dot{\beta}}V\lowest,
\end{eqnarray}
where 
\begin{equation}
-\,\frac{1}{6}M\,=\,R\lowest,\;\;
-\,\frac{1}{6}\bar{M}\,=\,R^{\dagger}\lowest,\;\;
-\,\frac{1}{3}b_{a}\,=\,G_{a}\lowest
\end{equation}
are the auxiliary components of supergravity multiplet. It is 
convenient to write the lowest components of $\Diff^{2}U$ and 
$\bar{\Diff}^{2}\bar{U}$ as follows:
\begin{equation}
-4F_{U}\,=\,\Diff^{2}U\lowest, \;\;\; 
-4\bar{F}_{\bar{U}}\,=\,\bar{\Diff}^{2}\bar{U}\lowest.
\end{equation}
$(F_{U}-\bar{F}_{\bar{U}})$ can be explicitly expressed as follows:
\begin{equation}
(F_{U}-\bar{F}_{\bar{U}})\,=\,4i\nabla^{m}\!B_{m}
\,+\,u\bar{M}\,-\,\bar{u}M.
\end{equation}
The expression for $(F_{U}+\bar{F}_{\bar{U}})$ contains the auxiliary 
field $D$. The bosonic components of $T^{I}$ and $\bar{T}^{I}$ are
\begin{eqnarray}
t^{I}\,&=&\,T^{I}\lowest,\;\;\;
-4F^{I}\,=\,\Diff^{2}T^{I}\lowest,\nonumber\\
\bar{t}^{I}\,&=&\,\bar{T}^{I}\lowest,\;\;\;
-4\bar{F}^{I}\,=\,\bar{\Diff}^{2}\bar{T}^{I}\lowest.
\end{eqnarray}
We leave the details of constructing the component lagrangian for this simple 
model (in the K\"{a}hler superspace formulation) to Section 2.3.2, and present 
here only the scalar potential obtained from eliminating the auxiliary fields 
in the boson Lagrangian given in (2.46) below:
\begin{equation}
V_{pot}\,=\,\frac{1}{16e^{2}\dilaton}
(\,1\,+\,2b\dilaton\,-\,2b^{2}\dilaton^{2}\,)\mu^{6}e^{-\,1/{b\dilaton}}.
\end{equation}
Eq. (2.25) agrees with the result obtained in~\cite{Binetruy95}, where the 
model defined by (2.17) was studied for the case of a single modulus using the 
superconformal formulation of supergravity. 

However, this simple model is not viable. As expected, the weak-coupling limit 
$\dilaton=0$ is always a minimum. As shown in Fig. 2.1, the scalar potential 
starts with $V_{pot}=0$ at $\dilaton=0$, first rises and then falls without 
limit as $\dilaton$ increases. 
\begin{figure}
\epsfxsize=12cm
\epsfysize=8cm
\epsfbox{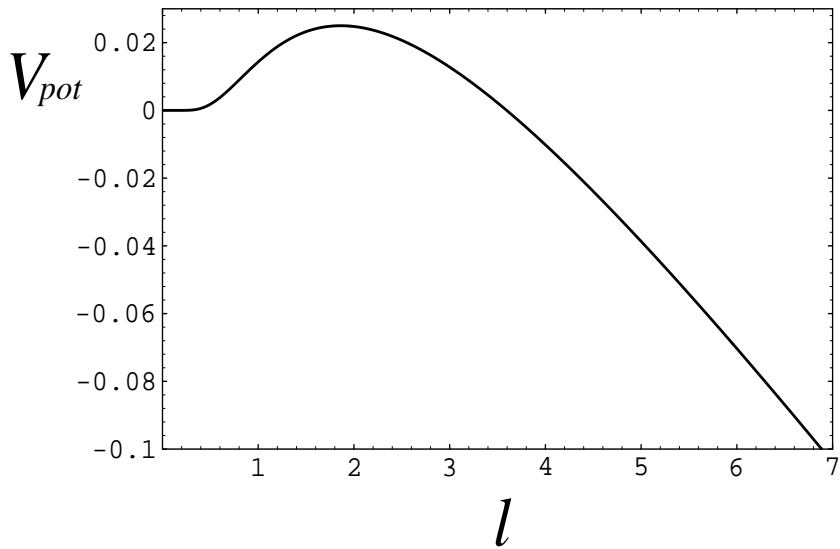}
\caption{The scalar potential $V_{pot}$ (in reduced Planck units)
            is plotted versus the dilaton $\dilaton$. $\mu$=1.}
\end{figure}
Therefore, $V_{pot}$ is unbounded from below, and this simple model has no 
well-defined vacuum. This may be somewhat surprising because the model 
defined by (2.17) superficially appears to be of the no-scale type: 
the Green-Schwarz counterterm, that destroys the no-scale property of chiral 
models and destabilizes the potential, is cancelled here by quantum effects 
that induce a potential for the condensate.  However the resulting quantum 
contribution to the Lagrangian (2.19), $bV\ln(U\bar{U}/V)$, has an implicit 
$T^I$-dependence through the superfield $U$ due to its nonvanishing K\"ahler 
weight: $w(U)= 2$. This implicit moduli-dependence is a consequence of the 
anomaly matching condition, and parallels the construction of the effective 
theory in the chiral multiplet formalism~\cite{vy,tom,bg89,fmtv} which is also not 
of the no-scale form once the Green-Schwarz counterterm is included.  

If we take a closer look at (2.25), it is clear that the unboundedness of 
$V_{pot}$ in the strong-coupling limit $\:\dilaton\,\rightarrow\,\infty\:$
is caused by a term of two-loop order: $\:-2b^{2}\dilaton^{2}$. This 
observation strongly suggests that the underlying reason for unboundedness
is our poor control over the model in the strong-coupling regime. 
The form of the superpotential $W_{VY}$ is completely fixed by the underlying 
anomaly structure.  However the K\"{a}hler potential is much less constrained,
and the choice (2.17) cannot be expected to be valid in the strong-coupling 
regime where the non-perturbative contributions should not be ignored. We
conclude that the unboundedness shown in Fig. 2.1 simply reflects the 
importance of non-perturbative contributions to the K\"{a}hler potential. In 
particular, it is natural to expect that the stringy non-perturbative effects 
conjectured by Shenker \cite{Shenker90,Banks94} are the non-perturbative 
contributions to the K\"{a}hler potential ignored in this simple model. 
In the absence of a better knowledge of the exact K\"{a}hler potential, we 
will consider models with generic K\"{a}hler potentials in the following 
sections. 
\subsection{General Static Model}
\hspace{0.8cm}
In this section, we show how to construct the component lagrangian
for generic linear multiplet models of static gaugino condensation in the 
K\"{a}hler superspace formulation. Further computational details can be 
found in \cite{Adamietz93,Binetruy90}. Although our results can probably be 
rephrased in the chiral multiplet formalism, the equivalent chiral multiplet
formalism are expected to be rather complicated because of the constraint on 
the gaugino condensate chiral superfield $U$.  Quite generally we do not 
expect a simple ansatz in one formalism to appear simple in the other.

As suggested in Section 2.3.1, we extend the simple model in (2.17) to linear 
multiplet models of static gaugino condensation with generic K\"{a}hler 
potentials defined as follows: 
\begin{eqnarray}
K\,&=&\,\ln V\,+\,g(V)\,+\,G, \nonumber \\
\Lag_{eff}\,&=&\,\superint\,E\,\{\,(\,-2\,+\,f(V)\,)\,+\, 
bVG\,+\,bV\ln(e^{-K}\bar{U}U/\mu^{6})\,\}.\hspace{1.5cm}
\end{eqnarray}
For convenience, we also write $\;\ln V\,+\,g(V)\,\equiv\,k(V).\;$ 
$g(V)$ and $f(V)$ represent quantum corrections to the tree-level 
K\"{a}hler potential. Here we have chosen to keep the K\"ahler potential under 
discussion as generic as possible. However, as suggested by \cite{Banks94}, 
stringy non-perturbative corrections to the K\"ahler potential are probably the 
most important non-perturbative corrections. And, as we have discussed in detail 
in Section 2.2.2, such stringy non-perturbative corrections can be nicely 
parametrized by $g(V)$ and $f(V)$ using the linear multiplet formalism. 
According to (2.8), $g(V)$ and $f(V)$ are unambiguously related to each other 
by the following first-order differential equation:
\begin{equation}
V\frac{\diff g(V)}{\diff V}\,=\,
-V\frac{\diff f(V)}{\diff V}\,+\,f,
\end{equation}
\begin{equation}
g(V=0)\,=\,0 \;\;\;\mbox{and}\;\;\; f(V=0)\,=\,0. 
\end{equation}
The boundary condition of $g(V)$ and $f(V)$ at $V=0$ (the weak-coupling limit) 
is fixed by the tree-level K\"{a}hler potential. Before trying to specify 
$g(V)$ and $f(V)$, it is reasonable to assume for the present that $g(V)$ and 
$f(V)$ are arbitrary but bounded.

In the construction of the component field lagrangian, we use the chiral 
density multiplet method~\cite{Binetruy90}, which provides us with the locally 
supersymmetric generalization of the $F$ term construction in global 
supersymmetry. The chiral density multiplet ${\bf r}$ and its hermitian 
conjugate ${\bf \bar{r}}$ for the generic model in (2.26) are: 
\begin{eqnarray}
{\bf r}\,&=&\,-\,\frac{1}{8}(\bar{\Diff}^{2}-8R)\{\,(\,-2\,+\,f(V)\,)
\,+\,bVG\,+\,bV\ln(e^{-K}\bar{U}U/\mu^{6})\,\}, \nonumber\\
{\bf \bar{r}}\,&=&\,-\,\frac{1}{8}(\Diff^{2}-8R^{\dagger})\{\,(\,-2\,
+\,f(V)\,)\,+\,bVG\,+\,bV\ln(e^{-K}\bar{U}U/\mu^{6})\,\}.
\hspace{1.5cm}
\end{eqnarray}
In order to obtain the component lagrangian $\Lag_{eff}$, we need 
to work out the following expression
\begin{eqnarray}
\frac{1}{e}\Lag_{eff}\,&=&\,-\,\frac{1}{4}\Diff^{2}{\bf r}
\lowest\,+\,\frac{i}{2}(\bar{\psi}_{m}\bar{\sigma}^{m})^{\alpha}
\Diff_{\alpha}{\bf r}\lowest\nonumber\\
& &\,-\,(\bar{\psi}_{m}\bar{\sigma}^{mn}\bar{\psi}_{n}+\bar{M})
{\bf r}\lowest \,\,+\,\, \mbox{h.c.}
\end{eqnarray}
An important point in the computation of (2.30) is the evaluation of the 
component field content of the K\"{a}hler supercovariant derivatives, a rather 
tricky process. The details of this computation have by now become general 
wisdom and we can to a large extent rely on the existing literature 
\cite{Wess}. In particular, the Lorentz transformation and the K\"{a}hler 
transformation are incorporated in a very similar way in the K\"{a}hler 
superspace formulation, and the Lorentz connection as well as the so-called 
K\"{a}hler connection $A_{M}$ are incorporated into the K\"{a}hler 
supercovariant derivatives in a concise and constructive way. The K\"{a}hler
connection $A_{M}$ is not an independent field but rather expressed in terms 
of the K\"{a}hler potential $K$ as follows:
\begin{equation}
A_{\alpha}\,=\,\frac{1}{4}E_{\alpha}^{\hs M}\partial_{M}K,\;\;\;
A_{\dot{\alpha}}\,=\,-\,\frac{1}{4}
E_{\dot{\alpha}}^{\hs M}\partial_{M}K,
\end{equation}
\begin{equation}
\sigma^{a}_{\alpha\dot{\alpha}}A_{a}\,=\,
\frac{3}{2}i\sigma^{a}_{\alpha\dot{\alpha}}G_{a}\,-\,
\frac{1}{8}i[\,\Diff_{\alpha},\Diff_{\dot{\alpha}}\,]K.
\end{equation}
In order to extract the explicit form of the various couplings, we choose to 
write out explicitly the vectorial part of the K\"{a}hler connection and keep 
only the Lorentz connection in the definition of covariant derivatives when we 
present the component expressions. In the following, we give the lowest 
component of the vectorial part of the K\"{a}hler connection 
$\: A_{m}\lowest\:$ for our generic static model.
\begin{equation}
A_{m}\,=\,e_{m}^{\hs a}A_{a}\,+\,
\frac{1}{2}\psi_{m}^{\hs\alpha}A_{\alpha}\,+\,
\frac{1}{2}\bar{\psi}_{m\dot{\alpha}}A^{\dot{\alpha}}.
\end{equation}

\begin{eqnarray}
A_{m}\lowest\,&=&\,-\,\frac{i}{4\dilaton}(\dilaton\dg+1)B_{m}\,+\,
\frac{i}{6}(\dilaton\dg-2)e_{m}^{\hs a}b_{a}\nonumber\\
& &\,+\,\sum_{I}\frac{1}{4(t^{I}+\bar{t}^{I})}
(\nabla_{\!m}\bar{t}^{I}\,-\,\nabla_{\!m}t^{I}).
\end{eqnarray}

\begin{eqnarray}
\dg\,&=&\,\frac{\diff g(V)}{\diff V}\lowest, \;\;\;\;\;
\dgg\,=\,\frac{\diff^2 g(V)}{\diff V^2}\lowest,\nonumber\\
\df\,&=&\,\frac{\diff f(V)}{\diff V}\lowest, \;\;\;\;\;
\dff\,=\,\frac{\diff^2 f(V)}{\diff V^2}\lowest.
\end{eqnarray}

Another hallmark of the K\"{a}hler superspace formulation are the chiral 
superfield $X_{\alpha}$ and the antichiral superfield $\bar{X}^{\dot{\alpha}}$.
They arise in complete analogy with usual supersymmetric abelian gauge theory 
except that now the corresponding vector superfield is replaced by the 
K\"{a}hler potential:
\begin{eqnarray}
X_{\alpha}\,&=&\,-\,\frac{1}{8}(\DbDb-8R)
\Diff_{\alpha}K,\nonumber\\
\bar{X}^{\dot{\alpha}}\,&=&\,-\,\frac{1}{8}(\DaDa-8R^{\dagger})
\Diff^{\dot{\alpha}}\!K.
\end{eqnarray}
In the computation of (2.30), we need to decompose the lowest components of 
the following six superfields:
$X_{\alpha}$, $\bar{X}^{\dot{\alpha}}$, $\Diff_{\alpha}R$,
$\Diff^{\dot{\alpha}}\!R^{\dagger}$, 
$(\Diff^{\alpha}\!X_{\alpha}+\Diff_{\dot{\alpha}}\bar{X}^{\dot{\alpha}})$
and $(\Diff^{2}\!R+\bar{\Diff}^{2}\!R^{\dagger})$ into component 
fields. This is done by solving the following six simple algebraic equations:
\begin{eqnarray}
(V\frac{\diff g}{\diff V}+1)\Diff_{\alpha}R\,+\,
X_{\alpha}\,&=&\,\Xi_{\alpha},\\
3\Diff_{\alpha}R\,+\,X_{\alpha}\,&=&\,
-2(\sigma^{cb}\epsilon)_{\alpha\varphi}T_{cb}^
{\hs\hspace{0.04cm}\varphi}.
\end{eqnarray}

\begin{eqnarray}
(V\frac{\diff g}{\diff V}+1)\Diff^{\dot{\alpha}}\!R^{\dagger}
\,+\,\bar{X}^{\dot{\alpha}}\,&=&\,\bar{\Xi}^{\dot{\alpha}},\\
3\Diff^{\dot{\alpha}}\!R^{\dagger}\,+\,\bar{X}^{\dot{\alpha}}\,&=&\,
-2(\bar{\sigma}^{cb}\epsilon)^{
\dot{\alpha}\dot{\varphi}}T_{cb\dot{\varphi}}.
\end{eqnarray}

\begin{eqnarray}
(V\frac{\diff g}{\diff V}+1)(\Diff^{2}\!R+\bar{\Diff}^{2}\!
R^{\dagger})\,+\,(\Diff^{\alpha}\!X_{\alpha}+
\Diff_{\dot{\alpha}}\bar{X}^{\dot{\alpha}})\,&=&\,\Delta,\\
3(\Diff^{2}\!R+\bar{\Diff}^{2}\!R^{\dagger})\,+\,(\Diff^{\alpha}
\!X_{\alpha}+\Diff_{\dot{\alpha}}\bar{X}^{\dot{\alpha}})\,&=&\,
-2R_{ba}^{\hs ba}\,+\,12G^{a}G_{a} \hspace{0cm}\nonumber\\
\,& &\,+\,96RR^{\dagger}. \hspace{0cm}
\end{eqnarray}
The identities (2.38), (2.40) and (2.42) arise solely from the structure of 
K\"{a}hler superspace. (2.38) and (2.40) involve the torsion superfields 
$T_{cb}^{\hs\hspace{0.04cm}\varphi}$ and $T_{cb\dot{\varphi}}$, which in their 
lowest components contain the curl of the Rarita-Schwinger field. The 
identities (2.37), (2.39) and (2.41) arise directly from the definitions of 
$X_{\alpha}$, $\bar{X}^{\dot{\alpha}}$, 
$(\Diff^{\alpha}\!X_{\alpha}+\Diff_{\dot{\alpha}}\bar{X}^{\dot{\alpha}})$,
and therefore they depend on the K\"{a}hler potential explicitly. 
Computing $X_{\alpha}$, $\bar{X}^{\dot{\alpha}}$ and 
$(\Diff^{\alpha}\!X_{\alpha}+\Diff_{\dot{\alpha}}\bar{X}^{\dot{\alpha}})$
according to (2.36) defines the contents of $\Xi_{\alpha}$, 
$\bar{\Xi}^{\dot{\alpha}}$ and $\Delta$ respectively. In the following, we 
present the component field expressions of the lowest components of 
$\Xi_{\alpha}$, $\bar{\Xi}^{\dot{\alpha}}$ and $\Delta$.
\begin{eqnarray}
& &\,\frac{i}{2}(\bar{\psi}_{m}\bar{\sigma}^{m})^{\alpha}
\Xi_{\alpha}\lowest\,-\,\frac{i}{2}\bar{\Xi}_{\dot{\alpha}}
(\bar{\sigma}^{m}\psi_{m})^{\dot{\alpha}}\lowest \nonumber\\
&=&\,-\,\frac{1}{8\dilaton}(\dilaton\dg+1)(\,\bar{u}\,+\,
\frac{4}{3}\dilaton\bar{M}\,)(\psi_{m}\sigma^{mn}\psi_{n})
\nonumber\\
& &\,-\,\frac{1}{8\dilaton}(\dilaton\dg+1)(\,u\,+\,\frac{4}{3}\dilaton M\,)
(\bar{\psi}_{m}\bar{\sigma}^{mn}\bar{\psi}_{n})
\nonumber\\
& &\,+\,\frac{i}{4\dilaton}(\dilaton\dg+1)
(\,\eta^{mn}\eta^{pq}\,-\,\eta^{mq}\eta^{np}\,)
(\bar{\psi}_{m}\bar{\sigma}_{n}\psi_{p})\,\nabla_{\!q}\dilaton
\nonumber\\
& &\,+\,\frac{i}{6}(\dilaton\dg+1)\epsilon^{mnpq}
(\bar{\psi}_{m}\bar{\sigma}_{n}\psi_{p})
e_{q}^{\hspace{0.12cm} a}b_{a}
\nonumber\\
& &\,-\,\frac{i}{4\dilaton}(\dilaton\dg+1)\epsilon^{mnpq}
(\bar{\psi}_{m}\bar{\sigma}_{n}\psi_{p})B_{\!q}
\nonumber\\
& &\,-\,\frac{1}{4}(\Diff^{a}\Diff^{\alpha}k)
\psi_{a\alpha}\lowest
\,-\,\frac{1}{4}\bar{\psi}_{a\dot{\alpha}}
(\Diff^{a}\Diff^{\dot{\alpha}}k)\lowest.
\end{eqnarray}
The way $\Xi_{\alpha}\lowest$ and $\bar{\Xi}^{\dot{\alpha}}\lowest$
are presented in (2.43) will be useful for the computation of (2.30).
\begin{eqnarray}
& &\,\Delta\lowest \nonumber\\
&=&\,-\,\frac{1}{\dilaton^{2}}(\dilaton^{2}\!\dgg-1)
\nabla^{m}\!\dilaton\,\nabla_{\!m}\!\dilaton 
\,+\,\frac{1}{\dilaton^{2}}(\dilaton^{2}\!\dgg-1)B^{m}\!B_{m}
\nonumber\\
& &\,+\,4\sum_{I}\frac{1}{(t^{I}+\bar{t}^{I})^{2}}
\nabla^{m}\bar{t}^{I}\,\nabla_{\!m}t^{I}
\,-\,\frac{4}{9}(\dilaton^{2}\!\dgg-\dilaton\dg-2)\bar{M}\!M
\nonumber\\
& &\,+\,\frac{4}{9}(\dilaton^{2}\!\dgg+2\dilaton\dg+1)b^{a}b_{a}
\,-\,4\sum_{I}\frac{1}{(t^{I}+\bar{t}^{I})^{2}}
\bar{F}^{I}F^{I}
\nonumber\\
& &\,-\,\frac{4}{3\dilaton}(\dilaton^{2}\!\dgg+\dilaton\dg)
B^{m}e_{m}^{\hs a}b_{a}
\,-\,\frac{1}{2\dilaton}(\dilaton\dg+1)(F_{U}+\bar{F}_{\bar{U}})
\nonumber\\
& &\,-\,\frac{1}{6\dilaton}(2\dilaton^{2}\!\dgg-\dilaton\dg-3)
(\,u\bar{M}\,+\,\bar{u}M\,)
\,-\,\frac{1}{4\dilaton^{2}}(\dilaton^{2}\!\dgg-1)\bar{u}u
\nonumber\\
& &\,+\,2\nabla^{m}\!\nabla_{\!m}k
\,-\,(\Diff^{a}\Diff^{\alpha}k)
\psi_{a\alpha}\lowest
\,-\,\bar{\psi}_{a\dot{\alpha}}
(\Diff^{a}\Diff^{\dot{\alpha}}k)\lowest.
\end{eqnarray}
It is unnecessary to decompose the last two terms in (2.43) and in 
(2.44) because they eventually cancel with one another. 

Eqs.(2.31--44) describe the key steps involved in the computation of (2.30). 
The rest of it is standard and will not be detailed here. In the following, we 
present the component field expression of $\Lag_{eff}$ as the sum of the 
bosonic part $\Lag_{B}$ and the gravitino part $\Lag_{\tilde{G}}$ as 
follows.\footnote{Only the bosonic and gravitino parts of the component field 
expressions are presented here.}
\begin{equation}
\Lag_{eff}\,=\,\Lag_{B}\,+\,\Lag_{\tilde{G}}.
\end{equation}

\begin{eqnarray}
\frac{1}{e}\Lag_{B}\,&=&\,-\,\frac{1}{2}{\cal R}
\,-\,\frac{1}{4\dilaton^{2}}(1+\dilaton\dg)
\nabla^{m}\!\dilaton\,\nabla_{\!m}\!\dilaton
\nonumber\\
& &\,+\,\frac{1}{4\dilaton^{2}}(\dilaton\dg+1)B^{m}\!B_{m}
\,-\,(1+b\dilaton)\sum_{I}\frac{1}{(t^{I}+\bar{t}^{I})^{2}}
\nabla^{m}\bar{t}^{I}\,\nabla_{\!m}t^{I}
\nonumber\\
& &\,+\,\frac{1}{9}(\dilaton\dg-2)\bar{M}\!M
\,-\,\frac{1}{9}(\dilaton\dg-2)b^{a}b_{a}
\nonumber\\
& &\,+\,(1+b\dilaton)\sum_{I}\frac{1}{(t^{I}+\bar{t}^{I})^{2}}
\bar{F}^{I}F^{I}
\nonumber\\
& &\,+\,\frac{1}{8\dilaton}\{\,1\,+\,f\,+\,b\dilaton\ln(e^{-k}\bar{u}u/
        \mu^{6})\,+\,2b\dilaton\,\}(F_{U}+\bar{F}_{\bar{U}})
\nonumber\\
& &\,-\,\frac{1}{8\dilaton}\{\,1\,+\,f\,+\,b\dilaton\ln(e^{-k}\bar{u}u/
\mu^{6})\,+\,\frac{2}{3}b\dilaton(\dilaton\dg+1)\,\}
(\,u\bar{M}\,+\,\bar{u}M\,)
\nonumber\\
& &\,-\,\frac{1}{16\dilaton^{2}}(1+2b\dilaton)(1+\dilaton\dg)\bar{u}u
\nonumber\\
& &\,-\,\frac{i}{2}b\ln(\frac{\bar{u}}{u})\nabla^{m}\!B_{m}
\,-\,\frac{i}{2}b\sum_{I}\frac{1}{(t^{I}+\bar{t}^{I})}
(\,\nabla^{m}\bar{t}^{I}\,-\,\nabla^{m}t^{I}\,)B_{m}.
\end{eqnarray}

\begin{eqnarray}
\frac{1}{e}\Lag_{\tilde{G}}\,&=&\,\frac{1}{2}\epsilon^{mnpq}
(\,\bar{\psi}_{m}\bar{\sigma}_{n}\!\nabla_{\!p}\psi_{q}\,-\,
\psi_{m}\sigma_{n}\!\nabla_{\!p}\bar{\psi}_{q}\,)
\nonumber\\
& &\,-\,\frac{1}{8\dilaton}\{\,1\,+\,f\,+\,b\dilaton\ln(e^{-k}\bar{u}u/
\mu^{6})\,\}\,\bar{u}\,(\psi_{m}\sigma^{mn}\psi_{n})
\nonumber\\
& &\,-\,\frac{1}{8\dilaton}\{\,1\,+\,f\,+\,b\dilaton\ln(e^{-k}\bar{u}u/
\mu^{6})\,\}\,u\,(\bar{\psi}_{m}\bar{\sigma}^{mn}\bar{\psi}_{n})
\nonumber\\
& &\,-\,\frac{1}{4}(1+b\dilaton)\sum_{I}\frac{1}{(t^{I}+\bar{t}^{I})}
\epsilon^{mnpq}(\bar{\psi}_{m}\bar{\sigma}_{n}\psi_{p})
(\,\nabla_{\!q}\bar{t}^{I}\,-\,\nabla_{\!q}t^{I}\,)
\nonumber\\
& &\,+\,\frac{i}{4\dilaton}(1+b\dilaton)(1+\dilaton\dg)
(\,\eta^{mn}\eta^{pq}\,-\,\eta^{mq}\eta^{np}\,)
(\bar{\psi}_{m}\bar{\sigma}_{n}\psi_{p})\,\nabla_{\!q}\dilaton
\nonumber\\
& &\,-\,\frac{i}{4}b\dilaton
(\,\eta^{mn}\eta^{pq}\,-\,\eta^{mq}\eta^{np}\,)
(\bar{\psi}_{m}\bar{\sigma}_{n}\psi_{p})\,
\nabla_{\!q}\ln(\bar{u}u)
\nonumber\\
& &\,+\,\frac{1}{4}b\dilaton\,\epsilon^{mnpq}
(\bar{\psi}_{m}\bar{\sigma}_{n}\psi_{p})\,
\nabla_{\!q}\ln(\frac{\bar{u}}{u}).
\end{eqnarray}
For completeness, we also give the definitions of covariant derivatives:
\begin{eqnarray}
\nabla_{\!m}\!\dilaton\,&=&\,\partial_{m}\!\dilaton,\;\;\;
\nabla_{\!m}t^{I}\,=\,\partial_{m}t^{I},\;\;\;
\nabla_{\!m}\bar{t}^{I}\,=\,\partial_{m}\bar{t}^{I},
\nonumber\\
\nabla_{\!m}\psi_{n}^{\hs\alpha}\,&=&\,
\partial_{m}\psi_{n}^{\hs\alpha}\,+\,
\psi_{n}^{\hs\beta}\omega_{m\beta}^{\hs\hs\alpha},
\;\;\;
\nabla_{\!m}\bar{\psi}_{n\dot{\alpha}}\,=\,
\partial_{m}\bar{\psi}_{n\dot{\alpha}}\,+\,
\bar{\psi}_{n\dot{\beta}}\,\omega_{m\hs\dot{\alpha}}
^{\hs\dot{\beta}}. \hspace{2cm}
\end{eqnarray}

To proceed further, we need to eliminate the auxiliary fields from 
$\Lag_{eff}$ through their equations of motion. The equation of motion of the 
auxiliary field $(F_{U}+\bar{F}_{\bar{U}})$ is
\begin{equation}
f\,+\,1\,+\,b\dilaton\ln(e^{-k}\bar{u}u/\mu^{6})\,+\,
2b\dilaton\,=\,0.
\end{equation}
Eq. (2.49) implies that in static models of gaugino condensation the auxiliary 
field $\bar{u}u$ is expressed in terms of dilaton $\dilaton$. The equations of 
motion of $F^{I}$, $\bar{F}^{I}$ and the auxiliary fields 
$b^{a}$, $M$, $\bar{M}$ of the supergravity multiplet are (if 
$\dilaton\dg-2 \neq 0$)
\begin{eqnarray}
F^{I}\,&=&\,0, \;\;\;
\bar{F}^{I}\,=\,0,\nonumber\\
b^{a}\,&=&\,0, \nonumber\\
M\,&=&\,\frac{\, 3\,}{\, 4\,}bu,\;\;\;
\bar{M}\,=\,\frac{\, 3\,}{\, 4\,}b\bar{u}.
\end{eqnarray}
Now we are left with only one auxiliary field to eliminate, where this 
auxiliary field can be either $\,\,i\ln({\bar{u}}/{u})$ or $B_{m}$. This 
corresponds to the fact that there are two ways to perform duality 
transformation. If we take $\,\,i\ln({\bar{u}}/{u})$ to be auxiliary, its 
equation of motion is
\begin{equation}
\nabla_{\!q}\{\,B^{q}\,-\,\frac{i}{2}\dilaton\,\epsilon^{mnpq}
(\bar{\psi}_{m}\bar{\sigma}_{n}\psi_{p})\,\}\,=\,0,
\end{equation}
which ensures that $\,\{B^{q}\,-\,\frac{i}{2}\dilaton\,\epsilon^{mnpq}
(\bar{\psi}_{m}\bar{\sigma}_{n}\psi_{p})\}\,$ is dual to the field strength of 
an antisymmetric tensor \cite{Binetruy95}. The term $B^{m}\!B_{m}$ in the 
lagrangian $\Lag_{eff}$ thus generates a kinetic term of this antisymmetric 
tensor field and its coupling to the gravitino. The other way to perform the 
duality transformation is to treat $B_{m}$ as an auxiliary field by rewriting 
the term $\,-\,\frac{i}{2}b\ln({\bar{u}}/{u})\nabla^{m}\!B_{m}$ in 
$\Lag_{eff}$ as $\,\frac{i}{2}bB^{m}\nabla_{\!m}\!\ln({\bar{u}}/{u})$, and 
then to eliminate $B_{m}$ from $\Lag_{eff}$ through its equation of motion 
as follows:
\begin{eqnarray}
B_{m}\,&=&\,-\,i\frac{b\dilaton^{2}}{(\dilaton\dg+1)}\nabla_{\!m}\!
\ln(\frac{\bar{u}}{u}) \nonumber\\
& &\,+\,i\frac{b\dilaton^{2}}{(\dilaton\dg+1)}
\sum_{I}\frac{1}{(t^{I}+\bar{t}^{I})}
(\,\nabla_{\!m}\bar{t}^{I}\,-\,\nabla_{\!m}t^{I}\,).
\end{eqnarray}
The terms $B^{m}\!B_{m}$ and 
$\,\frac{i}{2}bB^{m}\nabla_{\!m}\!\ln({\bar{u}}/{u})$ in $\Lag_{eff}$ will 
generate a kinetic term for $\,\,i\ln({\bar{u}}/{u})$. It is clear that 
$\,\,i\ln({\bar{u}}/{u})$ plays the role of the pseudoscalar dual to $B_{m}$ 
in the lagrangian obtained from the above after a duality transformation.  
With (2.49--52), it is then trivial to eliminate the auxiliary fields from 
$\Lag_{eff}$. The physics of $\Lag_{eff}$ will be investigated in the 
following sections.
\subsection{Gaugino Condensate and the Gravitino Mass}
\hspace{0.8cm}
Hidden-sector gaugino condensation has been a very attractive scheme 
\cite{Nilles82,Dine85} for supersymmetry breaking in the context of 
superstring. However, before we can make any progress in superstring 
phenomenology, two important questions must be answered: is the dilaton 
stabilized, and is supersymmetry broken? Past analyses have generally found 
that, in the absence of a second source of supersymmetry breaking, the dilaton 
is destabilized in the direction of vanishing gauge coupling constant (the 
so-called runaway dilaton problem) and supersymmetry is unbroken. To address 
the above questions in generic linear multiplet models of gaugino condensation, 
we first show how the three issues of supersymmetry breaking, gaugino 
condensation and dilaton stabilization are reformulated, and how they are 
interrelated, by examining the explicit expressions for the gravitino mass and 
the gaugino condensate. A detailed investigation of the vacuum will be 
presented in Section 2.4.

The explicit expression for the gaugino condensate in terms of the dilaton
$\dilaton$ is determined by (2.49):
\begin{equation}
\bar{u}u\,=\,\frac{1}{e^{2}}\dilaton\mu^{6}
e^{g\,-\,({f+1})/{b\dilaton}}.
\end{equation}
With $g(\dilaton)$=0 and $f(\dilaton)$=0, we recover the result of the simple 
model (2.17) \cite{Binetruy95}. For generic models, the dilaton dependence of 
the gaugino condensate involves $g(\dilaton)$ and $f(\dilaton)$ which 
represent stringy non-perturbative corrections to the tree-level K\"{a}hler 
potential. Recall that in the linear multiplet formalism the gauge coupling of
the superstring effective field theory is $g^{2}(M_{s})\,=\,\left\langle\,
2\dilaton/\left(1+f(\dilaton)\right)\,\right\rangle$. Therefore, it is easy to
see that the dependence on the gauge coupling constant $g(M_{s})$ of the 
gaugino condensate is indeed consistent with the usual results obtained by the
renormalization group equation arguments. According to our assumption of 
boundedness for $g(\dilaton)$ and $f(\dilaton)$ (especially at $\dilaton$ =0 
where following (2.28) we have the boundary conditions $g(\dilaton=0)$=0 and 
$f(\dilaton=0)$=0), $\dilaton$=0 is the only pole of 
$\:g\,-\,({f+1})/{b\dilaton}.\:$ Therefore, we can draw a simple and clear 
relation between $\langle\bar{u}u\rangle$ and $\langle\,\dilaton\,\rangle$: 
gauginos condense ({\it i.e.}, $\langle\bar{u}u\rangle\neq 0$) if and only if the 
dilaton is stabilized ({\it i.e.}, $\langle\,\dilaton\,\rangle\neq 0$.) Note that this
conclusion does not depend on the details of the quantum corrections $g$ and 
$f$.

Another physical quantity of interest is the gravitino mass $m_{\tilde{G}}$ 
which is the natural order parameter measuring supersymmetry breaking. The 
expression for $m_{\tilde{G}}$ follows directly from $\Lag_{\tilde{G}}$.
\begin{equation}
m_{\tilde{G}}\,=\,\frac{\,1\,}{\,4\,}b\,
\sqrt{\langle\bar{u}u\rangle},
\end{equation}
where we have used (2.49). This expression for the gravitino mass is simple 
and elegant even for generic linear multiplet models of static gaugino
condensation. From the viewpoint of superstring effective theories, an 
interesting feature of (2.54) is that the gravitino mass $m_{\tilde{G}}$ 
contains no explicit dependence on the modulus $T^{I}$, which provides a 
direct relation between $m_{\tilde{G}}$ and $\langle\bar{u}u\rangle$. This 
feature can be traced to the fact that the Green-Schwarz counterterm cancels 
the $T^{I}$ dependence of the superpotential completely, a unique feature of 
the linear multiplet formalism. As we will see in Section 4.5, this unique
feature is still true even in a generic string orbifold model. We recall that 
in the chiral multiplet formalism of gaugino condensation -- without the 
condition (2.12) -- that have been studied previously (with or without the 
Green-Schwarz cancellation mechanism), $m_{\tilde{G}}$ always involves a 
moduli-dependence, and therefore the relation between supersymmetry breaking 
({\it i.e.}, $m_{\tilde{G}}\neq 0$) and gaugino condensation 
({\it i.e.}, $\langle\bar{u}u\rangle\neq 0$) remains undetermined until the true 
vacuum can be found. By contrast, in generic linear multiplet models of 
gaugino condensation, there is a simple and direct relation, Eq.(2.54): 
supersymmetry is broken ({\it i.e.}, $m_{\tilde{G}}\neq 0$) if and only if 
gaugino condensation occurs ($\langle\bar{u}u\rangle\neq 0$). We wish to 
emphasize that the above features of the linear multiplet model are unique in 
the sense that they are simple only in the linear multiplet model. This is 
related to the fact pointed out in Sections 2.1 and 2.2.3 that, once the 
constraint (2.12) on the condensate field $U$ is imposed, the chiral 
counterpart of the linear multiplet model is in general very complicated, and 
it is more natural to work in the linear multiplet formalism. Our conclusion of 
this section is best illustrated by the following diagram:

\vspace{0.4cm}
\fbox{\rule[-0.2cm]{0cm}{1.1cm}{\bf\shortstack{Supersymmetry\\Breaking}}}
$\;\Longleftrightarrow\;$
\fbox{\rule[-0.2cm]{0cm}{1.1cm}{\bf\shortstack{Gaugino\\Condensation}}}
$\;\Longleftrightarrow\;$
\fbox{\rule[-0.2cm]{0cm}{1.1cm}{\bf\shortstack{Stabilized\\Dilaton}}}
\vspace{0.4cm}

The equivalence among the above three issues is obvious. Therefore, in the 
following section, we only need to focus on one of the three issues in the 
investigation of the vacuum, for example, the issue of dilaton stabilization.
\section{Supersymmetry Breaking and Stabilization of the Dilaton}
\hspace{0.8cm}
As argued in Section 2.3.1, non-perturbative contributions to the K\"{a}hler 
potential should be introduced to cure the unboundedness problem of the simple 
model (2.17). In the context of the generic model of static gaugino 
condensation (2.26), it is therefore interesting to address the question as to 
how the simple model (2.17) should be modified in order to obtain a viable 
theory ({\it i.e.}, with $V_{pot}$ bounded from below). We start with the scalar 
potential $V_{pot}$ arising from (2.46) after solving for the auxiliary fields 
(using (2.49), (2.50) and (2.52)). Recalling that (2.27) yields the identity 
$\:1+\dilaton\dg=1+f-\dilaton\df\:$, we obtain
\begin{equation}
V_{pot}\,=\,\frac{1}{16e^{2}\dilaton}\{\,
(1+f-\dilaton\df)(1+b\dilaton)^{2}\,-\,3b^{2}\dilaton^{2}\,\}
\mu^{6}e^{g\,-\,({1+f})/{b\dilaton}},
\end{equation}
which depends only on the dilaton $\dilaton$. The necessary and sufficient 
condition for $V_{pot}$ to be bounded from below is
\begin{eqnarray}
f-\dilaton\df\,&\geq&\,-\CO(\dilaton e^{{1}/{b\dilaton}})
\;\;\;\;\;\mbox{for}\;\;\;
\dilaton\,\rightarrow\, 0, \\
f-\dilaton\df\,&\geq&\,\hspace{0.34cm}2\;\;\;\;\;
\hspace{1.1cm}\mbox{for}\;\;\;
\dilaton\,\rightarrow\,\infty.
\end{eqnarray}
It is clear that condition (2.56) is not at all restrictive, and therefore has 
no nontrivial implication. On the contrary, condition (2.57) is quite 
restrictive; in particular the simple model (2.17) violates this condition. 
Condition (2.57) not only restricts the possible forms of the function $f$ in 
the strong-coupling regime but also has important implications for dilaton 
stabilization and for supersymmetry breaking. To make the above statement more 
precise, let us revisit the unbounded potential of Fig. 2.1, with the 
tree-level K\"{a}hler potential defined by $g(V)=f(V)=0$. Adding physically 
reasonable corrections $g(V)$ and $f(V)$ (constrained by (2.56--57)) to this 
simple model should not qualitatively alter its behavior in the weak-coupling 
regime. Therefore, as in Fig. 2.1, the potential of the modified model in the 
weak-coupling regime starts with $V_{pot}=0$ at $\dilaton=0$, first rises and 
then falls as $\dilaton$ increases. On the other hand, adding $g(V)$ and 
$f(V)$ completely alters the strong-coupling behavior of the original simple 
model. As guaranteed by condition (2.57), the potential of the modified model 
in the strong-coupling regime is always bounded from below, and in most cases 
rises as $\dilaton$ increases. Joining the weak-coupling behavior of the 
modified model to its strong-coupling behavior therefore strongly suggests 
that its potential has a non-trivial minimum (at $\dilaton\neq 0$). 
Furthermore, if this non-trivial minimum is global, then the dilaton is 
stabilized. We conclude that not only does (2.56--57) tell us how to modify the 
theory, but a large class of theories so modified have naturally a stabilized 
dilaton (and therefore broken supersymmetry by the argument of Section 2.3.3). 
In view of the fact that there is currently little knowledge of the exact 
K\"{a}hler potential, the above conclusion, which applies to generic 
K\"{a}hler potentials subject to (2.56--57), is especially important to the 
search for supersymmetry breaking and dilaton stabilization\footnote{Similar
points of view was advocated in \cite{Casas96} using the chiral multiplet
formalism. However, neither modular invariance nor the important constraint 
(2.12) was considered in \cite{Casas96}.}. As discussed in Sections 2.1 and 
2.2.2, the most interesting physical implication of this conclusion is that it 
is actually stringy non-perturbative effects that stabilize the dilaton and 
allow dynamical supersymmetry breaking via the field-theoretical 
non-perturbative effect of gaugino condensation. Furthermore, (2.57) can be 
interpreted as the necessary condition for stringy non-perturbative effects to 
stabilize the dilaton.\footnote{In the presence of significant stringy 
non-perturbative effects, (2.57) could have implications for gauge coupling 
unification. This is considered in the study of multi-gaugino and matter 
condensation \cite{multiple}.} 

Here we use a simple example only to illustrate the above important argument. 
A more detailed discussion of possible stringy non-perturbative corrections 
will be given in Chapters 4 and 5 where a generic and phenomenologically 
viable model is presented. Consider $\,f(V)\,=\,A e^{-\,{B}/{V}}$, where 
$A$ and $B$ are constants to be determined by the non-perturbative dynamics. 
The regulation conditions (2.56--57) require $A\geq 2$. In Fig. 2.2, $V_{pot}$ 
is plotted versus the dilaton $\dilaton$, where $A=6.92$, $B=1$ and $\mu$=1. 
\begin{figure}
\epsfxsize=12cm
\epsfysize=8cm
\epsfbox{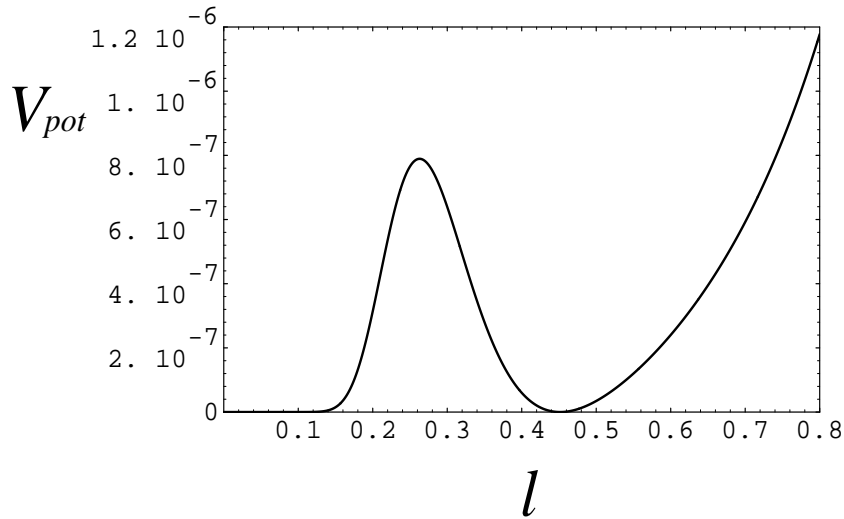}
\caption{The scalar potential $V_{pot}$ (in reduced Planck units)
            is plotted versus the dilaton $\dilaton$. $\,A=6.92$,
            $\,B=1\,$ and $\,\mu$=1.}
\end{figure}
Fig. 2.2 has two important features. Firstly, $V_{pot}$ of this modified theory 
is indeed bounded from below, and the dilaton is stabilized. Therefore, we 
obtain supersymmetry breaking, gaugino condensation and dilaton stabilization 
in this example. The gravitino mass is 
$m_{\tilde{G}}\,=\,7.6\times 10^{-5}$ in reduced Planck units. Secondly, 
the {\em vev} of dilaton is stabilized at the phenomenologically interesting 
range ($\langle\,\dilaton\,\rangle = 0.45$ in Fig. 2.2).
The above features involve no unnaturalness since they are insensitive to 
$A$. Furthermore, the dilaton is naturally stabilized in a weak coupling regime 
if $B$ is of order one. Fig. 2.2 is a nice realization of the argument in the 
preceding paragraph. It should be contrasted with the racetrack model where 
at least three gaugino condensates and large numerical coefficients are needed 
in order to achieve similar results. Besides, the racetrack model has a serious 
phenomenological problem of having a large negative cosmological constant. We 
can also consider possible stringy non-perturbative contributions to the 
K\"{a}hler potential suggested in \cite{Shenker90}. It turns out that we 
obtain the same general features as those of Fig. 2.2. This is not surprising
since, as argued in the preceding paragraph, the important features that we 
find in Fig. 2.2 are common to a large class of models. More such discussions
will be presented in Chapters 4 and 5 in conjunction with other issues.

Note that the value of the cosmological constant is irrelevant to the 
arguments presented here and in Section 2.3.3. In other words, the generic 
model (2.26) suffers from the usual cosmological constant problem,
although we can find a fine-tuned subset of models whose cosmological 
constants vanish. For example, the cosmological constant of Fig. 2.2 
vanishes by fine tuning $A$. It remains an open question as to whether or not 
the cosmological constant problem could be resolved within the context of the 
linear multiplet formalism of gaugino condensation if the exact K\"{a}hler 
potential were known.
\section{Concluding Remarks}
\hspace{0.8cm}
We have presented a concrete example of a solution to the infamous runaway 
dilaton problem, within the context of local supersymmetry and the linear 
multiplet formalism for the string dilaton. We considered models for a static 
condensate that reflect the modular anomaly of the effective field theory while
respecting the exact modular invariance of the underlying string theory.  The 
simplest such model~\cite{Binetruy95,sduality} has a nontrivial potential that 
is, however, unbounded in the direction of strong coupling. Including 
stringy non-perturbative corrections \cite{Shenker90,Banks94} to the K\"ahler
potential for the dilaton, the potential is stabilized, allowing a vacuum
configuration in which condensation occurs and supersymmetry is broken.  This 
is in contrast to previous analyses, based on the chiral multiplet formalism for 
the dilaton, in which supersymmetry breaking with a bounded vacuum energy was 
achieved only by introducing an additional source of supersymmetry breaking, 
such as a constant term in the superpotential \cite{Dine85,constant,chiral91}.

In further contrast to most of the models studied using the chiral multiplet
formalism, supersymmetry breaking arises from a nonvanishing vacuum 
expectation value of the auxiliary field associated with the dilaton rather 
than the moduli: roughly speaking, in the dual chiral multiplet formalism, 
$\langle F_S \rangle\ne 0$ rather than $\langle F^{I} \rangle\ne 0$. That 
is, only the dilaton participates in supersymmetry breaking (the so-called
dilaton-dominated scenario.) As we shall see in Chapter 4, this unique feature
is in fact true in generic string orbifold models, which therefore has 
non-trivial implications for FCNC. As a consequence, gaugino masses and 
$A$ terms are generated at tree level. Although scalar masses are still 
protected at tree level by a Heisenberg symmetry~\cite{heis}, they will be 
generated at one loop by renormalizable interactions\footnote{The situation is
more complicated in a generic orbifold model, and will be discussed in 
\mbox{Chapter 5}.}. For the model considered here, the hierarchy 
(about five orders of 
magnitude) between the Planck scale and the gravitino mass is insufficient to 
account for the observed scale of electroweak symmetry breaking. Of course, 
this is completely due to the large gauge content of the hidden E$_{8}$ gauge 
group under consideration in this chapter, and will certainly be improved when 
a generic string model with a product of smaller hidden gauge groups 
${\cal G} = \Pi_a{\cal G}_a$. In that case, we will have to generalize the 
studies of this chapter by considering multiple gaugino condensation as well 
as hidden matter condensation. Another unsatisfactory feature of the model
presented in Chapter 2 is that, according to (2.55), the moduli $T^{I}$ remain
flat directions of the scalar potential, and therefore the {\em vev} of $t^{I}$
is undetermined. Fortunately, this is a feature belonging only to string models
with hidden E$_{8}$ gauge group and no hidden matter. As we shall see in Chapter
4, in a generic string model where multiple gaugino condensation as well 
as hidden matter condensation occurs naturally, hidden matter condensation
together with string threshold corrections\footnote{Both are required by 
modular invariance.} generates a non-perturbative potential for the moduli 
$T^{I}$. Furthermore, the moduli are therefore stabilized at the self-dual 
point. The generalization of our formalism to generic string orbifold models, 
including models without universal anomaly cancellation, will be presented in 
Chapter 4.

As mentioned before, we have only dealt with generic models of static gaugino
condensation in this chapter, but in the context of supergravity or 
superstrings
it can be shown that models of dynamical gaugino condensation rather than 
models of static gaugino condensation occur. Therefore, in the next chapter we
will answer two questions: first, we show how to construct generic models of
dynamical gaugino condensation using the linear multiplet formalism. Secondly,
we study how the models of dynamical gaugino condensation are connected to the 
models of static gaugino condensation, and show that static gaugino 
condensation is indeed the appropriate effective description of dynamical
gaugino condensation and therefore justify the use of static gaugino
condensation in Chapter 2. Notice that the Kalb-Ramond field (or the
model-independent axion, in the dual description) remains massless in the 
static models considered here. It has recently been shown in the context of 
global supersymmetry \cite{Burgess95,Binetruy95} that an axion mass term 
is naturally generated in models of dynamical gaugino condensation. 
Again, as we shall see in Chapter 3, one of the axions does get a very large mass 
through dynamical gaugino condensation in the context of local supersymmetry. 
On the other hand, after this very heavy axion is integrated out, the resulting
axion content is in fact the same as that of static gaugino condensation, and 
we are still left with a massless model-independent axion. Furthermore, we 
will show in Chapters 4 and 5 that this model-independent axion axion will pick
up a very small mass through multiple gaugino condensation. It can escape the
cosmological bound on the axion decay constant and it has the desirable
properties to be the candidate for the QCD axion. 
\newpage
\setcounter{chapter}{2}
\chapter{Dynamical Gaugino Condensation}
\hspace{0.8cm}
\setcounter{equation}{0}
\setcounter{figure}{0}
\setcounter{footnote}{0}
\newpage

\section{Introduction}
\hspace{0.8cm}
In Chapter 2, we have studied models of static gaugino condensation using the
linear multiplet formalism. As mentioned before, one of the major motivations 
for studying models of dynamical gaugino condensation is the observation that 
kinetic terms of the gaugino condensate naturally arise from field-theoretical 
loop corrections \cite{sduality} as well as from classical string corrections 
\cite{Ant}. For example, the relevant field-theoretical one-loop correction 
has been computed using the chiral multiplet formalism \cite{sduality,Jain96}: 
\begin{equation}
\Lag_{one-loop}\;\ni\;\frac{N_{G}}{128\pi^{2}}
\superint\,E\,(S+\bar{S})^{2}\,({\cal W}^{\alpha}{\cal W}_{\alpha})\,
({\cal W}_{\dot{\alpha}}{\cal W}^{\dot{\alpha}})\,\ln\Lambda^{2},
\end{equation}
where $\Lambda$ is the effective cut-off and $N_{G}$ is the number of gauge 
degrees of freedom. Therefore, the confined theory using the linear multiplet
formalism should contain a term which corresponds to (3.1):
\begin{equation}
\Lag_{eff}\;\ni\;\superint\,E\,\frac{\bar{U}U}{V^{2}},
\end{equation}
as well as higher-order corrections $\,\left(\bar{U}U/V^{2}\right)^{2}$,
$\,\left(\bar{U}U/V^{2}\right)^{3},\cdots .\,$ These $D$ terms are corrections 
to the K\"ahler potential, and will generate the kinetic terms for the gaugino
condensate $U$. An interesting interpretation of these corrections is that they
are S-duality invariant in the sense defined by Gaillard and Zumino 
\cite{zumino}. This S-duality, which is an SL(2,R) symmetry among elementary
fields, is a symmetry of the equations of motion only of the 
dilaton-gauge-gravity sector in the limit of vanishing gauge coupling 
constants. The implication of this S-duality for gaugino condensation has 
recently been studied in \cite{sduality} using the chiral multiplet formalism. 

For studies of gaugino condensation in the past where the important constraint
(2.12) was not included, the connection between static and dynamical gaugino
condensation is very easy to see and trivial: static gaugino condensation is
just the low-energy limit of dynamical gaugino condensation after the gaugino
condensate is integrated out. However, it certainly becomes a non-trivial issue
once the constraint (2.12) is included, and it is necessary to settle this
issue in order to justify the use of static gaugino condensation in the context
of superstrings or supergravity. Therefore, in this chapter we would like to 
study generic models of dynamical gaugino condensation. In Section 3.2, the 
field component Lagrangian for the generic model of dynamical gaugino 
condensation is constructed, and its vacuum structure is analyzed. In Section 
3.3, the S-dual models of dynamical gaugino condensation are studied. In 
particular, we show that the model of static gaugino condensation is the 
appropriate effective description for the model of dynamical gaugino 
condensation and its implications. 
\section{Generic Model of Dynamical Gaugino Condensation}
\hspace{0.8cm}
It will be shown in this section how to construct the component field 
Lagrangian for the generic model of dynamical gaugino condensation using the 
K\"ahler superspace formulation of supergravity \cite{Binetruy90,Binetruy91}.
Similar to Chapter 2, we consider here string orbifold models with gauge groups 
E$_{8}\otimes$E$_{6}\otimes$U(1)$^{2}$, three untwisted (1,1) moduli $T^{I}$ 
($I\,=\,1,\;2,\;3$) \cite{Gaillard92,KL,Dixon90}, and universal modular anomaly
cancellation \cite{ant} ({\it e.g.}, the Z$_{3}$ and Z$_{7}$ orbifolds). The 
confined E$_{8}$ hidden sector is described by the following generic model of 
a single dynamical gaugino condensate $U$ with K\"ahler potential $K$:
\begin{eqnarray}
K\,&=&\,\ln V\,+\,g(V,\bar{U}U)\,+\,G, \nonumber \\
\Lag_{eff}\,&=&\,
\superint\,E\,\left\{\,\left(\,-2\,+\,f(V,\bar{U}U)\,\right)\,+\,bVG\,\right\} 
\,+\,\left\{\,\superint\,\frac{E}{R}\,e^{K/2}W_{VY}\,+\,\mbox{h.c.}\,\right\},
\nonumber \\
G\,&=&\,-\sum_{I}\ln(T^{I}+\bar{T}^{I}),
\end{eqnarray}
where $\;U\,=\,-(\DbDb-8R)V\,$, $\,\bar{U}\,=\,-(\DaDa-8R^{\dagger})V.\;$ We 
also write $\;\ln V\,+\,g(V,\bar{U}U)\,\equiv\,k(V,\bar{U}U).\;$ The term
$\,\left(\,-2\,+\,f(V,\bar{U}U)\,\right)\,$ of $\Lag_{eff}$ is the superspace 
integral which yields the kinetic actions for the linear multiplet, 
supergravity, matter, and gaugino condensate. The term $\,bVG\,$ is the
Green-Schwarz counterterm \cite{Gaillard92} which cancels the full modular 
anomaly here. $b\,=\,C/8\pi^{2}\,=\,2b_{0}/3$, and $C\,=\,30$ is the Casimir 
operator in the adjoint representation of $\mbox{E}_{8}$. $b_{0}$ is the 
$\mbox{E}_{8}$ one-loop $\beta$-function coefficient. $g(V,\bar{U}U)$ and 
$f(V,\bar{U}U)$ represent the quantum corrections to the tree-level K\"{a}hler 
potential. $\,g(V,\bar{U}U)\,$ and $\,f(V,\bar{U}U)\,$ are taken to be 
arbitrary but bounded here. The dynamical model (3.3) is the straightforward
generalization of the static model (2.26) by including the $\bar{U}U$ 
dependence in the K\"ahler potential. Using superspace partial integration 
(2.18), up to a total derivative we can also rewrite (3.3) as a single $D$ 
term:
\begin{eqnarray}
K\,&=&\,\ln V\,+\,g(V,\bar{U}U)\,+\,G, \nonumber \\
\Lag_{eff}\,&=&\,
\superint\,E\,\left\{\,\left(\,-2\,+\,f(V,\bar{U}U)\,\right)\,+\, 
bVG\,+\,bV\ln(e^{-K}\bar{U}U/\mu^{6})\,\right\}.\hspace{1.5cm}
\end{eqnarray}

Only the bosonic and gravitino parts of the component field Lagrangian will be 
presented here. In the following, for convenience and completeness we 
enumerate the definitions of the bosonic component fields:
\begin{eqnarray}
\dilaton\,&=&\,V\lowest,\nonumber\\
\sigma^{m}_{\alpha\dot{\alpha}}B_{m}\,&=&\,
\frac{1}{2}\left[\,\Diff_{\alpha},\Diff_{\dot{\alpha}}\,\right]V\lowest\,+\,
\frac{2}{3}\dilaton\sigma^{a}_{\alpha\dot{\alpha}} b_{a},\nonumber\\
u\,&=&\,U\lowest\,=\,-(\bar{\Diff}^{2}-8R)V\lowest,\nonumber\\
\bar{u}\,&=&\,\bar{U}\lowest\,=\,-(\Diff^{2}-8R^{\dagger})V\lowest,
\nonumber \\
-4F_{U}\,&=&\,\Diff^{2}U\lowest, \;\;\; 
-4\bar{F}_{\bar{U}}\,=\,\bar{\Diff}^{2}\bar{U}\lowest,
\nonumber \\
D\,&=&\,\frac{1}{8}\Diff^{\beta}(\bar{\Diff}^{2}-8R)
      \Diff_{\beta}V\lowest\nonumber\\
   &=&\,\frac{1}{8}\Diff_{\dot{\beta}}(\Diff^{2}-8R^{\dagger})
      \Diff^{\dot{\beta}}V\lowest,
\nonumber \\
t^{I}\,&=&\,T^{I}\lowest,\;\;\;
-4F^{I}\,=\,\Diff^{2}T^{I}\lowest,\nonumber\\
\bar{t}^{I}\,&=&\,\bar{T}^{I}\lowest,\;\;\;
-4\bar{F}^{I}\,=\,\bar{\Diff}^{2}\bar{T}^{I}\lowest,
\end{eqnarray}
where $\,b_{a}\,=\,-3G_{a}\lowest\,$, $\,M\,=\,-6R\lowest\,$, 
$\,\bar{M}\,=\,-6R^{\dagger}\lowest\,$ are the auxiliary components of the 
supergravity multiplet. $(F_{U}-\bar{F}_{\bar{U}})$ can be expressed as follows:
\begin{equation}
(F_{U}-\bar{F}_{\bar{U}})\,=\,4i\nabla^{m}\!B_{m}
\,+\,u\bar{M}\,-\,\bar{u}M,
\end{equation}
and $(F_{U}+\bar{F}_{\bar{U}})$ contains the auxiliary field $D$. We also 
write $\,Z\,\equiv\,\bar{U}U\,$, and its bosonic component 
$\,z\,\equiv\,Z\lowest\,=\,\bar{u}u\,$.

The construction of component field Lagrangian using chiral density 
multiplet method~\cite{Binetruy90} has been detailed in Chapter 2, and
therefore only the key steps are presented here. The chiral density multiplet 
${\bf r}$ and its hermitian conjugate ${\bf \bar{r}}$ for the generic model 
(3.3) are:
\begin{eqnarray}
{\bf r}\,&=&\,-\,\frac{1}{8}(\bar{\Diff}^{2}-8R)
\left\{\,\left(\,-2\,+\,f(V,\bar{U}U)\,\right)
\,+\,bVG\,+\,bV\ln(e^{-K}\bar{U}U/\mu^{6})\,\right\}, \nonumber\\
{\bf \bar{r}}\,&=&\,-\,\frac{1}{8}(\Diff^{2}-8R^{\dagger})
\left\{\,\left(\,-2\,+\,f(V,\bar{U}U)\,\right)
\,+\,bVG\,+\,bV\ln(e^{-K}\bar{U}U/\mu^{6})\,\right\},
\hspace{1.5cm}
\end{eqnarray}
and the component field Lagrangian $\Lag_{eff}$ is the same as (2.30).
The $\: A_{m}\lowest\:$ for the generic model (3.3) is:
\begin{eqnarray}
A_{m}\lowest\,&=&\,-\,\frac{i}{4\dilaton}\!\cdot\!
\frac{(1+\ldgl)}{(1-\zdgz)}B_{m}\,+\,\frac{i}{6}
\left[\,\frac{(1+\ldgl)}{(1-\zdgz)}\,-\,3\,\right]e_{m}^{\hs a}b_{a}
\nonumber\\
& &\,+\,\frac{1}{4(1-\zdgz)}\sum_{I}\frac{1}{(t^{I}+\bar{t}^{I})}
(\nabla_{\!m}\bar{t}^{I}\,-\,\nabla_{\!m}t^{I})
\nonumber\\
& &\,-\,\frac{\zdgz}{4(1-\zdgz)}\nabla_{\!m}\ln\!\left(\frac{\bar{u}}{u}\right).
\end{eqnarray}
The following are the simplified notations for partial derivatives of $g$:
\begin{equation}
g_{_{\dilaton}}\,\equiv\,\frac{\partial g(\dilaton,z)}{\partial \dilaton},
\;\;\;g_{_{z}}\,\equiv\,\frac{\partial g(\dilaton,z)}{\partial z},
\end{equation}
and similarly for other functions.

We need to decompose the lowest components of the following six superfields:
$X_{\alpha}$, $\bar{X}^{\dot{\alpha}}$, $\Diff_{\alpha}R$,
$\Diff^{\dot{\alpha}}\!R^{\dagger}$, 
$(\Diff^{\alpha}\!X_{\alpha}+\Diff_{\dot{\alpha}}\bar{X}^{\dot{\alpha}})$
and $(\Diff^{2}\!R+\bar{\Diff}^{2}\!R^{\dagger})$ into component 
fields, where
\begin{eqnarray}
X_{\alpha}\,&=&\,-\,\frac{1}{8}(\DbDb-8R)
\Diff_{\alpha}K,
\nonumber\\
\bar{X}^{\dot{\alpha}}\,&=&\,-\,\frac{1}{8}(\DaDa-8R^{\dagger})
\Diff^{\dot{\alpha}}\!K,
\nonumber\\
(\Diff^{\alpha}\!X_{\alpha}+\Diff_{\dot{\alpha}}\bar{X}^{\dot{\alpha}})\,
&=&\,-\,\frac{1}{8}\Diff^{2}\!\bar{\Diff}^{2}\!K
\,-\,\frac{1}{8}\bar{\Diff}^{2}\!\Diff^{2}\!K
\,-\,\Diff^{\alpha\dot{\alpha}}\!\Diff_{\alpha\dot{\alpha}}K
\nonumber\\
& &\,-\,G^{\alpha\dot{\alpha}}
\left[\,\Diff_{\alpha},\Diff_{\dot{\alpha}}\,\right]K
\,+\,2R^{\dagger}\bar{\Diff}^{2}\!K\,+\,2R\Diff^{2}\!K
\nonumber\\
& &\,-\,(\,\Diff^{\alpha}\!G_{\alpha\dot{\alpha}}
\,-\,2\Diff_{\dot{\alpha}}R^{\dagger}\,)\Diff^{\dot{\alpha}}\!K
\nonumber\\
& &\,+\,(\,\Diff^{\dot{\alpha}}\!G_{\alpha\dot{\alpha}}
\,+\,2\Diff_{\alpha}R\,)\Diff^{\alpha}\!K.
\end{eqnarray}
This is done by solving the following six algebraic equations: 
\begin{eqnarray}
\left(1+V\frac{\partial g}{\partial V}\right)\Diff_{\alpha}R\,+\,
\left(1-Z\frac{\partial g}{\partial Z}\right)X_{\alpha}\,&=&\,\Xi_{\alpha},\\
3\Diff_{\alpha}R\,+\,X_{\alpha}\,&=&\,
-2(\sigma^{cb}\epsilon)_{\alpha\varphi}T_{cb}^
{\hs\hspace{0.04cm}\varphi}.
\end{eqnarray}
\begin{eqnarray}
\left(1+V\frac{\partial g}{\partial V}\right)\Diff^{\dot{\alpha}}\!R^{\dagger}
\,+\,\left(1-Z\frac{\partial g}{\partial Z}\right)\bar{X}^{\dot{\alpha}}\,&=&\,
\bar{\Xi}^{\dot{\alpha}},\\
3\Diff^{\dot{\alpha}}\!R^{\dagger}\,+\,\bar{X}^{\dot{\alpha}}\,&=&\,
-2(\bar{\sigma}^{cb}\epsilon)^{
\dot{\alpha}\dot{\varphi}}T_{cb\dot{\varphi}}.
\end{eqnarray}
\begin{eqnarray}
\left(1+V\frac{\partial g}{\partial V}\right)(\Diff^{2}\!R+\bar{\Diff}^{2}\!
R^{\dagger})\,+\,\left(1-Z\frac{\partial g}{\partial Z}\right)
(\Diff^{\alpha}\!X_{\alpha}+
\Diff_{\dot{\alpha}}\bar{X}^{\dot{\alpha}})\,&=&\,\Delta,\\
3(\Diff^{2}\!R+\bar{\Diff}^{2}\!R^{\dagger})\,+\,(\Diff^{\alpha}
\!X_{\alpha}+\Diff_{\dot{\alpha}}\bar{X}^{\dot{\alpha}})\,&=&\,
-2R_{ba}^{\hs ba}\,+\,12G^{a}G_{a} \hspace{0cm}\nonumber\\
\,& &\,+\,96RR^{\dagger}. \hspace{0cm}
\end{eqnarray}
The computation of (3.10) defines the contents of $\Xi_{\alpha}$,
$\bar{\Xi}^{\dot{\alpha}}$ and $\Delta$. Eqs. (3.8--16) describe the key steps
in the computations of (2.30). In the following sections, several important 
issues of this construction will be discussed.
\subsection{Canonical Einstein Term}
\hspace{0.8cm}
In order to have the correctly normalized Einstein term in $\Lag_{eff}$, an
appropriate constraint should be imposed on the generic model (3.3). Therefore,
it is shown below how to compute the Einstein term for (3.3). According to 
(3.3), the following are those terms in $\Lag_{eff}$ that will contribute to 
the Einstein term:
\begin{eqnarray}
\frac{1}{e}\Lag_{eff}\;&\ni&\;\frac{1}{4}
\left[\,2-f+\ldfl-b\dilaton(1+\ldgl)\,\right]
(\Diff^{2}\!R+\bar{\Diff}^{2}\!R^{\dagger})\lowest
\nonumber\\
\;& &\;+\,\frac{1}{32}
\left[\,\zdfz+b\dilaton(1-\zdgz)\,\right]
\left(\,\frac{1}{\bar{u}}\Diff^{2}\!\bar{\Diff}^{2}\!\bar{U}
\,+\,\frac{1}{u}\bar{\Diff}^{2}\!\Diff^{2}\!U\,\right)\lowest.\hspace{1.5cm}
\end{eqnarray}
Note that the terms $\,\Diff^{2}\!\bar{\Diff}^{2}\!\bar{U}\,$ and 
$\,\bar{\Diff}^{2}\!\Diff^{2}\!U\,$ are related to 
$\,\Diff^{\alpha}\!X_{\alpha}\,$ and  
$\,\Diff_{\dot{\alpha}}\bar{X}^{\dot{\alpha}}\,$ through the following 
identities:
\begin{eqnarray}
\Diff^{2}\!\bar{\Diff}^{2}\!\bar{U}\,&=&\,
16\Diff^{a}\!\Diff_{a}\bar{U}\,+\,64iG^{a}\Diff_{a}\bar{U}
\,-\,48\bar{U}G^{a}\!G_{a}\,+\,48i\bar{U}\Diff^{a}\!G_{a} \nonumber\\
\,& &\,-\,8\bar{U}\Diff^{\alpha}\!X_{\alpha}
\,+\,16R^{\dagger}\bar{\Diff}^{2}\!\bar{U}
\,+\,8(\Diff^{\alpha}\!G_{\alpha\dot{\alpha}})(\Diff^{\dot{\alpha}}\!\bar{U}).
\nonumber\\
\bar{\Diff}^{2}\!\Diff^{2}\!U\,&=&\,
16\Diff^{a}\!\Diff_{a}U\,-\,64iG^{a}\Diff_{a}U
\,-\,48UG^{a}\!G_{a}\,-\,48iU\Diff^{a}\!G_{a} \nonumber\\
\,& &\,-\,8U\Diff_{\dot{\alpha}}\bar{X}^{\dot{\alpha}}
\,+\,16R\Diff^{2}\!U
\,-\,8(\Diff^{\dot{\alpha}}\!G_{\alpha\dot{\alpha}})(\Diff^{\alpha}\!U).
\end{eqnarray}
The contributions of $\,(\Diff^{2}\!R+\bar{\Diff}^{2}\!R^{\dagger})\lowest\,$
and $\,(\Diff^{\alpha}\!X_{\alpha}+\Diff_{\dot{\alpha}}\bar{X}^{\dot{\alpha}})
\lowest\,$ to the Einstein term are obtained by solving (3.15--16):
\begin{eqnarray}
(\Diff^{2}\!R+\bar{\Diff}^{2}\!R^{\dagger})\lowest\,&\ni&\,
-\,\frac{2(1-\zdgz)}{(2-\ldgl-3\zdgz)}R_{ba}^{\hs ba}\lowest.
\nonumber\\
(\Diff^{\alpha}\!X_{\alpha}+\Diff_{\dot{\alpha}}\bar{X}^{\dot{\alpha}})
\lowest\,&\ni&\,
+\,\frac{2(1+\ldgl)}{(2-\ldgl-3\zdgz)}R_{ba}^{\hs ba}\lowest.
\end{eqnarray}

By combining (3.17--19), it is straightforward to show that the Einstein term in 
$\Lag_{eff}$ is correctly normalized if and only if the following constraint 
is imposed:
\begin{equation}
(\,1\,+\,\zdfz\,)(\,1\,+\,\ldgl\,)\,=\,
(\,1\,-\,\zdgz\,)(\,1\,-\,\ldfl\,+\,f\,),
\end{equation}
which is a first-order partial differential equation. From now on, the study of 
the generic model (3.3) always assumes the constraint (3.20). (3.20) will be 
useful in simplifying the expression of $\Lag_{eff}$, and it turns out to 
be convenient to define $h$ as follows:
\begin{eqnarray}
h\,&\equiv&\,\frac{(\,1\,+\,\zdfz\,)}{(\,1\,-\,\zdgz\,)},\nonumber\\
&=&\,\frac{(\,1\,-\,\ldfl\,+\,f\,)}{(\,1\,+\,\ldgl\,)}.
\end{eqnarray}
Furthermore, the partial derivatives of $h$ satisfy the following consistency
condition:
\begin{equation}
(\,h\,-\,\ldhl\,)(\,\zdgz\,-\,1\,)\,+\,\zdhz(\,1\,+\,\ldgl\,)\,+\,1\,=\,0.
\end{equation}
Eqs. (3.21--22) will also be very useful in simplifying the expression of 
$\Lag_{eff}$. Notice that $h=1$ for generic models of static gaugino 
condensation, and (3.20) is reduced to (2.27). We will show in Section 3.3.2 
how to construct physically interesting solutions for this partial differential 
equation (3.20). 
\subsection{Component Field Lagrangian with Auxiliary Fields}
\hspace{0.8cm}
Once the issue of canonical Einstein term is settled, it is straightforward to
compute $\Lag_{eff}$ according to (3.6--13). The rest of it is 
standard and will not be detailed here. Because the component construction of
supergravity is well known for its complexity, here we try our best to
minimize irrelevant details. However, two important aspects of this 
construction using the linear multiplet formalism are worth emphasizing: how 
to solve the constraint (2.12) and how to perform a duality transformation for 
the vector component $B_{m}$ of $V$. As we shall see, they have non-trivial 
implications for the axions. Therefore, first we present the component 
Lagrangian with auxiliary fields, and in the next section we show how to 
perform a duality transformation for $B_{m}$. In the following, we present 
the component field expression of $\Lag_{eff}$ as the sum of the bosonic 
Lagrangian $\Lag_{B}$ and the gravitino Lagrangian $\Lag_{\tilde{G}}$.
\begin{equation}
\Lag_{eff}\,=\,\Lag_{B}\,+\,\Lag_{\tilde{G}}.
\end{equation}

\begin{eqnarray}
\frac{1}{e}\Lag_{B}\,&=&\,-\,\frac{1}{2}{\cal R}
\,-\,\frac{1}{4\dilaton^{2}}(h-\ldhl)(1+\ldgl)
\nabla^{m}\!\dilaton\,\nabla_{\!m}\!\dilaton
\nonumber\\
& &\,+\,\frac{1}{2\dilaton}\zdhz(1+\ldgl)\,
\nabla^{m}\!\ln(\bar{u}u)\,\nabla_{\!m}\!\dilaton
\nonumber\\
& &\,+\,\frac{u}{4\bar{u}}h_{_{z}}\!\cdot\!\zdgz
\frac{(2-\zdgz)}{(1-\zdgz)}
\nabla^{m}\bar{u}\,\nabla_{\!m}\bar{u}
\nonumber\\
& &\,-\,\frac{1}{2}h_{_{z}}
\left[\frac{(2-\zdgz)}{(1-\zdgz)}\,-\,\zdgz\right]
\nabla^{m}\bar{u}\,\nabla_{\!m}u
\nonumber\\
& &\,+\,\frac{\bar{u}}{4u}h_{_{z}}\!\cdot\!\zdgz
\frac{(2-\zdgz)}{(1-\zdgz)}
\nabla^{m}u\,\nabla_{\!m}u
\nonumber\\
& &\,-\,\frac{\zdhz}{2(1-\zdgz)}\sum_{I}\frac{1}{(t^{I}+\bar{t}^{I})}
(\,\nabla^{m}\bar{t}^{I}\,-\,\nabla^{m}t^{I}\,)\,
\nabla_{\!m}\!\ln\!\left(\frac{\bar{u}}{u}\right)
\nonumber\\
& &\,+\,\frac{\zdhz}{4(1-\zdgz)}
\sum_{I,J}\frac{1}{(t^{I}+\bar{t}^{I})(t^{J}+\bar{t}^{J})}
\nabla^{m}\bar{t}^{I}\,\nabla_{\!m}\bar{t}^{J}
\nonumber\\
& &\,-\,\frac{1}{2}\sum_{I,J}\left[2(h+b\dilaton)\delta_{IJ}\,+\,
\frac{\zdhz}{(1-\zdgz)}\right]
\frac{\nabla^{m}\bar{t}^{I}\,\nabla_{\!m}t^{J}}
{(t^{I}+\bar{t}^{I})(t^{J}+\bar{t}^{J})}
\nonumber\\
& &\,+\,\frac{\zdhz}{4(1-\zdgz)}
\sum_{I,J}\frac{1}{(t^{I}+\bar{t}^{I})(t^{J}+\bar{t}^{J})}
\nabla^{m}t^{I}\,\nabla_{\!m}t^{J}
\nonumber\\
& &\,+\,\frac{(2-\ldgl-3\zdgz)}{9(1-\zdgz)}\,b^{a}b_{a}
\nonumber\\
& &\,+\,\frac{(1+\ldgl)}{4\dilaton^{2}(1-\zdgz)}B^{m}\!B_{m}
\nonumber\\
& &\,+\,\frac{i}{2\dilaton}\left[h\,+\,b\dilaton\,-\,
\frac{1}{(1-\zdgz)}\right]\,B^{m}\,
\nabla_{\!m}\!\ln\!\left(\frac{\bar{u}}{u}\right)
\nonumber\\
& &\,-\,\frac{i}{2\dilaton}\left[h\,+\,b\dilaton\,-\,
\frac{1}{(1-\zdgz)}\right]\sum_{I}
\frac{(\,\nabla^{m}\bar{t}^{I}\,-\,\nabla^{m}t^{I}\,)}
{(t^{I}+\bar{t}^{I})}B_{m}
\nonumber\\
& &\,+\,4h_{_{z}}(1-\zdgz)(\nabla^{m}\!B_{m})^{2}
\nonumber\\
& &\,-\,2ih_{_{z}}\left[ 1\,-\,\zdgz\,-\,\frac{1}{3}(1+\ldgl)\right]
(\,u\bar{M}\,-\,\bar{u}M\,)\,\nabla^{m}\!B_{m}
\nonumber\\
& &\,-\,\frac{1}{4}h_{_{z}}\left[ 1\,-\,\zdgz\,-\,\frac{2}{3}(1+\ldgl)\right]
(\,u\bar{M}\,-\,\bar{u}M\,)^{2}
\nonumber\\
& &\,-\,\frac{1}{9}\left[\,3\,+\,(\ldhl-h)(1+\ldgl)\,\right]\bar{M}\!M
\nonumber\\
& &\,-\,\frac{1}{8\dilaton}
\left[\begin{array}{lll}&\,1\,+\,f\,+\,b\dilaton\ln(e^{-k}\bar{u}u/\mu^{6})&\\
&\,+\,\frac{2}{3}(\ldhl+b\dilaton)(1+\ldgl)& \end{array}\right]
(\,u\bar{M}\,+\,\bar{u}M\,)
\nonumber\\
& &\,+\,\frac{1}{4}h_{_{z}}(1-\zdgz)(F_{U}+\bar{F}_{\bar{U}})^{2}
\nonumber\\
& &\,+\,\left\{\begin{array}{lll} 
&\,\frac{1}{8\dilaton}\left[1\,+\,f\,+\,
b\dilaton\ln(e^{-k}\bar{u}u/\mu^{6})\right]& \\
&\,+\,\frac{1}{4\dilaton}(\ldhl+b\dilaton)(1-\zdgz)& \\
&\,-\,\frac{1}{6}h_{_{z}}(1+\ldgl)(\,u\bar{M}\,+\,\bar{u}M\,)&
\end{array}\right\}(F_{U}+\bar{F}_{\bar{U}})
\nonumber\\
& &\,+\,(h+b\dilaton)\sum_{I}\frac{1}{(t^{I}+\bar{t}^{I})^{2}}
\bar{F}^{I}F^{I}
\nonumber\\
& &\,-\,\frac{1}{16\dilaton^{2}}(\ldhl+h+2b\dilaton)(1+\ldgl)\bar{u}u.
\end{eqnarray}

\begin{eqnarray}
\frac{1}{e}\Lag_{\tilde{G}}\,&=&\,\frac{1}{2}\epsilon^{mnpq}
(\,\bar{\psi}_{m}\bar{\sigma}_{n}\!\nabla_{\!p}\psi_{q}\,-\,
\psi_{m}\sigma_{n}\!\nabla_{\!p}\bar{\psi}_{q}\,)
\nonumber\\
& &\,-\,\frac{1}{8\dilaton}\left[\,1\,+\,f\,+\,b\dilaton\ln(e^{-k}\bar{u}u/
\mu^{6})\,\right]\,\bar{u}\,(\psi_{m}\sigma^{mn}\psi_{n})
\nonumber\\
& &\,-\,\frac{1}{8\dilaton}\left[\,1\,+\,f\,+\,b\dilaton\ln(e^{-k}\bar{u}u/
\mu^{6})\,\right]\,u\,(\bar{\psi}_{m}\bar{\sigma}^{mn}\bar{\psi}_{n})
\nonumber\\
& &\,-\,\frac{1}{4}(h+b\dilaton)\sum_{I}\frac{1}{(t^{I}+\bar{t}^{I})}
\epsilon^{mnpq}(\bar{\psi}_{m}\bar{\sigma}_{n}\psi_{p})
(\,\nabla_{\!q}\bar{t}^{I}\,-\,\nabla_{\!q}t^{I}\,)
\nonumber\\
& &\,+\,\frac{i}{4\dilaton}(h+b\dilaton)(1+\ldgl)
(\,\eta^{mn}\eta^{pq}\,-\,\eta^{mq}\eta^{np}\,)
(\bar{\psi}_{m}\bar{\sigma}_{n}\psi_{p})\,\nabla_{\!q}\dilaton
\nonumber\\
& &\,-\,\frac{i}{4}\left[(1-\zdgz)(h+b\dilaton)-1\right]
(\,\eta^{mn}\eta^{pq}\,-\,\eta^{mq}\eta^{np}\,)
(\bar{\psi}_{m}\bar{\sigma}_{n}\psi_{p})\,
\nabla_{\!q}\ln(\bar{u}u)
\nonumber\\
& &\,+\,\frac{1}{4}(h-1+b\dilaton)\,\epsilon^{mnpq}
(\bar{\psi}_{m}\bar{\sigma}_{n}\psi_{p})\,
\nabla_{\!q}\!\ln\!\left(\frac{\bar{u}}{u}\right).
\end{eqnarray}

The bosonic Lagrangian $\Lag_{B}$ contains usual auxiliary fields and the
vector field $B_{m}$ which is dual to an axion. The details of this duality and
the structure of $\Lag_{B}$ will be discussed in the following sections. The
gravitino Lagrangian $\Lag_{\tilde{G}}$ is in its simplest form. An important 
physical quantity in $\Lag_{\tilde{G}}$ is the gravitino mass 
$m_{\tilde{G}}$ which is the natural order parameter measuring supersymmetry 
breaking. The expression of $m_{\tilde{G}}$ follows directly from 
$\Lag_{\tilde{G}}$:
\begin{equation}
m_{\tilde{G}}\,=\,\left\langle\,\left|\frac{1}{8\dilaton}\left[\,1\,+\,f\,
+\,b\dilaton\ln(e^{-k}\bar{u}u/\mu^{6})\,\right]u\right|\,\right\rangle.
\end{equation}
\subsection{Duality Transformation of $B_{m}$}
\hspace{0.8cm}
As pointed out in \cite{Binetruy95,Pillon}, the constraint (2.12) allows us to
interpret the degrees of freedom of $U$ as those of a 3-form supermultiplet, and
the vector field $B_{m}$ is dual to a 3-form $\Gamma^{npq}$. Since a 3-form is
dual to a 0-form in four dimensions, $B_{m}$ is also dual to a pseudoscalar 
$\axion$. In this section, we show explicitly how to rewrite the 
$B_{m}$ part of $\Lag_{B}$ in terms of the dual description using $\axion$.
According to (3.24), the $B_{m}$ terms in $\Lag_{B}$ are:
\begin{eqnarray}
\frac{1}{e}\Lag_{B}\;&\ni&\;
+\,\frac{(1+\ldgl)}{4\dilaton^{2}(1-\zdgz)}B^{m}\!B_{m}
\nonumber\\
\;& &\;+\,\frac{i}{2\dilaton}\left[h\,+\,b\dilaton\,-\,
\frac{1}{(1-\zdgz)}\right]\,B^{m}\,
\nabla_{\!m}\!\ln\!\left(\frac{\bar{u}}{u}\right)
\nonumber\\
\;& &\;-\,\frac{i}{2\dilaton}\left[h\,+\,b\dilaton\,-\,
\frac{1}{(1-\zdgz)}\right]\sum_{I}
\frac{(\,\nabla^{m}\bar{t}^{I}\,-\,\nabla^{m}t^{I}\,)}
{(t^{I}+\bar{t}^{I})}B_{m}
\nonumber\\
\;& &\;-\,2ih_{_{z}}\left[ 1\,-\,\zdgz\,-\,\frac{1}{3}(1+\ldgl)\right]
(\,u\bar{M}\,-\,\bar{u}M\,)\,\nabla^{m}\!B_{m}
\nonumber\\
\;& &\;+\,4h_{_{z}}(1-\zdgz)(\nabla^{m}\!B_{m})^{2}.
\end{eqnarray}
They are described by the following generic Lagrangian of $B_{m}$:
\begin{equation}
\frac{1}{e}\Lag_{B_{m}}\,=\,
\alpha B^{m}\!B_{m}\,+\,\beta\nabla^{m}\!B_{m}\,+\,\zeta^{m}\!B_{m}
\,+\,\tau(\nabla^{m}\!B_{m})^{2}.
\end{equation}
To find the dual description of $\Lag_{B_{m}}$, consider the following
Lagrangian $\Lag_{Dual}$.
\begin{equation}
\frac{1}{e}\Lag_{Dual}\,=\,
\alpha B^{m}\!B_{m}\,+\,\beta\nabla^{m}\!B_{m}\,+\,\zeta^{m}\!B_{m}
\,+\,\axion\nabla^{m}\!B_{m}\,-\,\frac{1}{4\tau}\axion^{2}.
\end{equation}
In $\Lag_{Dual}$, the auxiliary field $\axion$ acts like a Lagrangian
multiplier, and its equation of motion is:
\begin{equation}
\axion\,=\,2\tau\nabla^{m}\!B_{m}.
\end{equation}
Therefore, $\Lag_{B_{m}}$ follows directly from $\Lag_{Dual}$ using (3.30).
On the other hand, we can treat the $B_{m}$ in $\Lag_{Dual}$ as auxiliary, and 
write down the equation of motion for $B_{m}$ as follows:
\begin{equation}
B_{m}\,=\,\frac{1}{2\alpha}
\left(\,\nabla_{m}\axion\,+\,\nabla_{m}\beta\,-\,\zeta_{m}\,\right).
\end{equation}
Eliminating $B_{m}$ from $\Lag_{Dual}$ through (3.31) and then performing a 
field re-definition $\,\axion\,\Rightarrow\,\axion-\beta$, we obtain the 
Lagrangian $\Lag_{\axion}$ of $\axion$:
\begin{equation}
\frac{1}{e}\Lag_{\axion}\,=\,
-\,\frac{1}{4\alpha}\left(\,\nabla^{m}\!\axion\,-\,\zeta^{m}\,\right)
\left(\,\nabla_{m}\axion\,-\,\zeta_{m}\,\right)
\,-\,\frac{1}{4\tau}\left(\,\axion\,-\,\beta\,\right)^{2}.
\end{equation}
Therefore, $\Lag_{\axion}$ is the dual description of $\Lag_{B_{m}}$ in terms 
of $\axion$ which is interpreted as an axion. Notice that dynamical gaugino 
condensation naturally generates a mass term for the axion $\axion$ which
corresponds to the appearance of non-vanishing $\,(\nabla^{m}\!B_{m})^{2}\,$ 
in the dual description. The fact that $\axion$ is massive in dynamical 
gaugino condensation has already been observed in \cite{Binetruy95,Burgess95}. 
On the other hand, the $\,(\nabla^{m}\!B_{m})^{2}\,$ term vanishes in static 
gaugino condensation ({\it i.e.}, $\,h_{_{z}}=0\,$ in (3.27)), and it is found that 
the model-independent axion dual to $B_{m}$ is either massless or very light
\cite{dilaton,multiple,Binetruy95,Burgess95}. This issue of axion mass seems
to be a contradiction because we expect static gaugino condensation to be the 
appropriate effective description of dynamical gaugino condensation; the 
resolution is the following: In comparison with static gaugino condensation 
({\it e.g.}, \cite{dilaton,multiple}), dynamical gaugino condensation contains one 
more axionic degree of freedom $\axion$, and indeed $\axion$ is very massive 
({\it e.g.}, compared to the dilaton mass). As will be shown in Section 3.3.1,
after integrating out this massive axion $\axion$, the resulting axionic 
contents of dynamical gaugino condensation are identical to those of static 
gaugino condensation. Therefore, at low energy we are always left with a
massless or very light model-independent axion.

According to (3.27--28) and (3.32), the $\Lag_{eff}$ defined by (3.23--25) is 
rewritten in the dual description as follows:
\begin{equation}
\Lag_{eff}\,=\,\Lag_{kin}\,+\,\Lag_{pot}\,+\,\Lag_{\tilde{G}},
\end{equation}
where $\Lag_{kin}$ and $\Lag_{pot}$ refer to the kinetic part and the
non-kinetic part of the bosonic Lagrangian respectively. $\Lag_{\tilde{G}}$
is defined by (3.25). 

\begin{eqnarray}
\frac{1}{e}\Lag_{kin}\,&=&\,-\,\frac{1}{2}{\cal R}
\,-\,\frac{1}{4\dilaton^{2}}(h-\ldhl)(1+\ldgl)
\nabla^{m}\!\dilaton\,\nabla_{\!m}\!\dilaton
\nonumber\\
& &\,-\,\frac{(1-\zdgz)}{(1+\ldgl)}\dilaton^{2}
\nabla^{m}\!\axion\,\nabla_{\!m}\!\axion
\,+\,\frac{1}{2\dilaton}\zdhz(1+\ldgl)\,
\nabla^{m}\!\ln(\bar{u}u)\,\nabla_{\!m}\!\dilaton
\nonumber\\
& &\,+\,i\frac{(1-\zdgz)}{(1+\ldgl)}
\left[\,h\,+\,b\dilaton\,-\,\frac{1}{(1-\zdgz)}\,\right]
\dilaton\nabla^{m}\!\axion\,
\nabla_{\!m}\!\ln\!\left(\frac{\bar{u}}{u}\right)
\nonumber\\
& &\,-\,i\frac{(1-\zdgz)}{(1+\ldgl)}
\left[\,h\,+\,b\dilaton\,-\,\frac{1}{(1-\zdgz)}\,\right]
\sum_{I}\frac{(\,\nabla^{m}\bar{t}^{I}\,-\,\nabla^{m}t^{I}\,)}
{(t^{I}+\bar{t}^{I})}\,\dilaton\nabla_{\!m}\!\axion
\nonumber\\
& &\,+\,\frac{1}{4}
\left\{\begin{array}{lll}&\,\zdhz\!\cdot\!\zdgz\frac{(2-\zdgz)}{(1-\zdgz)}& \\
&\,+\,\frac{(1-\zdgz)}{(1+\ldgl)}\left[\,h\,+\,b\dilaton\,-\,
\frac{1}{(1-\zdgz)}\,\right]^{2}& \end{array}\right\}
\frac{1}{\bar{u}^{2}}\nabla^{m}\bar{u}\,\nabla_{\!m}\bar{u}
\nonumber\\
& &\,-\,\frac{1}{2}
\left\{\begin{array}{lll}
&\,\zdhz\left[\,\frac{(2-\zdgz)}{(1-\zdgz)}\,-\,\zdgz\,\right]& \\
&\,+\,\frac{(1-\zdgz)}{(1+\ldgl)}\left[\,h\,+\,b\dilaton\,-\,
\frac{1}{(1-\zdgz)}\,\right]^{2}& \end{array}\right\}
\frac{1}{\bar{u}u}\nabla^{m}\bar{u}\,\nabla_{\!m}u
\nonumber\\
& &\,+\,\frac{1}{4}
\left\{\begin{array}{lll}&\,\zdhz\!\cdot\!\zdgz\frac{(2-\zdgz)}{(1-\zdgz)}& \\
&\,+\,\frac{(1-\zdgz)}{(1+\ldgl)}\left[\,h\,+\,b\dilaton\,-\,
\frac{1}{(1-\zdgz)}\,\right]^{2}& \end{array}\right\}
\frac{1}{u^{2}}\nabla^{m}u\,\nabla_{\!m}u
\nonumber\\
& &\,-\,\frac{1}{2}
\left\{\begin{array}{lll}
&\,\frac{\zdhz}{(1-\zdgz)}& \\
&\,+\,\frac{(1-\zdgz)}{(1+\ldgl)}\left[\,h\,+\,b\dilaton\,-\,
\frac{1}{(1-\zdgz)}\,\right]^{2}& \end{array}\right\}
\sum_{I}\frac{(\,\nabla^{m}\bar{t}^{I}\,-\,\nabla^{m}t^{I}\,)}
{(t^{I}+\bar{t}^{I})}\nabla_{\!m}\!\ln\!\left(\frac{\bar{u}}{u}\right)
\nonumber\\
& &\,+\,\frac{1}{4}
\left\{\begin{array}{lll}
&\,\frac{\zdhz}{(1-\zdgz)}& \\
&\,+\,\frac{(1-\zdgz)}{(1+\ldgl)}\left[\,h\,+\,b\dilaton\,-\,
\frac{1}{(1-\zdgz)}\,\right]^{2}& \end{array}\right\}
\sum_{I,J}\frac{\nabla^{m}\bar{t}^{I}\,\nabla_{\!m}\bar{t}^{J}}
{(t^{I}+\bar{t}^{I})(t^{J}+\bar{t}^{J})}
\nonumber\\
& &\,-\,\frac{1}{2}\sum_{I,J}
\left\{\begin{array}{lll}
&\,2(h+b\dilaton)\delta_{IJ}\,+\,\frac{\zdhz}{(1-\zdgz)}& \\
&\,+\,\frac{(1-\zdgz)}{(1+\ldgl)}\left[\,h\,+\,b\dilaton\,-\,
\frac{1}{(1-\zdgz)}\,\right]^{2}& \end{array}\right\}
\frac{\nabla^{m}\bar{t}^{I}\,\nabla_{\!m}t^{J}}
{(t^{I}+\bar{t}^{I})(t^{J}+\bar{t}^{J})}
\nonumber\\
& &\,+\,\frac{1}{4}
\left\{\begin{array}{lll}
&\,\frac{\zdhz}{(1-\zdgz)}& \\
&\,+\,\frac{(1-\zdgz)}{(1+\ldgl)}\left[\,h\,+\,b\dilaton\,-\,
\frac{1}{(1-\zdgz)}\,\right]^{2}& \end{array}\right\}
\sum_{I,J}\frac{\nabla^{m}t^{I}\,\nabla_{\!m}t^{J}}
{(t^{I}+\bar{t}^{I})(t^{J}+\bar{t}^{J})}.
\end{eqnarray}

\begin{eqnarray}
\frac{1}{e}\Lag_{pot}\,&=&\,
\,\frac{h_{_{z}}(1+\ldgl)^{2}}{36(1-\zdgz)}(\,u\bar{M}\,-\,\bar{u}M\,)^{2}
\nonumber\\
& &\,-\,\frac{1}{9}\left[\,3\,+\,(\ldhl-h)(1+\ldgl)\,\right]\bar{M}\!M
\nonumber\\
& &\,-\,\frac{1}{8\dilaton}
\left[\begin{array}{lll}&\,1\,+\,f\,+\,b\dilaton\ln(e^{-k}\bar{u}u/\mu^{6})&\\
&\,+\,\frac{2}{3}(\ldhl+b\dilaton)(1+\ldgl)& \end{array}\right]
(\,u\bar{M}\,+\,\bar{u}M\,)
\nonumber\\
& &\,-\,\frac{i}{4}\left[\,1\,-\,\frac{(1+\ldgl)}{3(1-\zdgz)}\,\right]
\axion (\,u\bar{M}\,-\,\bar{u}M\,)
\nonumber\\
& &\,+\,\frac{1}{4}h_{_{z}}(1-\zdgz)(F_{U}+\bar{F}_{\bar{U}})^{2}
\nonumber\\
& &\,+\,\left\{\begin{array}{lll} 
&\,\frac{1}{8\dilaton}\left[1\,+\,f\,+\,
b\dilaton\ln(e^{-k}\bar{u}u/\mu^{6})\right]& \\
&\,+\,\frac{1}{4\dilaton}(\ldhl+b\dilaton)(1-\zdgz)& \\
&\,-\,\frac{1}{6}h_{_{z}}(1+\ldgl)(\,u\bar{M}\,+\,\bar{u}M\,)&
\end{array}\right\}(F_{U}+\bar{F}_{\bar{U}})
\nonumber\\
& &\,+\,(h+b\dilaton)\sum_{I}\frac{1}{(t^{I}+\bar{t}^{I})^{2}}
\bar{F}^{I}F^{I}
\nonumber\\
& &\,-\,\frac{1}{16\dilaton^{2}}(\ldhl+h+2b\dilaton)(1+\ldgl)\bar{u}u
\nonumber\\
& &\,-\,\frac{\bar{u}u}{16\zdhz(1-\zdgz)}\,\axion^{2}.
\end{eqnarray}
The $\,b^{a}b_{a}\,$ term has been eliminated by its equation of motion, 
$\,b^{a}=0$, and $\Lag_{kin}$ is in its simplest form. Note that the kinetic
terms of those axionic degrees of freedom $\axion$, $\,i\ln(\bar{u}/u)\,$ and
$\,i(\bar{t}^{I}-t^{I})\,$ are more complicated, which essentially reflects
the non-trivial constraint (2.12) satisfied by $U$ and $\bar{U}$. An important
issue is the structure of $\Lag_{pot}$, and it will be discussed in the next
section.
\subsection{The Scalar Potential}
\hspace{0.8cm}
It is straightforward to solve the equations of motion for the auxiliary fields
$\,b^{a}$, $\,F^{I}$, $\,\bar{F}^{I}$, $\,M$, $\,\bar{M}\,$ and 
$(F_{U}+\bar{F}_{\bar{U}})$ respectively as follows:
\begin{eqnarray}
b^{a}\,&=&\,0,
\nonumber\\
F^{I}\,&=&\,0, \;\;\;
\bar{F}^{I}\,=\,0,
\nonumber\\
M\,&=&\,-\,\frac{3}{8\dilaton}\left[\,1\,+\,f\,
+\,b\dilaton\ln(e^{-k}\bar{u}u/\mu^{6})\,\right]u
\,-\,\frac{3iu}{4}\axion,
\nonumber\\
\bar{M}\,&=&\,-\,\frac{3}{8\dilaton}\left[\,1\,+\,f\,
+\,b\dilaton\ln(e^{-k}\bar{u}u/\mu^{6})\,\right]\bar{u}
\,+\,\frac{3i\bar{u}}{4}\axion,
\nonumber\\
(F_{U}+\bar{F}_{\bar{U}})\,&=&\,\frac{(\ldhl-h)}{4\zdhz}
\left[\,1\,+\,f\,+\,b\dilaton\ln(e^{-k}\bar{u}u/\mu^{6})\,\right]
\frac{\bar{u}u}{\dilaton} \nonumber\\
& &\,-\,\frac{(\ldhl+b\dilaton)}{2\zdhz}\!\cdot\!\frac{\bar{u}u}{\dilaton}.
\end{eqnarray}
Note that $\,\langle\,|M|\,\rangle\,=\,3m_{\tilde{G}}\,$ because 
$\,\langle\,\axion\,\rangle=0$ always. To obtain the scalar potential, 
the auxiliary fields are eliminated from $\Lag_{eff}$ defined by (3.33), and 
$\Lag_{eff}$ is then rewritten as follows:
\begin{equation}
\frac{1}{e}\Lag_{eff}\,=\,\frac{1}{e}\Lag_{kin}\,-\,V_{pot}
\,+\,\frac{1}{e}\Lag_{\tilde{G}},
\end{equation}
where $V_{pot}$ is the scalar potential. $\,\Lag_{kin}\,$ and 
$\,\Lag_{\tilde{G}}\,$ are defined by (3.34) and (3.25) respectively. 

\begin{eqnarray}
V_{pot}\,&=&\,
\,\frac{1}{16}(\ldhl+h+2b\dilaton)(1+\ldgl)\frac{\bar{u}u}{\dilaton^{2}}
\nonumber\\
& &\,+\,\frac{1}{64\zdhz(1-\zdgz)}
\left\{\begin{array}{lll} 
&\,1\,+\,f\,+\,b\dilaton\ln(e^{-k}\bar{u}u/\mu^{6})& \\
&\,+\,2(\ldhl+b\dilaton)(1-\zdgz)&
\end{array}\right\}^{2}\frac{\bar{u}u}{\dilaton^{2}}
\nonumber\\
& &\,-\,\frac{(2-\ldgl-3\zdgz)}{64(1-\zdgz)}
\left[\,1\,+\,f\,+\,b\dilaton\ln(e^{-k}\bar{u}u/\mu^{6})\,\right]^{2}
\frac{\bar{u}u}{\dilaton^{2}}
\nonumber\\
& &\,+\,\frac{(h-\ldhl-3\zdhz)\bar{u}u}{16\zdhz}\,\axion^{2}.
\end{eqnarray}

Several interesting aspects of $V_{pot}$ can be uncovered. Firstly, 
there is always a trivial vacuum with 
$\,\left\langle\,V_{pot}\,\right\rangle=0\,$
in the specific weak-coupling limit defined as follows:
\begin{equation}
\dilaton\,\rightarrow\,0,\;\;\;
z\,\rightarrow\,\frac{1}{e^{2}}\dilaton\mu^{6}e^{-1/b\dilaton}
\,\rightarrow\,0,\;\;\;\mbox{and}\;\;\;
g(\dilaton,z),\;f(\dilaton,z)\,\rightarrow\,0.
\end{equation}
Note that quantum corrections to the K\"ahler potential, $g$ and $f$, should
vanish in this limit. As expected, this is consistent with the well-known 
runaway behavior of the dilaton near the weak-coupling limit. 

To proceed further, in the following of this section we only study 
$\,V_{pot}\,$ in the $\,z\ll 1\,$ regime. Since a physically interesting model 
of dynamical gaugino condensation should predict a small scale of condensation 
({\it i.e.}, $\,\langle\,z\,\rangle\ll 1$), there is no loss of generality in 
this choice. Note that in the $\,z\ll 1\,$ regime we have $\,h\approx 1$, 
$\,\ldhl\approx 0$, $\,\zdhz\approx 0\,$ and $\,\zdgz\approx 0\,$ up to small 
corrections that depend on $z$. The structure of $\,V_{pot}\,$ can be analyzed 
as follows: The only axion-dependent term in $\,V_{pot}\,$ is the effective 
axion mass term, the last term in $\,V_{pot}$. In order to avoid a tachyonic
axion, the sign of the effective axion mass term must be positive. Therefore,
the absence of a tachyonic axion requires $\,\zdhz >0$, which is the first 
piece of information about the $\,\bar{U}U$-dependence of the dynamical model. 
Furthermore, $\,\langle\,\axion\,\rangle=0$ always, and therefore the last 
term in $\,V_{pot}\,$ is of no significance in discussing the vacuum structure. 
Because of $\,\zdhz >0$, the second term in $\,V_{pot}\,$ is always positive. 
The signs of the first term and the third term in $\,V_{pot}\,$ remain 
undetermined in general; however, near the weak-coupling limit the first term 
is positive and the third term is negative (which is expected because the 
third term is the contribution of auxiliary fields $M$ and $\bar{M}$). Notice 
that the second term in $\,V_{pot}\,$ contains a factor $\,1/\zdhz\,$ 
($1/\zdhz\gg 1$), and therefore it is the dominant contribution to 
$\,V_{pot}\,$ except near the path $\,\gamma\,$ defined by 
$\,\left\{1+f+b\dilaton\ln(e^{-k}\bar{u}u/\mu^{6})
+2(\ldhl+b\dilaton)(1-\zdgz)\right\}=0$. Hence, the vacuum always sits close 
to the path $\,\gamma$. This observation will be essential to the following 
discussion of vacuum structure. 

The second piece of information about the $\,\bar{U}U$-dependence of the 
dynamical model can be obtained as follows. For $\,0<\dilaton<\infty$, the 
first term and the third term in $\,V_{pot}\,$ vanish in the limit 
$\,z\rightarrow 0\,$ generically. If $\,h_{_{z}}\,$ has a pole at $\,z=0$, 
then the second term in $\,V_{pot}\,$ also vanishes for $\,z\rightarrow 0\,$
and $\,0<\dilaton<\infty$. Therefore, for those dynamical models whose 
$\,h_{_{z}}\,$ has a pole at $\,z=0$, there exists a continuous family of 
degenerate vacua (parametrized by $\langle\,\dilaton\,\rangle$) with 
$\,\langle\,z\,\rangle=0$ (no gaugino condensation), $\,m_{\tilde{G}}=0$ 
(unbroken supersymmetry) and $\,\left\langle V_{pot}\right\rangle=0$. In other
words, in the vicinity of $z=0$ those models always exhibit runaway of $z$ 
toward the degenerate vacua at $z=0$ which do not have the desired physical 
features; whether those models may possess other non-trivial vacuum or not is 
outside the scope of this simple analysis.

On the other hand, the dynamical models whose $\,h_{_{z}}\,$ has no pole at 
$\,z=0\,$ are much more interesting. If $\,h_{_{z}}\,$ has no pole at $\,z=0$, 
then $\,V_{pot}\rightarrow\infty\,$ for $\,z\rightarrow 0\,$ and 
$\,0<\dilaton<\infty$. Therefore, these dynamical models exhibit no runaway of 
$z$ toward $\,z=0\,$ except for the weak-coupling limit (3.39). Furthermore, 
the equation of motion for $z$ is
\begin{equation}
1+f+b\dilaton\ln(e^{-k}\bar{u}u/\mu^{6})
+2(\ldhl+b\dilaton)(1-\zdgz)\,=\,0\,+\,\CO\(\zdhz\).
\end{equation}
Impose (3.40), and from (3.26) we have the gravitino mass 
$\,m_{\tilde{G}}\,=\,\frac{1}{4}b\langle\,|u|\,\rangle\,+\,
\CO\(z^{3/2}h_{_{z}}\)$. To the lowest order, it is identical to the 
$m_{\tilde{G}}$ of static gaugino condensation, (2.54); therefore,
similar to Section 2.3 we can argue that {\em supersymmetry is broken 
if and only if the dilaton is stabilized} for dynamical gaugino condensation.
In fact, for dynamical models whose $\,h_{_{z}}\,$ has no pole at $\,z=0$,
it can be shown that they are effectively described by static gaugino 
condensation of Chapter 2. As pointed out in Section 3.1, kinetic terms of 
the gaugino condensate $U$ naturally arise in generic string models, where
these terms are S-duality invariant and correspond to corrections 
$\,\bar{U}U/V^{2}$, $\,\left(\bar{U}U/V^{2}\right)^{2},\cdots$ to the K\"ahler 
potential. This interesting class of S-dual dynamical gaugino condensation
obviously belongs to dynamical models whose $\,h_{_{z}}\,$ has no pole at 
$\,z=0$ discussed here. In Section 3.3, S-dual dynamical gaugino condensation 
will be studied in detail.
\section{S-Dual Model of Dynamical Gaugino Condensation} 
\hspace{0.8cm}
As discussed in Section 3.1, we consider in this section models of dynamical 
gaugino condensation where the kinetic terms for gaugino condensate arise from 
the S-dual loop corrections defined by (3.2). More precisely, we consider the 
following dynamical model:
\begin{eqnarray}
K\,&=&\,\ln V\,+\,g(V,X)\,+\,G, 
\nonumber \\
\Lag_{eff}\,&=&\,
\superint\,E\,\left\{\,\left(\,-2\,+\,f(V,X)\,\right)\,+\,bVG\,
+\,bV\ln(e^{-K}\bar{U}U/\mu^{6})\,\right\},\hspace{1.5cm}
\end{eqnarray}
\begin{equation}
\left(2+X\frac{\partial f}{\partial X}\right)
\left(1-V\frac{\partial g}{\partial V}\right)=
\left(2-X\frac{\partial g}{\partial X}\right)
\left(1-f+V\frac{\partial f}{\partial V}\right).
\end{equation}
For convenience, we have written the S-dual combination
$\,(\bar{U}U)^{\frac{1}{2}}/V\,$ as a vector superfield $X$, and
therefore its lowest component $\,x\,=\,X\lowest\,$ is 
$x\,=\,(\bar{u}u)^{\frac{1}{2}}/\dilaton\,=\,\sqrt{z}/\dilaton$.
Eq. (3.42) guarantees the correct normalization of the Einstein term. 
$\,g(V,X)\,$ and $\,f(V,X)\,$ satisfy the boundary condition in the
weak-coupling limit defined by (3.39). We also assume that $\,g(V,X)\,$ and 
$\,f(V,X)\,$ have the following power-series representations\footnote{It should
be noted that one can actually start with a more generic dynamical model by
considering more generic $\,g(V,X)\,$ and $\,f(V,X)\,$, and the discussions of
Section 3.3 remain valid.} in terms of $X^{2}$:
\begin{eqnarray}
g(V,X)\,&\equiv&\,g^{(0)}(V)\,+\,g^{(1)}(V)\!\cdot\!X^{2}
\,+\,g^{(2)}(V)\!\cdot\!X^{4}\,+\,\cdots.
\nonumber\\
f(V,X)\,&\equiv&\,f^{(0)}(V)\,+\,f^{(1)}(V)\!\cdot\!X^{2}
\,+\,f^{(2)}(V)\!\cdot\!X^{4}\,+\,\cdots.
\end{eqnarray}
Furthermore, $\,g^{(n)}(V)\,$ and $\,f^{(n)}(V)\,$ ($n\geq 0$) are assumed to 
be arbitrary but bounded here. The interpretation of each term in (3.43) is 
obvious: As has been discussed in Section 2.2.2, in the linear multiplet
formalism $\,g^{(0)}(V)\,$ and $\,f^{(0)}(V)\,$ are to be identified as 
stringy (non-perturbative) corrections to the K\"ahler potential. 
$\,g^{(n)}(V)\!\cdot\!X^{2n}\,$ and \mbox{$\,f^{(n)}(V)\!\cdot\!X^{2n}\,$} 
($n\geq 1$) are therefore S-dual loop corrections to the K\"ahler potential 
in the presence of stringy (non-perturbative) effects. 

It is also more convenient to use the coordinates $\,(\,\dilaton,\,x\,)\,$ 
instead of $\,(\,\dilaton,\,z\,)\,$ for the field configuration space. The 
component field expressions constructed in Section 3.2 can easily be 
rewritten in the new coordinates $\,(\,\dilaton,\,x\,)\,$ according to the 
following rules:
\begin{eqnarray}
\ldgl\;&\rightarrow&\;\ldgl\,-\,\xdgx \nonumber\\
\zdgz\;&\rightarrow&\;\frac{1}{2}\xdgx,
\end{eqnarray}
where
\begin{equation}
g_{_{\dilaton}}\,\equiv\,\frac{\partial g(\dilaton,x)}{\partial \dilaton},
\;\;\;g_{_{x}}\,\equiv\,\frac{\partial g(\dilaton,x)}{\partial x}
\end{equation}
on the right-hand side of (3.44) are to be understood as partial derivatives in 
the coordinates $\,(\,\dilaton,\,x\,)$. The scalar potential of this generic 
model follows directly from (3.38):
\begin{eqnarray}
V_{pot}\,&=&\,\frac{1}{16}(\,1\,+\,\ldgl\,-\,\xdgx\,)
(\,h\,+\,\ldhl\,-\,\xdhx\,+\,2b\dilaton\,)\,x^{2}
\nonumber\\
& &\,+\,\frac{1}{16\xdhx(\,2\,-\,\xdgx\,)}
\left\{\begin{array}{lll} 
&\,1\,+\,f\,+\,b\dilaton\ln(e^{-k}\bar{u}u/\mu^{6})& \\
&\,+\,(\,2\,-\,\xdgx\,)(\,\ldhl\,-\,\xdhx\,+\,b\dilaton\,)&
\end{array}\right\}^{2}x^{2}
\nonumber\\
& &\,-\,\frac{(\,4\,-\,2\ldgl\,-\,\xdgx\,)}{64(\,2\,-\,\xdgx\,)}
\left[\,1\,+\,f\,+\,b\dilaton\ln(e^{-k}\bar{u}u/\mu^{6})\,\right]^{2}x^{2}
\nonumber\\
& &\,+\,\frac{(\,2h\,-\,2\ldhl\,-\,\xdhx\,)\bar{u}u}{16\xdhx}\,\axion^{2}.
\end{eqnarray}
The kinetic terms also follow directly from (3.34). The absence of a tachyonic 
axion requires $\,\xdhx>0$. 
\subsection{Effective Description of Dynamical Gaugino Condensation}
\hspace{0.8cm}
As discussed in Section 3.1, one of the major motivations for studying
dynamical
gaugino condensation is to understand how static gaugino condensation could
emerge as the effective description of dynamical gaugino condensation after all
the heavy modes belonging to dynamical gaugino condensation are integrated out.
Unlike studies in the past where the important constraint (2.12) on the gaugino
condensate chiral superfield $U$ is ignored, proving the above connection is
certainly non-trivial. From this point of view, our construction in Section 3.2
can be regarded as efforts to solve (2.12) in the context of dynamical gaugino
condensation using the linear multiplet formalism, and the above connection is 
actually obvious after (2.12) is explicitly solved. In order to make the 
following discussion as explicit as possible, in this section we choose to 
study S-dual dynamical gaugino condensation. However, we 
would like to emphasize that our discussion is actually valid for any 
dynamical model whose $\,h_{_{z}}\,$ has no pole at $\,z=0$. 

Firstly, the axionic contents of dynamical gaugino condensation are $\axion$,
$\,i\ln(\bar{u}/u)\,$ and $\,i(\bar{t}^{I}-t^{I})$.
Since a physically interesting model of dynamical gaugino condensation should 
predict a small scale of condensation ({\it i.e.}, 
$\,\langle\,x\,\rangle\ll 1$), it is clear from (3.46) that generally the 
condensate $x$ and the axion $\axion$ are much heavier than the other fields, 
and therefore should be integrated out. It is straightforward to integrate out 
$\axion$ and $x$ through their equations of motion: The equation of motion for 
$\axion$ is $\,\axion=0$. The equation of motion for $x$ is:
\begin{equation}
1\,+\,f\,+\,b\dilaton\ln(e^{-k}\bar{u}u/\mu^{6})\,+\,(\,2\,-\,\xdgx\,)
(\,\ldhl\,-\,\xdhx\,+\,b\dilaton\,)\,=\,0\,+\,\CO(x^2).
\end{equation}
(3.47) can be re-written in a more instructive form:
\begin{equation}
x^{2}\;=\;\frac{\mu^{6}}{e^{2}\dilaton}\,
e^{g^{(0)}\,-\,\left({1+f^{(0)}}\right)/{b\dilaton}}\,+\,
\CO(x^4),
\end{equation}
where we have used the fact that \mbox{$\,g\approx g^{(0)}$}, 
\mbox{$\,f\approx f^{(0)}$}, \mbox{$\,h\approx 1$}, 
\mbox{$\,\ldgl\approx \dilaton g^{(0)}_{_{\dilaton}}$}, 
\mbox{$\,\ldfl\approx \dilaton f^{(0)}_{_{\dilaton}}$}, 
\mbox{$\,\ldhl\approx 0$}, \mbox{$\,\xdgx\approx 0$}, 
\mbox{$\,\xdfx\approx 0\,$} and \mbox{$\,\xdhx\approx 0\,$} up to corrections 
of order $\CO(x^2)$. The (bosonic) effective Lagrangian, 
$\,\Lag_{eff}\,=\,\Lag_{kin}\,-\,eV_{pot}$, of the dynamical model (3.34,46) 
after integrating out $\axion$ and $x$ is as follows:

\begin{eqnarray}
\frac{1}{e}\Lag_{kin}\,&=&\,-\,\frac{1}{2}{\cal R}
\,-\,\frac{1}{4\dilaton^{2}}\left(1+\dilaton g^{(0)}_{_{\dilaton}}\right)
\nabla^{m}\!\dilaton\,\nabla_{\!m}\!\dilaton
\nonumber\\
& &\,-\,(1+b\dilaton)\sum_{I}\frac{1}{(t^{I}+\bar{t}^{I})^{2}}
\nabla^{m}\bar{t}^{I}\,\nabla_{\!m}t^{I}
\,+\,\frac{1}{4\dilaton^{2}}\left(1+\dilaton g^{(0)}_{_{\dilaton}}\right)
\tilde{B}^{m}\!\tilde{B}_{m}
\nonumber\\
& &\,+\,\CO(x^2),
\end{eqnarray}
where
\begin{eqnarray}
\tilde{B}_{m}\,&\equiv&
\,-\,i\frac{b\dilaton^{2}}{\left(1+\dilaton g^{(0)}_{_{\dilaton}}\right)}
\nabla_{\!m}\!\ln(\frac{\bar{u}}{u}) \nonumber\\
& &\,+\,i\frac{b\dilaton^{2}}{\left(1+\dilaton g^{(0)}_{_{\dilaton}}\right)}
\sum_{I}\frac{1}{(t^{I}+\bar{t}^{I})}
(\,\nabla_{\!m}\bar{t}^{I}\,-\,\nabla_{\!m}t^{I}\,).
\end{eqnarray}

\begin{eqnarray}
V_{pot}\,&=&\,\frac{1}{16e^{2}\dilaton}\left\{\,
\left(1+f^{(0)}-\dilaton f^{(0)}_{_{\dilaton}}\right)
\left(1+b\dilaton\right)^{2}\,-\,3b^{2}\dilaton^{2}\,\right\}\mu^{6}
e^{g^{(0)}\,-\,\left({1+f^{(0)}}\right)/{b\dilaton}} 
\nonumber\\
& &\,+\,\CO(x^4).
\end{eqnarray}
Furthermore, (3.42) leads to $\,\dilaton g^{(0)}_{_{\dilaton}}=
f^{(0)}-\dilaton f^{(0)}_{_{\dilaton}}\,$ to the lowest order in $x^2$.

In comparison with static gaugino condensation studied in Chapter 2, 
it is clear that the effective Lagrangian of dynamical gaugino condensation
after integrating out the heavy fields are indeed identical to the Lagrangian
of the static model, (2.46), to the lowest order in $x^2$. 
Note that, in (3.51), the $\CO(x^4)$ terms do not depend on the remaining 
axionic degrees of freedom (i.e., $\,i\ln(\bar{u}/u)\,$ and 
$\,i(\bar{t}^{I}-t^{I})$), and therefore these remaining axions are massless 
as they should be in static gaugino condensation\footnote{As pointed out in 
\cite{multiple} as well as in Chapter 4 here, 
these axionic degrees of freedom naturally acquire different 
masses in scenarios of multiple gaugino condensation.} \cite{dilaton}.
In conclusion, after integrating out the heavy modes the axions left in the 
effective theory of dynamical gaugino condensation are identical to those of
static gaugino condensation. Consistently there is always a massless (or
very light in multiple gaugino condensation \cite{multiple}) model-independent 
axion. According to the equation of motion for $x$, (3.48), $\,x^{2}\ll 1\,$ 
actually holds for any value of $\dilaton$. It implies that only the 
lowest-order terms of (3.49) and (3.51) are important, and, as we have expected 
and now prove here, the static model of gaugino condensation is indeed the 
appropriate effective description of the dynamical model. This proof therefore 
justifies the use of static gaugino condensation in Chapter 2. 

This proof also implies that the necessary and sufficient condition for 
$V_{pot}$ of dynamical gaugino condensation to be bounded from below is 
exactly the same as that of static gaugino condensation (2.57),
\begin{equation}
f^{(0)}-\dilaton f^{(0)}_{_{\dilaton}}\,\geq\,2\;\;\;\;\;\;
\mbox{for}\;\;\;\dilaton\,\rightarrow\,\infty,
\end{equation}
which depends only on stringy non-perturbative effects $g^{(0)}$ and 
$f^{(0)}$. (3.52) does not depend on the details of S-dual loop corrections, 
and therefore it holds for generic S-dual dynamical models. Furthermore, 
(3.52) implies that only stringy non-perturbative effects are important in
stabilizing the dilaton, and therefore allowing supersymmetry breaking via 
gaugino condensation. S-dual loop corrections play no role in this issue, 
and S-dual loop corrections alone cannot stabilize the dilaton. As discussed 
in Section 2.4, (3.52) can also be interpreted as the necessary condition 
for the dilaton to be stabilized.
\subsection{Solving for Dynamical Gaugino Condensation}
\hspace{0.8cm}
In the previous section, the dynamical model of gaugino condensation is 
analyzed through its effective Lagrangian after integrating out the heavy 
modes. One can also analyze the dynamical model directly, and obtain the same 
conclusion. Here, we would like to present a typical example of dynamical 
gaugino condensation as a concrete supplement to the analysis of Section 3.3.1. 
Solving for dynamical gaugino condensation is generically difficult due to the 
partial differential equation, (3.20) or (3.42), which guarantees the correct 
normalization of the Einstein term. On the other hand, only those solutions of 
(3.20) which are of physical interest deserve study. Therefore, in the 
following we show explicitly how to construct the solution for the interesting 
S-dual model of dynamical gaugino condensation defined by (3.41--43). In order 
to simplify the presentation but leave the generality of our conclusion 
unaffected, we choose a specific form for $f(V,X)$ in the following discussion: 
\mbox{$\,f(V,X)\,=\,f^{(0)}(V)\,+\,\ep X^{2}$,} where $\ep$ is a constant and 
$|\ep|$ is in principle a small number because $X$-dependent terms arise from 
loop corrections. In this restricted solution space, (3.42) together with the 
boundary condition (3.39) can be re-expressed as an infinite number of 
ordinary differential equations with appropriate boundary conditions 
(evaluated at \mbox{$\theta =\bar{\theta}=0$}) as follows:
\begin{eqnarray}
\dilaton g^{(0)}_{_{\dilaton}}&=&f^{(0)}-\dilaton f^{(0)}_{_{\dilaton}}.
\nonumber\\
\dilaton g^{(1)}_{_{\dilaton}}-
\left(1-f^{(0)}+\dilaton f^{(0)}_{_{\dilaton}}\right)g^{(1)}&=&
-\ep\!\cdot\!\dilaton g^{(0)}_{_{\dilaton}}+2\ep.
\nonumber\\
\dilaton g^{(n)}_{_{\dilaton}}-
n\left(1-f^{(0)}+\dilaton f^{(0)}_{_{\dilaton}}\right)g^{(n)}&=&
-\ep\!\cdot\!\dilaton g^{(n-1)}_{_{\dilaton}}-\ep(n-1)g^{(n-1)},
\nonumber\\
& &\mbox{for}\;\;\;n\geq 2.
\end{eqnarray}
The associated boundary conditions in the weak-coupling limit are:
\begin{eqnarray}
& &g^{(0)}(\dilaton=0)\,=\,0,\;\;\;f^{(0)}(\dilaton=0)\,=\,0,
\nonumber\\
& &g^{(1)}(\dilaton=0)\,=\,-2\ep,
\nonumber\\
& &g^{(n)}(\dilaton=0)\,=\,-\,\frac{2}{n}\,\ep^{n}
\;\;\;\;\;\mbox{for}\;\;\;n\geq 2.
\end{eqnarray}
Therefore, $\,g(V,X)\,$ is unambiguously\footnote{In fact, there is one free
parameter $\beta$ involved due to the fact that 
$\,g^{(n)}_{_{\dilaton}}(\dilaton=0)\,$ is not well-defined in (4.15); this 
ambiguity can be parametrized by 
$\,g^{(n)}_{_{\dilaton}}(\dilaton=0)\,=\,n\ep^{n-1}\beta$. We take $\beta=0$
here.} related to $\,f(V,X)\,$ in this interesting solution space. 

Firstly, notice that the boundedness of $g^{(n)}$ and $f^{(n)}$ can be 
guaranteed if (3.52) is satisfied. Therefore, the solution defined by 
(3.53--54)\footnote{The generalization to generic $f(V,X)$ is straightforward.} 
exists at least for viable dynamical models in the sense of (3.52). Secondly, 
$g^{(n)}$ is suppressed by a small factor $|\ep|^{n}$, which is obvious from 
(3.53--54). Therefore, the solution defined by (3.53--54) converges for 
$\,x^{2}<\CO\left(1/\ep\right)$. Since a physically interesting model 
of gaugino condensation should predict a small scale of condensation 
({\it i.e.}, $\,\langle\,x^{2}\,\rangle\ll 1$), this solution does cover the 
regime of physical interest.\footnote{This solution can in principle be 
extended into the $\,x^{2}>\CO\left(1/\ep\right)\,$ regime using the method of 
characteristics.} 

(3.52) is the necessary condition for stringy non-perturbative effects to
stabilize the dilaton. By looking into the details of the scalar potential, it
can also be argued \cite{dilaton} that stringy non-perturbative corrections to 
the K\"ahler potential may naturally stabilize the dilaton if (3.52) is 
satisfied. In the following, the solution defined by (3.53--54) is used to 
construct a typical realization of this argument. Furthermore, it is the 
typical feature of this example rather than the specific form of $g(V,X)$ and 
$f(V,X)$ assumed in this example that we want to emphasize. 
\begin{figure}
\epsfxsize=12cm
\epsfysize=8cm
\epsfbox{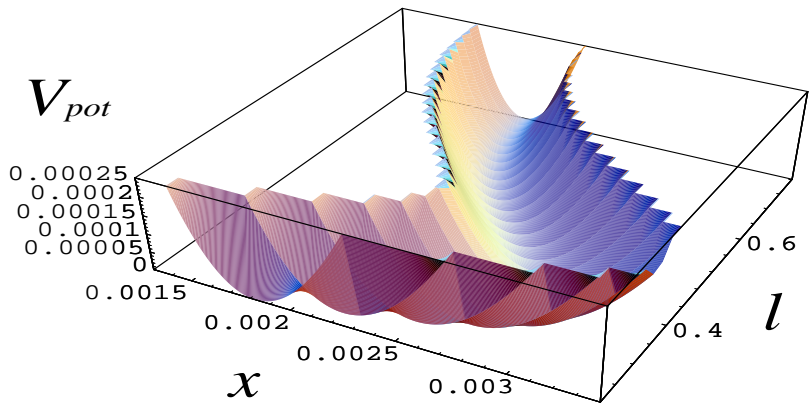}
\caption{The scalar potential $\,V_{pot}\,$ (in reduced Planck units)
            is plotted versus $\,\dilaton\,$ and $\,x$. $\,A=6.8$, $\,B=1$, 
            $\,\ep=-0.1\,$ and $\,\mu$=1. (The rippled surface of $V_{pot}$ 
            is simply due to discretization of the $\dilaton$-axis.)}
\end{figure}
In \mbox{Fig. 3.1}, the scalar potential $\,V_{pot}\,$ is plotted versus 
$\,\dilaton\,$ and $\,x\,$ for an example with 
\mbox{$\,f(V,X)=f^{(0)}(V)+\ep X^{2}\,$} and 
\mbox{$\,f^{(0)}(V)=A\!\cdot\!e^{-B/V}$}. There is a non-trivial vacuum with 
the dilaton stabilized at $\langle\,\dilaton\,\rangle=0.52$, $\,x\,$ stabilized 
at $\,\langle\,x\,\rangle=\langle\,\sqrt{\bar{u}u}/\dilaton\,\rangle=0.0024$, 
and (fine-tuned) vanishing vacuum energy 
$\,\left\langle\,V_{pot}\,\right\rangle=0$. 
Supersymmetry is broken at the vacuum and the gravitino mass 
$\,m_{\tilde{G}}=4\times 10^{-4}\,$ in reduced Planck units. 
To uncover more details of dilaton stabilization in \mbox{Fig. 3.1}, a cross 
section of $\,V_{pot}\,$ is presented in \mbox{Fig. 3.3}. More precisely, with 
the value of $\dilaton\,$ fixed, $\,V_{pot}\,$ is minimized only with respect 
to $\,x$; the location of this minimum is denoted as 
\mbox{($\dilaton$, $x_{min}(\dilaton)$).} The path defined by 
\mbox{($\dilaton$, $x_{min}(\dilaton)$)} is shown in \mbox{Fig. 3.2}. 
\begin{figure}
\epsfxsize=12cm
\epsfysize=8cm
\epsfbox{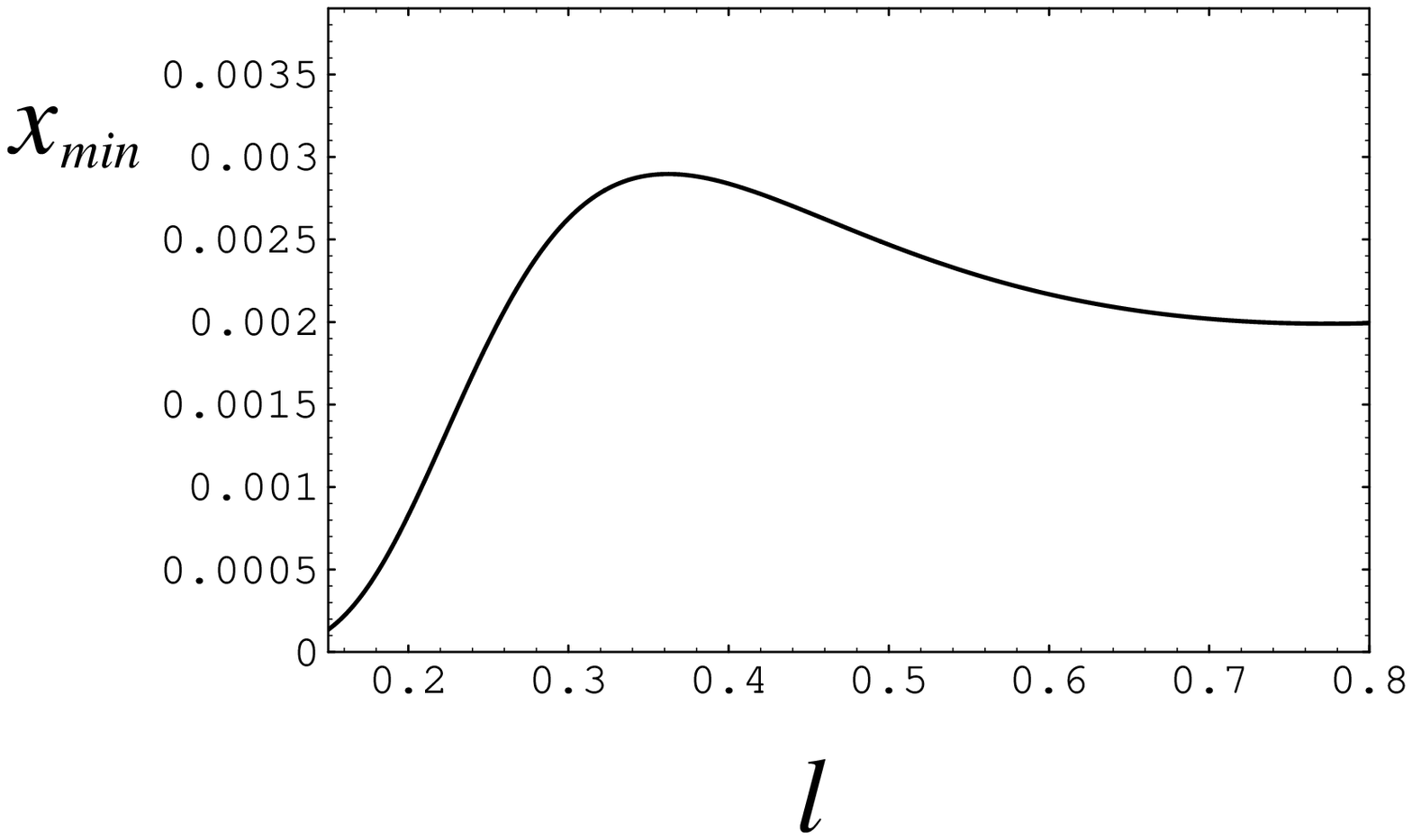}
\caption{$\,x_{min}(\dilaton)\,$ is plotted versus $\,\dilaton\,$ for 
\mbox{Figure 3.1.}}
\end{figure}
The cross section of $\,V_{pot}\,$ is obtained by making a cut along 
\mbox{($\dilaton$, $x_{min}(\dilaton)$)}; that is, the cross section of 
$\,V_{pot}\,$ is defined as \mbox{$\,V'_{pot}(\dilaton)\equiv 
V_{pot}\left(\dilaton,x_{min}(\dilaton)\right)$}. 
\begin{figure}
\epsfxsize=12cm
\epsfysize=8cm
\epsfbox{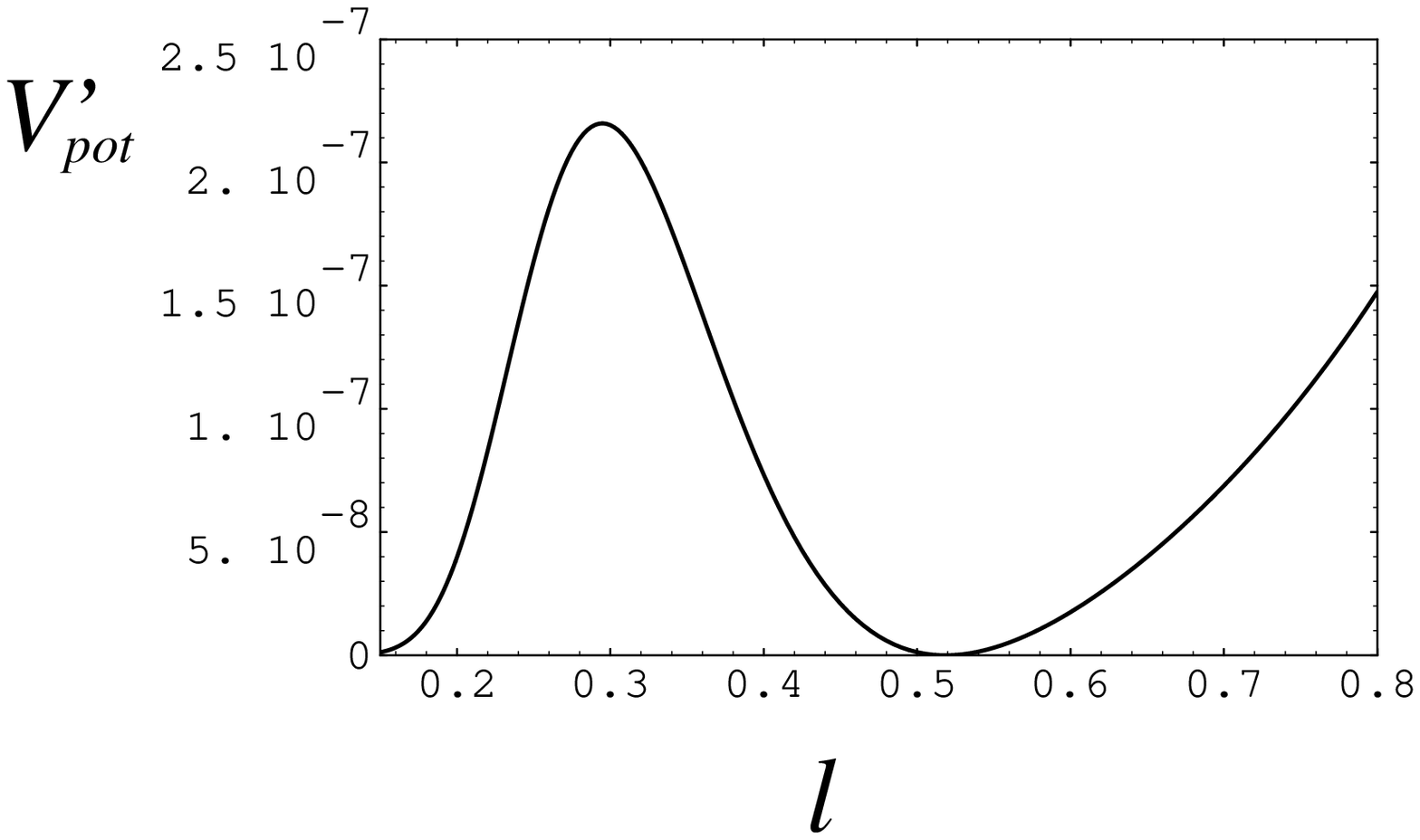}
\caption{The cross section of the scalar potential, 
   $V'_{pot}(\dilaton)\equiv V_{pot}\left(\dilaton,x_{min}(\dilaton)\right)\,$ 
   (in reduced Planck units), is plotted versus $\,\dilaton\,$ for 
   \mbox{Figure 3.1.}}
\end{figure}
\mbox{Fig. 3.3} shows that the dilaton is indeed stabilized at 
\mbox{$\langle\,\dilaton\,\rangle=0.52$.} Therefore, we have presented a 
concrete example with stabilized dilaton, broken supersymmetry, and 
(fine-tuned) vanishing cosmological constant. As pointed out in Sections 2.1 
and 2.5, this is in contrast with condensate models studied previously 
\cite{racetrack,Dine85,constant} which either need the assistance of an
additional source of supersymmetry breaking or have a large and negative 
cosmological constant problem. 
\section{Concluding Remarks}
\hspace{0.8cm}
The field component Lagrangian for the linear multiplet formalism of generic 
dynamical gaugino condensation is constructed and studied. A major conclusion 
of this chapter is that static gaugino condensation is indeed the appropriate
effective description of dynamical gaugino condensation after the heavy modes
are integrated out. Some issues about the axions are also clarified. This 
justifies our studies in Chapter 2, and allows us to use static gaugino 
condensation in constructing more realistic models in Chapter 4.
\newpage
\setcounter{chapter}{3} 
\chapter{Gaugino and Matter Condensation in Generic String Models}
\hspace{0.8cm}
\setcounter{equation}{0} 
\setcounter{figure}{0} 
\setcounter{footnote}{0}
\newpage
 
\section{Introduction}
It was recently shown how to formulate gaugino condensation using the linear 
multiplet \cite{Gates,Linear} formalism for the dilaton superfield, both in 
global supersymmetry~\cite{Binetruy95,Burgess95} and in the superconformal 
formulation of supergravity~\cite{Binetruy95}. Using the K\"ahler superspace 
formulation of supergravity~\cite{Binetruy90,Binetruy91}, which we use 
throughout this study, it was subsequently shown~\cite{sduality} how to 
include the Green-Schwarz term for a string model with a pure Yang-Mills $E_8$ 
hidden sector. In this case there are no moduli-dependent threshold corrections
and there is a single constant -- the $E_8$ Casimir $C$ -- that governs both 
the Green-Schwarz counterterm and the coupling renormalization. This model of 
gaugino condensation has been studied in detail in Chapters 2 and 3, where it 
was found that the dilaton can be stabilized at a phenomenologically 
acceptable value with broken supersymmetry if stringy non-perturbative
corrections \cite{Shenker90,Banks94} to the K\"ahler potential are included.
However, the model studied in Chapters 2 and 3 has several drawbacks from the
viewpoint of phenomenology. As discussed in Section 2.5, due to the large 
gauge content of $E_{8}$ a sufficiently large gauge hierarchy is not generated. 
Furthermore, the string moduli $T^{I}$ remain flat directions. As we have 
pointed out, these unsatisfactory features belong only to the specific string 
model with with a pure Yang-Mills $E_8$ hidden sector, and therefore are not
generic at all. As we will see, in a generic string model the hidden sector 
contains a product of smaller gauge groups. Therefore, a large enough gauge 
hierarchy could be generated naturally. Furthermore, a generic string model 
contains hidden matter, and together with string threshold corrections the 
hidden matter condensation lifts the flat directions associated with the 
moduli. 

Consider a generic string model whose hidden sector gauge group is a product
of simple groups: $\G=\prod_a\G_a$. One immediate difficulty is the following: 
since we need to describe several gaugino condensates $U_a\simeq{\WaWa}_a$ and
each gaugino condensate $U_a$ is constrained by (2.12) separately, therefore 
according to (2.13) we need to introduce several vector superfields $V_a$.
However, since the theory has a single dilaton $\ell$, it must be identified 
with the lowest component of $V = \sum_a V_a$. What should we do with the other
components $\ell_a = V_a|_{\theta = \bar \theta = 0}$? We will see that, in 
our description, these are non-propagating degrees of freedom which actually do 
not appear in the Lagrangian. Similarly only one antisymmetric tensor field 
(also associated with $V=\sum_a V_a$) is dynamical. This allows us to 
generalize our approach to the case of multiple gaugino condensation.

Let us stress that the goal is very different from the so-called ``racetrack'' 
ideas~\cite{racetrack} where resorting to multiple gaugino condensation is 
necessary in order to get supersymmetry breaking. Here supersymmetry is broken 
already for a single gaugino condensate. Indeed, we will see that the picture 
which emerges from multiple gaugino condensation (complete with threshold 
corrections and Green-Schwarz mechanism) is very different from the standard 
``racetrack'' description: indeed, the scalar potential is largely dominant 
by the condensate with the largest one-loop beta-function coefficient.

To be more precise, we generalize in this chapter the Lagrangian (2.26) studied
in Chapter 2 to string models with arbitrary hidden sector gauge groups and 
with three untwisted (1,1) moduli $T^{I}$. We take the K\"ahler potential for 
the effective theory at the condensation scale to be:
\begin{eqnarray} 
K &=& k(V) + \sum_Ig^I,\;\;\;\; 
g^I = -\ln(T^I + \bar{T}^I), \;\;\;\; V = \sum_{a=1}^nV_a, 
\end{eqnarray}
where the $V_a$ are vector superfields and $n$ is the number of (asymptotically 
free) nonabelian gauge groups $\G_a$ in the hidden sector:
\begin{equation}
\G_{\rm hidden} = \prod_{a=1}^n\G_a\otimes U(1)^m. 
\end{equation}
We will take $\,\G_{\rm hidden}\,$ to be a subgroup of E$_{8}$. In general,
there will be hidden matter associated with the hidden sector gauge groups.

We introduce both gaugino condensate superfields $U_a$ and hidden 
matter condensate superfields $\Pi^\alpha$ that are non-propagating:
\begin{equation}
U_a\simeq {\WaWa}_a,\quad \Pi^\alpha\simeq 
\prod_A\(\Phi^A\)^{n^A_\alpha},
\end{equation}
where ${\cal W}_a$ and $\Phi^A$ are the gauge and matter chiral superfields,
respectively. The matter condensate $\Pi^\alpha$ is a chiral superfield of 
K\"ahler weight $w=0$, while the gaugino condensate $U_a$ associated with 
gauge subgroup $\G_a$ is a chiral superfield of K\"ahler weight $w=2$, and is 
identified with the chiral projection of $V_a$:
\begin{equation}
U_a\,=\,-(\DbDb-8R)V_a, \;\;\;\; \bar{U}_a\,=\,
-(\DaDa-8R^{\dagger})V_a.
\end{equation}
We are thus introducing $n$ scalar fields $\ell_a = V_a\lowest$. However only 
one of these is physical, namely $\ell = \sum_a\ell_a$; the others do not 
appear in the effective component Lagrangian constructed below.

The effective Lagrangian for multiple gaugino condensation is constructed and 
analyzed in Sections 4.2--4.5. In an appendix we discuss a parallel 
construction using the chiral supermultiplet representation for the dilaton 
and unconstrained chiral supermultiplets for the gaugino condensates in order 
to illustrate the differences between the two approaches and the significance 
of including the constraints (4.4). 
\section{Construction of the Effective Lagrangian}
We adopt the following superfield Lagrangian:
\begin{equation}
\Lag_{eff} = \Lag_{KE} + \Lag_{GS} + \Lag_{th} + 
\Lag_{VY} + \Lag_{pot},
\end{equation}
where
\begin{equation}
\Lag_{KE} = \superint\,E \[-2 + f(V)\], \quad k(V) = \ln\,V + g(V),  
\end{equation}
is the kinetic energy term for the dilaton, chiral and gravity superfields. 
The functions $f(V),g(V)$ parameterize stringy nonperturbative effects. 
According to (2.8), they are related by the following first-order differential
equation:
\begin{equation}
V\frac{\diff g(V)}{\diff V}\,=\,
-V\frac{\diff f(V)}{\diff V}\,+\,f,
\end{equation}
which ensures that the Einstein term has canonical form~\cite{dilaton}. In the 
classical limit $g=f=0$; we therefore impose the boundary condition at the 
weak-coupling limit:
\begin{equation}
g(V=0)\,=\,0 \;\;\;\mbox{and}\;\;\; f(V=0)\,=\,0. 
\end{equation}
Two counterterms are introduced to cancel the modular anomaly 
\cite{Gaillard92}, namely the Green-Schwarz counterterm 
\cite{Derendinger92,Ovrut93}:
\begin{equation}
\Lag_{GS} = b\superint\,E V\sum_Ig^I,\quad 
b = {C\over8\pi^2}, 
\end{equation}
and the term induced by string loop corrections~\cite{Dixon90}:
\begin{eqnarray}
\Lag_{th} &=& -\sum_{a,I}{b_a^I\over64\pi^2}\superint\,{E\over R}U_a
\ln\eta^2(T^I) + {\rm h.c.}.
\end{eqnarray}
The parameters
\begin{equation}
b^I_a = C - C_a + \sum_A\(1 - 2q^A_I\)C^A_a, \quad C = C_{E_{8}},
\end{equation}
vanish for orbifold compactifications with no $N=2$ supersymmetry sector 
\cite{ant}. Here $C_a$ and $C_a^A$ are quadratic Casimir operators in the 
adjoint and matter representations, respectively. $\;q^A_I$ are the modular 
weights of the matter superfields $\Phi^A$ of the underlying hidden sector. 
The term
\begin{equation}
\Lag_{VY} = 
\sum_a{1\over8}\superint\,{E\over R}U_a\[b'_a\ln(e^{-K/2}U_a/\mu^{3})
 + \sum_\alpha b^\alpha_a\ln\Pi^\alpha\] + {\rm h.c.}, 
\end{equation}
where $\mu$ is a mass parameter naturally of order one in reduced Planck units 
(which we will set to unity hereafter), is the generalization to supergravity 
\cite{tom,bg89} of the Veneziano-Yankielowicz superpotential term generated by
condensation, including~\cite{matter} the gauge invariant composite matter 
fields $\Pi^\alpha$ introduced in eq. (4.3) (one can also take linear 
combinations of such gauge invariant monomials that have the same modular 
weight).  Finally
\begin{eqnarray}
\Lag_{pot} &=& {1\over2}\superint\,{E\over R}e^{K/2}W(\Pi^\alpha,T^I) + 
{\rm h.c.} 
\end{eqnarray}
is a superpotential for the hidden matter condensates $\Pi^\alpha$ that 
respects the symmetries of the superpotential $W(\Phi^A,T^I)$ of the 
underlying theory.

The coefficients $b'_a$ and $b^\alpha_a$ in (4.12) are dictated by the chiral 
and conformal anomalies of the underlying field theory. Under modular 
transformations, we have:
\begin{eqnarray}
T^{I}&\rightarrow&\frac{aT^{I}-ib}{icT^{I}+d},\;\;\;\;
ad-bc=1,\;\;\;a,b,c,d\;\in\mbox{Z},\nonumber \\
g^I &\to& g^I + H^I + \bar{H}^I, \quad H^I = \ln(icT^I + d), \quad
\nonumber \\
\Phi^A &\to& e^{-\sum_IH^Iq_I^A}\Phi^A,\nonumber \\
\lambda_a &\to& e^{-{i\over2}\sum_I{\rm Im}H^I}\lambda_a,\quad
\chi^A\to e^{{1\over2}\sum_I(i{\rm Im}H^I - 2q^A_IH^I)}\chi^A, \quad
\theta\to e^{-{i\over2}\sum_I{\rm Im}H^I}\theta,\nonumber \\
U_a &\to& e^{-i\sum_I{\rm Im}H^I}U_a, \quad \Pi^\alpha \to 
e^{-\sum_IH^Iq_I^\alpha}\Pi^\alpha,\quad 
\nonumber \\
q^\alpha_I &=& \sum_A n^A_\alpha q^A_I .
\end{eqnarray}
The field-theoretical loop corrections to the effective Yang-Mills Lagrangian 
from orbifold compactification have been determined \cite{Gaillard92,KL} using 
supersymmetric regularization procedures that ensure a supersymmetric form for 
the modular anomaly. Matching the variation under (4.14) of that contribution 
to the Yang-Mills Lagrangian with the variation of the effective Lagrangian 
(4.12) we require
\begin{equation}
\delta\Lag_{VY} = - {1\over64\pi^2}\sum_{a,I}\int d^4\theta 
{E\over R}U_a\[C_a - \sum_{A,I}C^A_a\(1 - 2q_I^A\)\]H^I + {\rm h.c.}, 
\end{equation}
which implies
\begin{equation}
b'_a + \sum_{\alpha,A} b^\alpha_an^A_\alpha q_I^A =
{1\over8\pi^2}\[C_a - \sum_AC^A_a\(1 - 2q^A_I\)\]\;\;\;\;\;\;
\forall \;\;I.
\end{equation}
In the flat space limit where the reduced Planck 
mass\footnote{The reduced Planck mass ${M'}_{P}=M_{P}/\sqrt{8\pi}$, where 
$M_{P}$ is the Planck mass.} ${M'}_P\to\infty$, under a canonical scale 
transformation 
$$\lambda \to e^{{3\over2}\sigma}\lambda,\;\;\;\; 
U\to e^{3\sigma}U,\;\;\;\;
\Phi^A\to e^{\sigma}\Phi^A,\;\;\;\;
\Pi^\alpha\to e^{\sum_An^A_\alpha\sigma}\Pi^\alpha,\;\;\;\;
\theta\to e^{-{1\over2}\sigma}\theta,$$ 
we have the standard trace anomaly as determined by the $\beta$-functions:
\begin{equation}
\delta\Lag_{eff} = {1\over64\pi^2}\sigma\sum_{a}\int d^4\theta 
{E\over R}U_a\(3C_a - \sum_AC^A_a\) + {\rm h.c.} + \CO({M'}_{P}^{-1}),
\end{equation}
which requires
\begin{equation}
3b'_a + \sum_{\alpha,A}b^\alpha_an^A_\alpha =
{1\over8\pi^2}\(3C_a - \sum_AC^A_a\) + \CO({M'}_{P}^{-1}). 
\end{equation}
Eqs. (4.16) and (4.18) are solved by~\cite{matter} (up to $\CO({M'}_{P}^{-1})$ 
corrections)
\begin{eqnarray}
b'_a &=& {1\over8\pi^2}\(C_a - \sum_AC_a^A\) ,
\nonumber \\
\sum_{\alpha,A} b^\alpha_a n^A_\alpha q_I^A &=& 
\sum_A{C^A_a\over4\pi^2}q_I^A,
\quad \sum_{\alpha,A} b^\alpha_an^A_\alpha = 
\sum_A{C^A_a\over4\pi^2}.  
\end{eqnarray}
Note that the above arguments do not completely fix $\Lag_{eff}$ since we can 
{\it a priori} add chiral and modular invariant terms of the form:
\begin{equation}
\Delta\Lag = \sum_{a,\alpha}b'_{a\alpha}\superint EV_a\ln
\(e^{\sum_Iq^\alpha_Ig^I}\Pi^\alpha\bar{\Pi}^\alpha\).
\end{equation}
For specific choices of the $b'_{a\alpha}$ the matter condensates $\Pi^\alpha$
can be eliminated from the effective Lagrangian.  However the resulting 
component Lagrangian has a linear dependence on the unphysical scalar fields 
$\ell_a - \ell_b$, and their equations of motion impose physically unacceptable 
constraints on the moduli supermultiplets. To ensure that $\Delta\Lag$ contains 
the fields $\ell_a$ only through the physical combination $\sum_a\ell_a$, we 
have to impose $b'_{a\alpha} = b'_\alpha$ independent of $a$. If these terms 
were added, the last condition in (4.19) would become 
\begin{equation}
\sum_{\alpha,A}b^\alpha_an^A_\alpha + \sum_Ab'_{\alpha}n^A_\alpha
= \sum_A{C^A_a\over4\pi^2}.
\end{equation}
We shall not include such terms here. 

Combining (4.11) with (4.19) gives 
$b_a^I = 8\pi^2\(b - b'_a - \sum_\alpha b^\alpha_aq_I^\alpha\)$. 
Combining the terms (4.6)--(4.13) by superspace partial integration (2.18), 
the ``Yang-Mills'' part of the Lagrangian (4.5) can be expressed -- up to a 
total derivatives that we drop in the subsequent analysis -- as a modular 
invariant $D$ term:
\begin{eqnarray}
\Lag_{eff} &=& \superint\,E \Bigg( -2 + f(V) + \sum_aV_a\Bigg\{
b'_a\ln(\bar{U}_aU_a/e^gV) + \sum_\alpha
b^\alpha_a\ln\(\Pi_r^\alpha\bar{\Pi}_r^\alpha\)
\nonumber \\ & & \;\; 
- \sum_I{b_a^I\over8\pi^2}\ln\[\(T^I + \bar{T}^I\)
|\eta^2(T^I)|^2\]\Bigg\}\Bigg) + \Lag_{pot},
\end{eqnarray}
where 
\begin{equation}
\Pi_r^\alpha = \prod_A(\Phi_r^A)^{n^A_\alpha} = 
e^{\sum_Iq^\alpha_Ig^I/2}
\Pi^\alpha,\;\;\;\; \Phi_r^A = e^{\sum_Iq^A_Ig^I/2}\Phi^A, 
\end{equation}
is a modular invariant field composed of elementary fields that are canonically
normalized in the vacuum. The interpretation of this result in terms of
renormalization group running will be discussed below. We have implicitly 
assumed affine level-one compactification. The generalization to higher affine 
levels is trivial.  

The construction of the component field Lagrangian obtained from (4.22) 
parallels that given in Section 2.3.2 for the case $\;\G=E_8$. Since the 
superfield Lagrangian is a sum of $F$ terms that contain only spinorial 
derivatives of the superfield $V_a$, and the Green-Schwarz and kinetic terms 
that contain $V_a$ only through the sum $V$, the unphysical scalars $\ell_a$ 
appear in the component Lagrangian only through the physical dilaton $\ell$. 
The result for the bosonic Lagrangian is:
\begin{eqnarray}
\frac{1}{e}\Lag_{B}\,&=&\,-\,\frac{1}{2}{\cal R}
\,-\,(1+b\ell)\sum_{I}\frac{1}{(t^{I}+\bar{t}^{I})^{2}}
\(\pp^{m}\bar{t}^{I}\,\pp_{\!m}t^{I} - \bar{F}^{I}F^{I}\) 
\nonumber\\
& &\,-\,\frac{1}{16\dilaton^{2}}\(1+\dilaton\dg\)
\[4\(\pp^{m}\!\dilaton\,
\pp_{\!m}\!\dilaton - B^{m}\!B_{m}\) + \bar{u} u 
- 4e^{K/2}\ell\(W\bar{u} + 
u\bar{W}\)\]\nonumber\\
& &\,+\,\frac{1}{9}\(\dilaton\dg-2 \)\[\bar{M}\!M 
- b^{m}b_{m} - {3\over4}\lbr
\bar{M}\(\sum_bb'_bu_b - 4We^{K/2}\) + {\rm h.c.}\rbr
\]\nonumber\\
& &\,+\,\frac{1}{8}\sum_a\Bigg\{{f+1\over\ell} + 
\,b'_a\ln(e^{2-K}\bar{u}_au_a) 
+ \sum_\alpha b^\alpha_a\ln(\pi^\alpha\bar{\pi}^\alpha) 
\nonumber \\ & & \qquad 
+ \sum_I\[bg^I - {b_a^I\over4\pi^2}\ln|\eta(t^I)|^2\]\Bigg\}
\(F_a - u_a\bar{M} + {\rm h.c.}\)\nonumber\\ 
& &\,-\,\frac{1}{16\dilaton}\sum_a\[b'_a\(1+\dilaton\dg\)
\bar{u}u_a -4\ell u_a
\(\sum_\alpha b^\alpha_a{F^\alpha\over\pi^\alpha} 
+ (b'_a-b){F^I\over2{\rm
Re}t^I}\) + {\rm h.c.} \] \nonumber\\
& &\,+\,\frac{i}{2}\sum_a\[b'_a\ln(\frac{u_a}{\bar{u}_a}) 
+ \sum_\alpha 
b^\alpha_a\ln(\frac{\pi^\alpha}{\bar{\pi}^\alpha})\]
\nabla^{m}\!B_{m}^a
\,-\,{b\over2}\sum_{I}
\frac{\pp^{m}{\rm Im}{t}^{I}}{{\rm Re}{t}^{I}}B_{m}, 
\nonumber \\ & &\,
+\,\sum_{I,a}{b^I_a\over16\pi^2}\[\zeta(t^I)\(2iB_a^m\nabla_m
t^I - u_aF^I\) + {\rm h.c.}\]
\nonumber \\ & &\,+\,e^{K/2}\[\sum_IF^I\(W_I + K_IW\) + 
\sum_\alpha
F^\alpha W_\alpha + {\rm h.c.}\],
\end{eqnarray}
where 
\begin{eqnarray}
\zeta(t)\,&=&\,{1\over\eta(t)}{\pp\eta(t)\over\pp t}, \quad
\eta(t) = e^{-\pi t/12}\prod_{m=1}^{\infty}\(1-e^{-2m\pi t}\), 
\nonumber \\
\dilaton\,&=&\,V\lowest,\nonumber\\
\sigma^{m}_{\alpha\dot{\alpha}}B^a_{m}\,&=&\,
\frac{1}{2}[\,\Diff_{\alpha},\Diff_{\dot{\alpha}}\,]
V_a\lowest\,+\,
\frac{2}{3}\dilaton_a\sigma^{m}_{\alpha\dot{\alpha}} b_{m},
\quad B^m = \sum_aB^m_a,\nonumber\\
u_a\,&=&\,U_a\lowest\,=\,-(\bar{\Diff}^{2}-8R)V_a\lowest,
\quad u = \sum_au_a,\nonumber\\
\bar{u}_a\,&=&\,\bar{U}_a\lowest\,=\,-(\Diff^{2}-8R^{\dagger})
V_a\lowest,
\quad \bar{u} = \sum_a\bar{u}_a,\nonumber \\
-4F^a\,&=&\,\Diff^{2}U_a\lowest, \;\;\;\;
-4\bar{F}^a\,=\,\bar{\Diff}^{2}\bar{U}_a\lowest, 
\quad F_U = \sum_aF^a,\nonumber\\
\pi^\alpha\,&=&\,\Pi^\alpha\lowest\,
\quad \bar{\pi}^\alpha\,=\,\bar{\Pi}^\alpha\lowest\,\nonumber \\
-4F^\alpha\,&=&\,\Diff^{2}\Pi^\alpha\lowest, \;\;\;\;
-4\bar{F}^\alpha\,=\,\bar{\Diff}^{2}\bar{\Pi}^\alpha\lowest, 
\nonumber \\
t^{I}\,&=&\,T^{I}\lowest,\;\;\;\;
-4F^{I}\,=\,\Diff^{2}T^{I}\lowest,\nonumber\\
\bar{t}^{I}\,&=&\,\bar{T}^{I}\lowest,\;\;\;\;
-4\bar{F}^{I}\,=\,\bar{\Diff}^{2}\bar{T}^{I}\lowest,
\end{eqnarray}
$b_m$ and $M = (\bar{M})^{\dag} = -6R\lowest$ are auxiliary components
of the supergravity multiplet~\cite{Binetruy90}. Notice that $\zeta(t)$ defined
in (4.25) is related to the Einstein function $\hat{G}_{2}(t)$ \cite{Einstein} 
as follows: $\,\hat{G}_{2}(t) = -\pi\(1+4\zeta(t){\rm Re}t\)/{\rm Re}t$. For 
$n=1,\;u_a=u,\;etc.,$ (4.24) reduces to (2.46) of Section 2.3.2.

The equations of motion for the auxiliary fields $b_m, M,F^I,F^a+\bar{F}^a$ 
and $F^\alpha$ give, respectively:
\begin{eqnarray}
b_m\,&=&\,0, \;\;\;\; 
M\,=\,\frac{\, 3\,}{\, 4\,}\,\(\sum_ab'_au_a - 4We^{K/2}\) ,
\nonumber\\
F^{I}\,&=&\,{{\rm Re}t^I\over2(1 + b\ell)}\lbr
\sum_a\bar{u}_a\[(b - b'_a) +
{b_a^I\over2\pi^2}\zeta(\bar{t}^I){\rm Re}t^I\]- 4
e^{K/2}\(2{\rm Re}t^I\bar{W}_I - \bar{W}\)\rbr, \nonumber \\ 
\bar{u}_au_a\,&=&\,\frac{\ell}{e^{2}}
e^{g\,-\,({f+1})/{b'_a\dilaton}-\sum_I
b^I_ag^I/8\pi^2b'_a}\prod_I|\eta(t^I)|^{b^I_a/2\pi^2b'_a}
\prod_\alpha(\pi^\alpha_r\bar{\pi}^\alpha_r)^{-b^\alpha_a/b'_a},
\quad \pi^\alpha_r = \Pi^\alpha_r\lowest, \nonumber \\ 
0\,&=&\,\sum_ab^\alpha_au_a\,+\,4\pi^\alpha e^{K/2}W_\alpha 
\quad\;\;\; \forall \;\;\alpha.
\end{eqnarray}
Using these, the Lagrangian (4.24) reduces to
\begin{eqnarray}
\frac{1}{e}\Lag_{B}\,&=&\,-\,\frac{1}{2}{\cal R}
\,-\,(1+b\ell)\sum_I\frac{\pp^{m}\bar{t}^{I}\,\pp_{\!m}t^{I}}
{(t^{I}+\bar{t}^{I})^{2}}\,-\,\frac{1}{4\dilaton^{2}}\(1+\dilaton\dg\)
\(\pp^{m}\!\dilaton\,\pp_{\!m}\!\dilaton - B^{m}\!B_{m}\)  \nonumber\\
& &\,-\,\sum_a\(b'_a\omega_a + \sum_\alpha b^\alpha_a\phi^\alpha\)
\nabla^{m}\!B_{m}^a\,-\,{b\over2}\sum_{I}
\frac{\pp^{m}{\rm Im}{t}^{I}}{{\rm Re}{t}^{I}}B_{m} \nonumber \\ & &
+\,i\sum_{I,a}{b^I_a\over8\pi^2}\[\zeta(t^I)B_a^m\nabla_mt^I 
- {\rm h.c.}\] - V_{pot},
\nonumber\\ V_{pot}\,&=&\,
\frac{\(1+\dilaton\dg\)}{16\dilaton^{2}}\lbr\bar{u}u +\ell
\[\bar{u}\(\sum_ab'_au_a - 4e^{K/2}W\)  + {\rm h.c.} \]\rbr\nonumber \\ 
& &\,+\,{1\over 16(1+b\ell)}\sum_I\left|\sum_au_a\(b-b'_a 
+ {b^I_a\over2\pi^2}
\zeta(t^I){\rm Re}t^I\)  - 4e^{K/2}\(2{\rm Re}t^IW_I - W\)\right|^2
\nonumber\\ & &\,+\,\frac{1}{16}\(\dilaton\dg-2 \)\left|\sum_bb'_bu_b
- 4We^{K/2}\right|^2 ,
\end{eqnarray}
where we have introduced the notation
\begin{eqnarray}
u_a &=& \rho_a e^{i\omega_a}, \quad \pi^\alpha = \eta^\alpha
e^{i\phi^\alpha}, 
\end{eqnarray}
and 
\begin{eqnarray}
2\phi^\alpha &=& 
-i\ln\(\sum_a b^\alpha_au_a\bar{W}_\alpha\over\sum_a b^\alpha_a\bar{u}_a
W_\alpha\) \quad\;\; {\rm if} \;\;\;\;\; W_\alpha\ne 0.
\end{eqnarray}
To go further we have to be more specific. Assume\footnote{For, {\it e.g.,} 
$\G= E_6\otimes SU(3),$ we take $\Pi\simeq (27)^3$ of $E_6$ or $(3)^3$ of 
$SU(3)$.} that for fixed $\alpha,\; b^\alpha_a\ne 0$ for only one value of 
$a$. For example, we allow no representations $(n,m)$ with both $n$ and 
$m\ne 1$ under $\G_a\otimes\G_b$. Then $u_a= 0$ unless $W_\alpha\ne 0$ for 
every $\alpha$ with $b^\alpha_a\ne 0$. We therefore assume that 
$b^\alpha_a \ne 0$ only if $W_\alpha\ne 0$.  

Since the $\Pi^\alpha$ are gauge invariant operators, we may take $W$ linear 
in $\Pi$:
\begin{equation}
W(\Pi,T) = \sum_\alpha W_\alpha(T)\Pi^\alpha, \;\;\;\;  
W_\alpha(T) =  c_\alpha\prod_I[\eta(T^I)]^{2(q^\alpha_I - 1)},
\end{equation}
where $\eta(T)$ is the Dedekind function. If there are gauge singlets $M^i$
with modular weights $q^i_I$, then the constants $c_\alpha$ are replaced by 
modular invariant functions: 
$$c_\alpha\to w_\alpha(M,T) = c_\alpha\prod_i(M^i)^{n^\alpha_i}
\prod_I[\eta(T^I)]^{2n^\alpha_iq^i_I}.$$  
In addition if some $M^i$ have gauge invariant couplings to vector-like 
representations of the gauge group 
$$W(\Phi,T,M)\ni c_{iAB}M^i\Phi^A\Phi^B
\prod_I[\eta(T^I)]^{2(q_I^A + q_I^B + q_I^i)},$$ 
one has to introduce condensates $\Pi^{AB}\simeq \Phi^A\Phi^B$ of dimension 
two, and corresponding terms in the effective superpotential: 
$$W(\Pi,T,M)\ni 
c_{iAB}M^i\Pi^{AB}\prod_I[\eta(T^I)]^{2(q_I^A + q_I^B + q_I^i)}.$$  
Since the $M^i$ are unconfined, they cannot be absorbed into the composite 
fields $\Pi$. The case with only vector-like representations has been considered
in~\cite{matter}.  To simplify the present discussion, we ignore this type of 
coupling and assume that the composite operators that are invariant under the 
gauge symmetry (as well as possible discrete global symmetries) are at least 
trilinear in the nonsinglets under the confined gauge group. We further assume 
that there are no continuous global symmetries--such as a flavor 
$SU(N)_L\otimes SU(N)_R$ whose anomaly structure has to be considered 
\cite{matter}. With these assumptions the equations of motion 
(4.26) give, using $\sum_\alpha b^\alpha_a q^\alpha_I + {b_a^I/8\pi^2} 
= b-b'_a$,
\begin{eqnarray}
\rho^2_a &=& e^{-2b_a'/b_a}e^Ke^{-(1+f)/b_a\ell - b\sum_Ig^I/b_a}
\prod_I|\eta(t^I)|^{4(b-b_a)/b_a}
\prod_\alpha|b^\alpha_a/4c_\alpha|^{-2b_a^\alpha/b_a}, \nonumber \\
\pi_r^\alpha &=& - e^{-{1\over2}[k+ \sum_I(1-q^\alpha_I)g^I]}
{b^\alpha_a\over4W_\alpha}u_a, 
\quad  b_a \equiv b'_a + \sum_\alpha b^\alpha_a. 
\end{eqnarray}
Note that promoting the second equation above to a superfield relation, and 
substituting the expression on the right hand side for $\Pi$ in (4.22) gives
\begin{eqnarray}
\Lag_{eff} &=& \superint\,E \Bigg( -2 + f(V) + \sum_aV_a\Bigg\{
b_a\ln(\bar{U}_aU_a/e^gV) \nonumber \\ & & \quad - 
\sum_\alpha b_a^\alpha
\ln\(e^{\sum_Ig^I(1 - q^\alpha_I)}
\left|4W_\alpha/b_a^\alpha\right|^2\)
\nonumber \\ & & \quad - \sum_I{b_a^I\over8\pi^2}
\ln\[\(T^I + \bar{T}^I\)
|\eta^2(T^I)|^2\]\Bigg\}\Bigg) + \Lag_{pot}. 
\end{eqnarray}
It is instructive to compare this result with the effective Yang-Mills 
Lagrangian found~\cite{Gaillard92,KL} by matching field-theoretical and string 
loop calculations. Making the identifications $V\to L, \;U_a\to{\WaWa}_a$, the 
effective Lagrangian at scale $\mu$ obtained from those results can be written
as follows:
\begin{eqnarray}
\Lag^{YM}_{eff}(\mu) &=& \superint\,E \Bigg( -2 + f(V) + 
\sum_aV_a\Bigg\{
{1\over 8\pi^2}\(C_a - {1\over3}\sum_AC^A_a\)
\ln\[{M^6_s g_{s}^{-4}\over\mu^6g_a(\mu)^{-4}}\] 
\nonumber \\ & & \quad
- {1\over4\pi^2}\sum_AC^A_a\ln\[g_s^{2\over3}
Z_A(M_s)/g_a^{2\over3}(\mu) Z_A(\mu)\] \nonumber \\ & & \quad 
- \sum_I{b_a^I\over8\pi^2}\ln\[\(T^I + \bar{T}^I\)
|\eta^2(T^I)|^2\]\Bigg\}\Bigg), 
\end{eqnarray}
with $M^2_s \approx g^2_s \approx 2\langle\,\ell\,\rangle$ ($g_s\equiv g(M_s)$) 
in the string perturbative limit, $f(V) = g(V) = 0$. The first term in the 
brackets in (4.32) can be identified with the corresponding term (4.33) provided
\begin{equation}
\sum_\alpha b^\alpha_a = {1\over12\pi^2}\sum_AC^A_a,\;\;\;\; b_a =
{1\over8\pi^2}\(C_a - {1\over3}\sum_AC^A_a\). 
\end{equation}
In fact, this constraint follows from (4.19) if the $\Pi^\alpha$ are all of 
dimension three, which is consistent with the fact that only dimension-three 
operators survive in the superpotential in the limit ${M'}_P\to\infty$. Then
$b_a$ is proportional to the $\beta$-function for $\G_a$, and 
$\,\langle\,\rho_a\,\rangle\,\approx\,
\langle\,|\lambda_{a}^{\alpha}\lambda_{a\alpha}|\,\rangle\,$ has the correct  
exponential suppression factor for a small gauge coupling constant as expected 
by a RGE analysis. In the absence of (stringy) nonperturbative corrections to 
the K\"ahler potential ($f(V) = g(V) = 0$), 
$\,\,2\langle\,V\,\lowest\rangle = 2\langle\,\ell\,\rangle = g^2_s = M_s^2$ is 
the string scale in reduced Planck units and also the gauge coupling at that
scale~\cite{Gaillard92,KL}. Therefore, the argument of the logarithm in (4.33),
\begin{equation}
\left\langle\,{\bar{U}_a U_a\over V}\,\right\rangle^{1/3} \,\approx\, 
{\left\langle\,\left|\lambda_{a}^{\alpha}\lambda_{a\alpha}\right|\,
\right\rangle^{2/3}\over g_s^{2/3}} \,=\, 
{\left\langle\,\left|\lambda_{a}^{\alpha}\lambda_{a\alpha}\right|\,
\right\rangle^{2/3}\over M_s^2 g_s^{-{4/3}}} ,
\end{equation}
gives the exact two-loop result for the coefficient of $C_a$ in the 
renormalization group running from the string scale to the appropriate 
condensation scale~\cite{Gaillard92,KL,chiral91}. The relation between 
$\langle\,\pi^\alpha\,\rangle$ and $\langle\,u_a\,\rangle$, and hence the 
appearance of the gaugino condensate as the effective infra-red cut-off for 
massless matter loops, is related to the Konishi anomaly~\cite{kon}. The 
matter loop contributions have additional two-loop corrections involving 
matter wave-function renormalization~\cite{Jain96,barb,Gava,gjs}:
\begin{eqnarray}
{\pp\ln Z_A(\mu)\over\pp\ln\mu^2} &=& -
{1\over 32\pi^2}\Bigg[\ell e^g\sum_{BC}e^{\sum_Ig^I\(1- q^A_I - q^B_I - 
q^C_I\)}Z^{-1}_A(\mu)Z^{-1}_B(\mu)Z^{-1}_C(\mu)|W_{ABC}|^2
\nonumber \\ & &
- 4\sum_ag^2_a(\mu)C_2^a(R_A) \Bigg] + \CO(g^4) + \CO(\Phi_{A}^{2}),
\end{eqnarray}
where $C_2^a(R_A) = ({\rm dim}\G_a/{\rm dim}R_A)C_a^A, \;R_A$ is the 
representation of $\G_a$ on $\Phi_A$. The boundary condition on $Z_A$ 
\cite{Gaillard92} is $Z_A(\mu_s) = (1 - p_A\ell)^{-1}$, where $p_A$ is the 
coefficient of $e^{\sum_Iq^A_Ig^I}|\Phi^A|^2$ in the Green-Schwarz counterterm 
of the underlying theory: 
$V = \sum_Ig^I + p_Ae^{\sum_Iq^A_Ig^I}|\Phi^A|^2 +\CO(|\Phi^A|^4)$. The second 
line of (4.32) can be interpreted as a rough parameterization of the second 
line of (4.33).

In the following analysis, we retain only dimension three operators in the
superpotential, and do not include any unconfined matter superfields in the
effective condensate Lagrangian. The potential $V_{pot}$ takes the form:
\begin{eqnarray}
V_{pot} &=& {1\over16\ell^2}\sum_{a,b}\rho_a\rho_b\cos\omega_{ab}
R_{ab}(t^I),
\quad \omega_{ab} = \omega_a - \omega_b, \nonumber 
\\ R_{ab} &=& \(1+\ell\dg\)\(1 + b_a\ell\)\(1 + b_b\ell\) 
- 3\ell^2b_ab_b
+ {\ell^2\over(1 + b\ell)}\sum_Id_a(t^I)d_b(\bar{t}^I), 
\nonumber \\ d_a(t^I) &=& b - b'_a + {b^I_a\over2\pi^2}
\zeta(t^I){\rm Re}t^I
- \sum_\alpha b^\alpha_a\[\,1 - 4(q^\alpha_I - 1)\zeta(t^I){\rm Re}t^I\,\] 
\nonumber \\ &=& \(b - b_a\)\(1 + 4\zeta(t^I){\rm Re}t^I\)
\nonumber \\ &=& -\(b - b_a\)\frac{{\rm Re}t^I}{\pi}\hat{G}_{2}(t^I).
\end{eqnarray}
Note that $\;d_a(t^I)\,\propto\,F^I\,\propto\,\hat{G}_{2}(t^I){\rm Re}t^I\;$ 
vanishes at the self-dual point $t^I=1$, where $\zeta(t^I) = -1/4,
\;\hat{G}_{2}(t^I) = 0,\;\eta(t^I)\approx 0.77$. For 
Re$t^I \stackrel{\textstyle{>}}{\sim}1$ we have, to a very good approximation, 
$\zeta(t^I)\approx -\pi/12,\;\eta(t^I)\approx e^{-\pi t/12}$. Note that also 
$\rho_a$ -- and hence the potential $V_{pot}$ -- vanishes in the limits of 
large and small radii; from (4.31) we have
\begin{eqnarray}
\lim_{t^I\to \infty}\rho_a^2 &\sim  & (2{\rm Re}t^I)^{(b-b_a)/b_a}
e^{-\pi(b-b_a){\rm Re}t^I/3b_a}, \nonumber \\
 \lim_{t^I\to 0}\rho_a^2 & \sim & (2{\rm Re}t^I)^{(b_a -b)/b_a} 
e^{-\pi (b-b_a)/3b_a{\rm Re}t^I}, 
\end{eqnarray}
where the second line follows from the first by the duality invariance of
$\rho^2_a$. So there is potentially a ``runaway moduli problem''. However, as 
will be shown in Section 4.4, the moduli are stabilized at a physically 
acceptable vacuum, namely the self-dual point.

\section{Axion Content of the Effective Theory}
Next we consider the axion states of the effective field theory. If all 
$W_\alpha\ne 0$, the equations of motion for $\omega_a$ obtained from (4.27) 
read:
\begin{equation}
{\pp\Lag\over\pp\omega_a} = -b'_a\nabla^mB^a_m - {1\over2}
\sum_{\alpha,b}b^\alpha_b\({b^\alpha_a u_a\over\sum_c b^\alpha_cu_c} + 
{\rm h.c.}\)\nabla^mB^b_m - {\pp V_{pot}\over\pp\omega_a}=0. 
\end{equation}
These give, in particular,
\begin{eqnarray}
\sum_a{\pp\Lag\over\pp\omega_a} &=& - \sum_ab_a\nabla^mB^a_m = 0. 
\end{eqnarray}
The one-forms $B_m^a$ are {\it a priori} dual to 3-forms:
\begin{equation}
B_m^a = {1 \over 2} \epsilon_{mnpq} \left( {1 \over 3!4} 
\Gamma^{npq}_a + \partial^n b_a^{pq}\right),
\end{equation}
where $\Gamma^{npq}_a$ and $b_a^{pq}$ are 3-form and 2-form potentials,
respectively; (4.41) assures the constraints (2.10) for $\WaWa\to U_a$; 
explicitly
\begin{equation}
(\DaDa-24R^{\dagger})U_a\,-\,(\DbDb-24R)\bar{U}_a \,=\,-2i{}^* 
\Phi_a = 
-{2i \over 3!} \epsilon_{mnpq} \partial^m \Gamma^{npq}_a = 
-16i\nabla^mB_m^a.
\end{equation}
We obtain 
\begin{equation}
-b'_a{}^*\Phi_a - {1\over2}\sum_{\alpha,b}b^\alpha_b
\({b^\alpha_a u_a\over\sum_c b^\alpha_cu_c} + 
{\rm h.c.}\){}^*\Phi_b =
8{\pp V_{pot}\over\pp\omega_a}, \quad \sum_ab_a{}^*\Phi_a = 0. 
\end{equation}
If $\Gamma^{npq}\ne 0,\;b^{pq}$ can be removed by a gauge
transformation $\Gamma^{npq}\to \Gamma^{npq} + 
\pp^{[n}\Lambda^{pq]}$.  Thus
\begin{equation}
B_m^a = {1\over 2nb_a} \epsilon_{mnpq}\partial^{n}\tb^{pq} +
{1 \over 3!8} \epsilon_{mnpq}\Gamma^{npq}_a,\quad 
\sum_ab_a\Gamma^{npq}_a = 0, 
\quad \tb^{pq} = \sum_ab_ab^{pq}_a,
\end{equation}
and we have the additional equations of motion:
\begin{equation}
{\delta\over\delta\tb_{pq}}\Lag_B = 0, \;\;\;\;
\({1\over b_a}{\delta\over\delta\Gamma^{a}_{npq}} - {1\over b_b}
{\delta\over\delta\Gamma^{b}_{npq}}\)\Lag_B = 0,\quad
{\delta\over\delta\phi}\Lag_B \equiv {\pp\Lag_B\over\pp\phi} - 
\nabla^m\({\pp\Lag_B\over\pp(\nabla^m\phi)}\), 
\end{equation}
which are equivalent, respectively, to
\begin{equation}
\epsilon_{mnpq}\sum_a{1\over b_a}
\nabla^n{\delta\over\delta B_a^m}\Lag_B = 
0, \quad \({1\over b_a}{\delta\over\delta B_a^m} - 
{1\over b_b}{\delta\over\delta B_b^m}\)\Lag_B = 0, 
\end{equation}
with  
\begin{eqnarray}
{1\over e}{\delta\over\delta B^a_m}\Lag_B &=& 
{\(1+\ell\dg\)\over2\ell^2}B^m + b'_a\pp^m\omega_a  + {1\over2}
\sum_{\alpha,b}b^\alpha_a\({b^\alpha_b u_b\over\sum_c b^\alpha_cu_c}
+ {\rm h.c.}\)\pp^m\omega_b \nonumber \\ & &
+ \sum_\alpha b^\alpha_a\[\pp^m\ell{\pp\phi^\alpha\over\pp\ell} +
\sum_I\(\pp^m t^I{\pp\phi^\alpha\over\pp t^I} + {\rm h.c.}\)\]
\nonumber \\ & &
+ i\sum_{a,I}{b^I_a\over8\pi^2}\[\zeta(t^I)\pp^mt^I - {\rm h.c.}\]
- {b\over2}\sum_I{\pp^m{\rm Im}t^I\over {\rm Re}t^I} .
\end{eqnarray}
Combining these with (4.39) and the equations of motion for $\ell$ and $t^I$, 
one can eliminate $B^a_m$ to obtain the equations of motion for an equivalent
scalar-axion Lagrangian.

Again, these equations simplify considerably if we assume that for fixed 
$\alpha,\;b^\alpha_a\ne 0$ for only one value of $a$. In this case, (4.39) 
reduces to 
\begin{equation}
\nabla^mB^a_m = - {1\over b_a}{\pp V\over\pp\omega_a}, 
\end{equation}
and we have 
\begin{equation}
{\pp\phi^\alpha\over\pp\ell} = 0,\quad 
{\pp\phi^\alpha\over\pp t^I} =
i\zeta(t^I)\(q^\alpha_I -1\),
\end{equation}
if we restrict the potential to terms of dimension three with no gauge 
singlets $M^i$. Using 
$\;\sum_\alpha b^\alpha_a\(q_I^\alpha - 1\) + b^I_a/8\pi^2 = b-b_a\;$ gives:
\begin{eqnarray}
{1\over e}{\delta\over\delta B^a_m}\Lag_B &=& 
{\(1+\ell\dg\)\over2\ell^2}
B^m + b_a\pp^m\omega_a + i\sum_I\lbr\pp^mt^I\[\zeta(t^I)\(b-b_a\) 
+ {b\over4{\rm Re}t^I}\] - {\rm h.c.}\rbr
\nonumber 
\end{eqnarray}
\begin{equation}
\approx {\(1+\ell\dg\)\over2\ell^2}B^m + 
b_a\pp^m\omega_a + \sum_I\pp^m{\rm Im}t^I\[\(b - b_a\){\pi\over6}
- {b\over2{\rm Re}t^I}\] ,
\end{equation}
where the last line corresponds to the approximation 
$\zeta(t^I)\approx-\pi/12$. In the following we illustrate these equations 
using specific cases.

\subsection{Single Gaugino Condensate}
As we have seen in Section 2.3.2, for the case of a single gaugino condensate
there is an axion $\omega = \omega_a + (\pi/6)(b/b_a - 1)\sum_I{\rm Im}t^I$ 
that has no potential, and, setting
\begin{equation}
B_a^m = {1\over2}\epsilon^{mnpq}\pp_nb_{pq} = -
{2\ell^2\over\(1+\ell\dg\)}\(b_a\pp^m\omega 
- {b\over2}\sum_I{\pp^m{\rm Im}t^I\over{\rm Re}t^I}\),
\end{equation}
the equations of motion derived from (4.27) are equivalent to those of the 
effective bosonic Lagrangian:
\begin{equation}
\frac{1}{e}\Lag_{B}\,=\,-\,\frac{1}{2}{\cal R}
\,-\,(1+b\ell)\sum_I\frac{\pp^{m}\bar{t}^{I}\,\pp_{\!m}t^{I}}
{(t^{I}+\bar{t}^{I})^{2}}\,
-\,\frac{1}{4\dilaton^{2}}\(1+\dilaton\dg\)
\pp^{m}\!\dilaton\,\pp_{\!m}\!\dilaton  
- V(\ell,t^I,\bar{t}^I) $$ $$ 
\,-\,{\ell^2\over\(1+\ell\dg\)}\(b_a\pp^m\omega 
- {b\over2}\sum_I{\pp^m
{\rm Im}t^I\over{\rm Re}t^I}\)\(b_a\pp_m\omega 
- {b\over2}\sum_I{\pp_m
{\rm Im}t^I\over{\rm Re}t^I}\) . 
\end{equation}

\subsection{Two Gaugino Condensates: $b_1\ne b_2$}
Making the approximation $\eta(t)\approx e^{-\pi t/12}$, the Lagrangian (4.27) 
can be written as follows:
\begin{eqnarray}
\frac{1}{e}\Lag_{B}\,&=&\,-\,\frac{1}{2}{\cal R}
\,-\,(1+b\ell)\sum_I\frac{\pp^{m}\bar{t}^{I}\,\pp_{\!m}t^{I}}
{(t^{I}+\bar{t}^{I})^{2}}\,-\,\frac{1}{4\dilaton^{2}}\(1+\dilaton\dg\)
\(\pp^{m}\!\dilaton\,\pp_{\!m}\!\dilaton - B^{m}\!B_{m}\) 
\nonumber \\ & &
\,-\,\omega\nabla^{m}\!\tB_{m} -
\omega'\nabla^{m}\!B_{m}\,-\,{b\over2}\sum_{I}\frac{\pp^{m}
{\rm Im}{t}^{I}}{{\rm Re}{t}^{I}}B_{m} - V_{pot}, 
\end{eqnarray}
where
\begin{eqnarray}
\omega &=& {b_1\omega_1 - b_2\omega_2\over b_1 - b_2}  - 
{\pi\over 6}\sum_I{\rm Im}t^I, \quad \omega' = -{\omega_{12} 
\over\beta}
+ {b\pi\over 6}\sum_{I}{\rm Im}{t}^{I},\nonumber \\
\beta &=& {b_1 - b_2\over b_1b_2}, \quad \tB^m = \sum_a b_aB^m_a . 
\end{eqnarray}
We have 
\begin{eqnarray}
\omega_1 &=& \omega + {\pi\over6}\sum_I{\rm Im}t^I + {1\over b_1}
\(\omega'- {b\pi\over6}\sum_I{\rm Im}t^I\), \nonumber \\
\omega_2 &=& \omega + {\pi\over6}\sum_I{\rm Im}t^I 
+ {1\over b_2}\(\omega'
- {b\pi\over6}\sum_I{\rm Im}t^I\), \nonumber \\
{\pp V_{pot}\over\pp\omega_1} &=& -{\pp V_{pot}\over\pp\omega_2} = 
{\pp V_{pot}\over\pp\omega_{12}}.
\end{eqnarray}
Then taking $\omega,\omega'$ and $t^I$ as independent variables, the equations
of motion for $\omega$ and $\omega'$ are:
\begin{eqnarray}
\nabla^m\tB_m &=& 0,\quad \tB_m = 
{1\over 2}\epsilon_{mnpq}\partial^{n}\tb^{pq}, \nonumber \\
\nabla^mB_m &=& {1\over8}{}^*\Phi = \beta{\pp V\over\pp\omega_{12}}, 
\quad B_m = {1 \over 3!8}\epsilon_{mnpq}\Gamma^{npq}. 
\end{eqnarray}
Substituting the first of these into the Lagrangian (4.53), we see that the
axion $\omega$ and the three-form $\tB_m$ drop out because they appear only
linearly in the Lagrangian; hence they play the role of Lagrange multipliers.
The equation of motion for $\tb_{mn}$ implies the constraint on the phase 
$\omega$ as follows:
\begin{equation}
\nabla_m\pp^m\omega = 0.
\end{equation}
The equations of motion for Im$t^I$ and $\Gamma_{mnp}$ read:
\begin{eqnarray}
0 &=& \nabla_m\[\(1 + b\ell\){\pp^m{\rm Im}{t}^{I}\over2
\({\rm Re}{t}^{I}\)^2} + {b\over2{\rm Re}{t}^{I}}B^{m}\] 
-  i\({\pp V\over\pp t^I} - {\rm h.c.}\) - 
{b\pi\over48}{}^*\Phi, \nonumber \\ 
0 &=& {\(1+\ell\dg\)\over2\ell^2}B^m + \pp^m\omega' - {b\over2}
\sum_I{\pp^m{\rm Im}t^I\over{\rm Re}t^I},
\end{eqnarray}
and the equivalent bosonic Lagrangian is:
\begin{eqnarray}
\frac{1}{e}\Lag_{B}\,&=&\,-\,\frac{1}{2}{\cal R}
\,-\,(1+b\ell)\sum_I\frac{\pp^{m}\bar{t}^{I}\,\pp_{\!m}t^{I}}
{(t^{I}+\bar{t}^{I})^{2}}\,-\,\frac{1}{4\dilaton^{2}}\(1+\dilaton\dg\)
\pp^{m}\!\dilaton\,\pp_{\!m}\!\dilaton  \nonumber \\ & & 
\,-\,{\ell^2\over\(1+\ell\dg\)}\(\pp^m\omega' -
{b\over2}\sum_I{\pp^m{\rm Im}t^I\over{\rm Re}t^I}\)\(\pp_m
\omega' - {b\over2}\sum_I{\pp_m{\rm Im}t^I\over{\rm Re}t^I}\) 
\nonumber \\ & &
\,-\,V_{pot}(\ell,t^I,\bar{t}^I,\omega_{12}) . 
\end{eqnarray}
As in Section 4.3.1, there is a single dynamical axion $\omega'$ -- or, 
via a duality transformation, ${}^*\Phi$ -- but there is now a potential for 
the axion in the multi-condensate case.
\subsection{General Case} 
We introduce $n$ linearly independent vectors 
$\tB_m,B_m,\hB^i_m,\;i= 1\ldots n-2$, and decompose the $B^m_a$ as follows:
\begin{eqnarray}
B^m_a &=& a_a\tB^m + c_aB^m + \sum_id^i_a\hB^m_i, \quad \hB^m_i =
\sum_a e_i^a B^m_a. 
\end{eqnarray}
Then
\begin{equation}
\sum_a\[b_a\omega_a + (b-b_a){\pi\over6}\sum_I{\rm Im}t^I\]
\nabla_mB^m_a = \omega\nabla_m\tB^m + \omega'\nabla_mB^m
+ \sum_i\omega^i\nabla_m\hB^m_i, $$ $$
\omega_a = \omega + {\pi\over6}\sum_I{\rm Im}t^I 
+ {1\over b_a}\(\omega' - 
{b\pi\over6}\sum_I{\rm Im}t^I\) + \sum_i{e_i^a\over b_a}\omega^i,
\end{equation}
and the Lagrangian can be written as in (4.53) with an additional term:
\begin{equation}
{1\over e}\Lag_B \to {1\over e}\Lag_B - \sum_i\omega^i\nabla_m\hB^m_i, 
\end{equation}
The equations of motion for the phases $\omega$, $\omega'$ and $\omega^i$ are:
\begin{eqnarray}
\nabla_m\tB^m &=& -
{\pp V_{pot}\over\pp\omega} = -\sum_a{\pp V_{pot}\over\pp\omega_{a}} = 0,
\nonumber \\
\nabla_mB^m &=& - {\pp V_{pot}\over\pp\omega'} = 
-\sum_a{1\over b_a}{\pp V_{pot}\over\pp\omega_{a}} = 
{1\over2}\sum_{ab}\beta_{ab}
{\pp V_{pot}\over\pp\omega_{ab}} ={1\over8}{}^*\Phi, 
\quad \beta_{ab}\equiv {b_a - b_b\over b_ab_b}
\nonumber \\ 
\nabla_m\hB^m_i &=& -{\pp V_{pot}\over\pp\omega^i} = 
-\sum_a{e_i^ a\over b_a}{\pp V_{pot}\over\pp\omega_{a}} = {1\over8}{}^*\Phi_i,
\end{eqnarray}
and the equations for $\Gamma_{mnp}^i= 8\epsilon_{mnpq}\hB^q_i$ give
$\pp^m\omega^i = 0$. Hence
\begin{equation}
\omega_{ab} = - \beta_{ab}\(\omega' - 
{b\pi\over6}\sum_I{\rm Im}t^I\) + \theta_{ab}, \;\;\;\; 
\theta_{ab} =
{\rm constant}. 
\end{equation}
Therefore, as in the two-condensate case of Section 4.3.2, there is one 
dynamical axion with a potential. The dual bosonic Lagrangian is the same as 
(4.59), with $V_{pot} = V_{pot}(\ell,t^I,\bar{t}^I,\omega_{ab})$.

\section{The Effective Potential}
The potential (4.37) can be written in the form 
\begin{eqnarray}
V_{pot} &=& {1\over16\ell^2}\(v_1 - v_2 + v_3\),\nonumber \\ v_1 &=& 
\(1+\ell\dg\)\left|\sum_a\(1+b_a\ell\)u_a\right|^2,  \quad 
v_2 = 3\ell^2\left|\sum_ab_au_a\right|^2,\nonumber \\ 
v_3 &=& {\ell^2\over(1+b\ell)}\sum_I\left|\sum_ad_a(t^I)u_a\right|^2
= 4\ell^2\(1+b\ell\)\sum_I\left|{F^I\over{\rm Re}t^I}\right|^2. 
\end{eqnarray}
In the strong coupling limit
\begin{equation}
\lim_{\ell\to\infty}V_{pot} = \(\ell\dg - 2\)\left|\sum_ab_au_a\right|^2 ,
\end{equation}
giving the exactly same condition on the functions $f$, $g$ as (2.57) to assure
boundedness of the scalar potential. Therefore (2.57), the necessary 
condition for stringy non-perturbative effects to stabilize the dilaton, is
indeed true in general. Note however that if $v_1 = v_3 = 0$ has a solution 
with $v_2\ne 0$, the vacuum energy is always negative. $\;v_3 = 0$ is solved 
by $t^I=1$, {\it i.e.} the self-dual point. As explained below, this is the 
only nontrivial minimum if the cosmological constant is fine-tuned to vanish. 
In the case of two condensates, there is no solution to $v_1 = 0,\;v_2\ne 0$, 
for $f\ge0$, and the cosmological constant can be fine-tuned to vanish, as will 
be illustrated below in a toy example. More generally, the scalar potential 
$V_{pot}$ is dominated by the gaugino condensate with the largest one-loop
$\beta$-function coefficient, so the general case is qualitatively very similar 
to the single condensate case, and it appears that positivity of the scalar 
potential can always be imposed. Otherwise, one would have to appeal to another 
source of supersymmetry breaking to cancel the cosmological constant, such as a
fundamental 3-form potential~\cite{Pillon,3form} whose field strength is dual 
to a constant that has been previously introduced in the superpotential 
\cite{Dine85}, and/or an anomalous $U(1)$ gauge symmetry \cite{messenger}.

In the following we study $Z_3$-inspired toy models with $E_6$ and/or $SU(3)$ 
gauge groups in the hidden sector, and $3N_f$ matter superfields \cite{iban}
in the fundamental representation $f$. Asymptotic freedom requires 
$N_{27}\le3$ and $N_3 \le 5$. For a true $Z_3$ orbifold there are no 
moduli-dependent threshold corrections: $b^I_a=0$. In this case, universal 
anomaly cancellation determines the average value of the matter modular 
weights in these toy models as: $\langle\;2q_I^{27}-1\;\rangle = 2/N_{27}$,
$\;\langle\;2q_I^{3}-1\;\rangle = 18/N_3$. In some models Wilson line breaking 
of the hidden sector $E_8$ generates vector-like representations that could 
acquire masses above the condensation scale, so that the universal anomaly 
cancellation sum rule is not saturated by light states alone.  In this case 
the $q^\alpha_I$ no longer drop out of the equations, so some of the above 
formulae would be slightly modified.  In addition, one would have to include 
threshold effects~\cite{KL}, unless the masses of the heavy states are pushed 
to the string scale. Here we assume for simplicity that the sum rule is 
saturated by the light states. Denoting the fundamental matter fields by 
$\Phi_f^{I\alpha},\;\alpha = 1, \ldots,N_f$, the hidden matter condensates can 
be constructed as 
$$\Pi_f^\alpha = \prod_{I=1}^3\Phi_f^{I\alpha}, \quad b^\alpha_{E_6} = 
{3\over4\pi^2}, \quad b^\alpha_{SU(3)} = {1\over8\pi^2}, $$
where gauge indices have been suppressed.

In the analysis of the models described below, we assume -- for obvious
phenomenological reasons -- that the vacuum energy vanishes at the minimum
$\langle\; V_{pot} \;\rangle = 0$. Thus we solve the following equations: 
\begin{equation}
V_{pot} = {\pp V_{pot}\over\pp x} = 0, \quad x = \ell,t^I,\omega_a.
\end{equation}
For $x = \ell,t^I$, we have
\begin{eqnarray}
{\pp \rho_a\over\pp x} &=& {1\over2}\(A_x + {1\over b_a}B_x\)\rho_a, 
\quad B_\ell = {(1+\ell\dg)\over\ell^2},\quad B_I = 
{b\over2{\rm Re}t^I}\[1 + 4
\zeta(t^I){\rm Re}t^I\], \nonumber \\
{\pp V_{pot}\over\pp x} &=& \(A_x -{2\over\ell}\delta_{x\ell}\)V_{pot} + 
{1\over16\ell^2}
\sum_{ab}\rho_a\rho_b\cos\omega_{ab}\({B_x\over b_a}R_{ab} + 
{\pp\over \pp x}R_{ab}\) 
\nonumber \\ &=& {1\over16\ell^2}\sum_{ab}\rho_a\rho_b\cos\omega_{ab}
\({B_x\over n}\sum_{c}\beta_{ca}R_{ab} + {\pp\over\pp x}R_{ab}\) 
\nonumber \\ & & \qquad + 
\(A_x -{2\over\ell}\delta_{x\ell} + {B_x\over n}\sum_{a}{1\over b_a}\)V_{pot},
\end{eqnarray}
where $\beta_{ab}$ is defined in (4.63). By assumption, the last term in 
(4.68) vanishes in the vacuum. Note that the self-dual point, 
$d_a(t^I) = B_I = 0,\;t^I = 1,$ is always a solution to the minimization 
equations for $t^I$. It is the only solution for the single condensate case. 
For the multi-condensate case, if we restrict our analysis to the (relatively) 
weak coupling region, $\ell < 1/b_-$, where $b_-$ is the smallest 
$\beta$-function coefficient, the scalar potential $V_{pot}$ is dominated by 
the gaugino condensate with the largest $\beta$-function coefficient
$b_+:\;V_{pot}\approx\rho^2_+R_{++}/16\ell^2$. Moreover, since $\pi b/3b_a >1$, 
the scalar potential $V_{pot}$ is always dominated by this term for 
Re$t^{I}>1$ ({\it c.f.} Eq. (4.38)), so the only minimum for Re$t^I>1$ is 
Re$t^I \to\infty,\;\rho_a\to 0$. By duality the only minimum for Re$t^I <1$ is 
Re$t^I \to 0,\;\rho_a\to 0$, so the self-dual point is the only nontrivial 
solution. Since our scalar potential is always dominated by one gaugino 
condensate, the picture is very different from the ``race-track'' models 
studied previously~\cite{racetrack}.

At the self-dual point with $V_{pot}=0$, we have 
\begin{eqnarray}
{\pp^2 V_{pot}\over\pp(t^I)^2} &\approx& {1\over32\ell^2}
\sum_{ab}\rho_a\rho_b
\cos\omega_{ab}\({\pi^2\over9}{\ell^2\over(1+b\ell)}(b-b_a)(b-b_b) 
- {b\pi\over6n}\sum_{c}\beta_{ca}R_{ab}\)\nonumber \\ 
&\approx& {\rho^2_+\over32}\({\pi^2\over9}{(b-b_+)^2\over(1+b\ell)}
 - {b\pi\over6n\ell^2}\sum_{c}\beta_{c+}R_{++}\) . 
\end{eqnarray}
Positivity of the potential requires $R_{++}\ge 0$, and $\beta_{c+}\le 0$ by
definition, so the extremum at the self-dual point with 
$V_{pot}=0,\;\rho_+\ne0$ is a true minimum. In practice, the last term is 
negligible, and the normalized moduli squared mass is:
\begin{equation}
m^2_{t^I} \,\approx\,
\left<\,{1\over4}{(b-b_+)^2\over(1+b\ell)^2}\,\rho^2_+\,\right>.
\end{equation}

\subsection{Single Gaugino Condensate with Hidden Matter}
In this case $\beta_{ab} = 0$, and the minimization equations for $t^I$ require 
$${\pp\over\pp t^I}\left|1 + 4\zeta(t^I){\rm Re}t^I\right|^2 = 0,$$ 
which is solved by $1 + 4\zeta(t^I){\rm Re}t^I = 0,\; t^I=1.$ Then $v_3=F^I=0,$ 
and the scalar potential $V_{pot}$ is qualitatively the same as in the $E_8$ 
case studied in Chapter 2 -- except for the fact that here the string moduli 
are stabilized at the self-dual point. (Note however that if $\beta_{ab}= 0$ 
one can choose the $b'_{a\alpha}$ in (4.20) such that the matter condensates 
drop out of the effective Lagrangian; then $R_{aa}$ is independent of the 
moduli which remain undetermined.) The quantitative difference from the $E_8$ 
case is the value of the $\beta$-function coefficient: 
$b_{E_6} = {\(12 - 3N_{27}\)/8\pi^2},\;b_{SU(3)} = {\(6-N_3\)/16\pi^2}.$  
As in Chapter 2, two possible choices for the function $f$ are 
$f = A e^{-B/V}$ \cite{Banks94} and 
$f=A_p (\sqrt{V})^{-p} e^{-B/\sqrt{V}}$ \cite{Shenker90}, where we
fine tune the parameter $A$ (or $A_{p}$) to get a vanishing cosmological 
constant. 

Attention has been drawn to the leading correction for small coupling that is 
of the form $f = A e^{-B/\sqrt{V}}$ \cite{Shenker90}. If we restrict $f$ to 
this form, we have to require a rather large value for the parameter $A$: 
$A\simeq 40$ in order to cancel the cosmological constant. On the other hand, 
the important feature of $f$ here is its behaviour in the strong coupling 
regime; if $f$ contains terms of the form $A e^{-B/V^{n\over 2}}$, the strong
coupling limit will be dominated by the term with the largest value of $n$.
In the numerical analysis we take $f = Ae^{-B/V}$; adding to this a term of 
the form $f = A'e^{-B'/\sqrt{V}}$ will not significantly affect the analysis.
We find that the {\em vev} of $\dilaton$ is insensitive to the content of the 
hidden sector; it is completely determined by stringy non-perturbative effects,
provided a potential for $\dilaton$ is generated by the strongly coupled hidden
Yang-Mills sector.  More specifically, taking $f = Ae^{-B/V}$ we find that 
$\langle\,V_{pot}\,\rangle = 0$ requires $A\approx e^2 \approx 7.4$, and the 
dilaton is stabilized at a value $\langle\,\dilaton\,\rangle \approx B/2$.  
Taking $B = 1$ gives $\langle\,\dilaton\,\rangle \approx0.5,\; 
\langle\,f(\dilaton)\,\rangle \approx 1,$ and the squared gauge coupling at 
the string scale is $g_s^2 = \langle\,2\dilaton/(1+f)\,\rangle \approx 0.5.$  
If instead we use $f = Ae^{-B/\sqrt{V}}$, the corresponding numbers are 
$A\approx2e^3 \approx 40,\;\langle\,\dilaton\,\rangle\approx B^2/9,\; g_s^2 
\approx 2B^2/27.$ Therefore, the {\em vev} of the dilaton $\dilaton$ 
completely determined by stringy non-perturbative effects, and the dilaton is
naturally stabilized at a weak coupling regime if, for example, the parameter
$B$ in the function $f$ considered here is of order one. 

One may look more closely at the second choice which is a genuine stringy
nonperturbative effect\footnote{We do not consider here the case where the
coefficient $B$ in the exponent is moduli-dependent \cite{eva}. Such stringy 
nonperturbative contributions would perturb the moduli ground state away from 
the self-dual point. However, one has to worry about the problem of modular
invariance for this type of stringy nonperturbative contributions 
\cite{Dine89}}. Taking for illustrative purposes 
$f=\(A_0 + A_1/\sqrt{\ell}\)\,e^{-B/\sqrt{\ell}}$, 
where the condition (4.66) or (2.57) requires $A_0$ to be larger than $2$, 
one finds a realistic minimum for $\CO(1)$ values of the parameters: 
$B\langle\,\ell\,\rangle^{-1/2}\,\approx\, 1.1$ to $1.3$, 
$A_0 \,\approx\, 2.7$ to $5.3$ and $A_1 \,\approx\, -3.1$ to $-4.6$. 
Therefore, the previous problem of a rather large value of $A$ ($A\approx 40$) 
for $f = Ae^{-B/\sqrt{V}}$ does not exist in general. From now on we take 
$f = Ae^{-1/V}$ in the numerical analysis, but notice that the major 
conclusions of the analysis apply to more generic choices for $f$.

The scalar potential $V_{pot}$ for $\G_a = E_6,\; N_{27} = 1$, is plotted in 
Figures 4.1--4.3. Fig. 4.1 shows the scalar potential in the $\ell,\ln t$ 
plane, where we have set $t^I=t$, Im$t=0$; with this choice of variables the 
$T$-duality invariance of the scalar potential is manifest. 
\begin{figure}
\epsfxsize=12cm
\epsfysize=8cm
\epsfbox{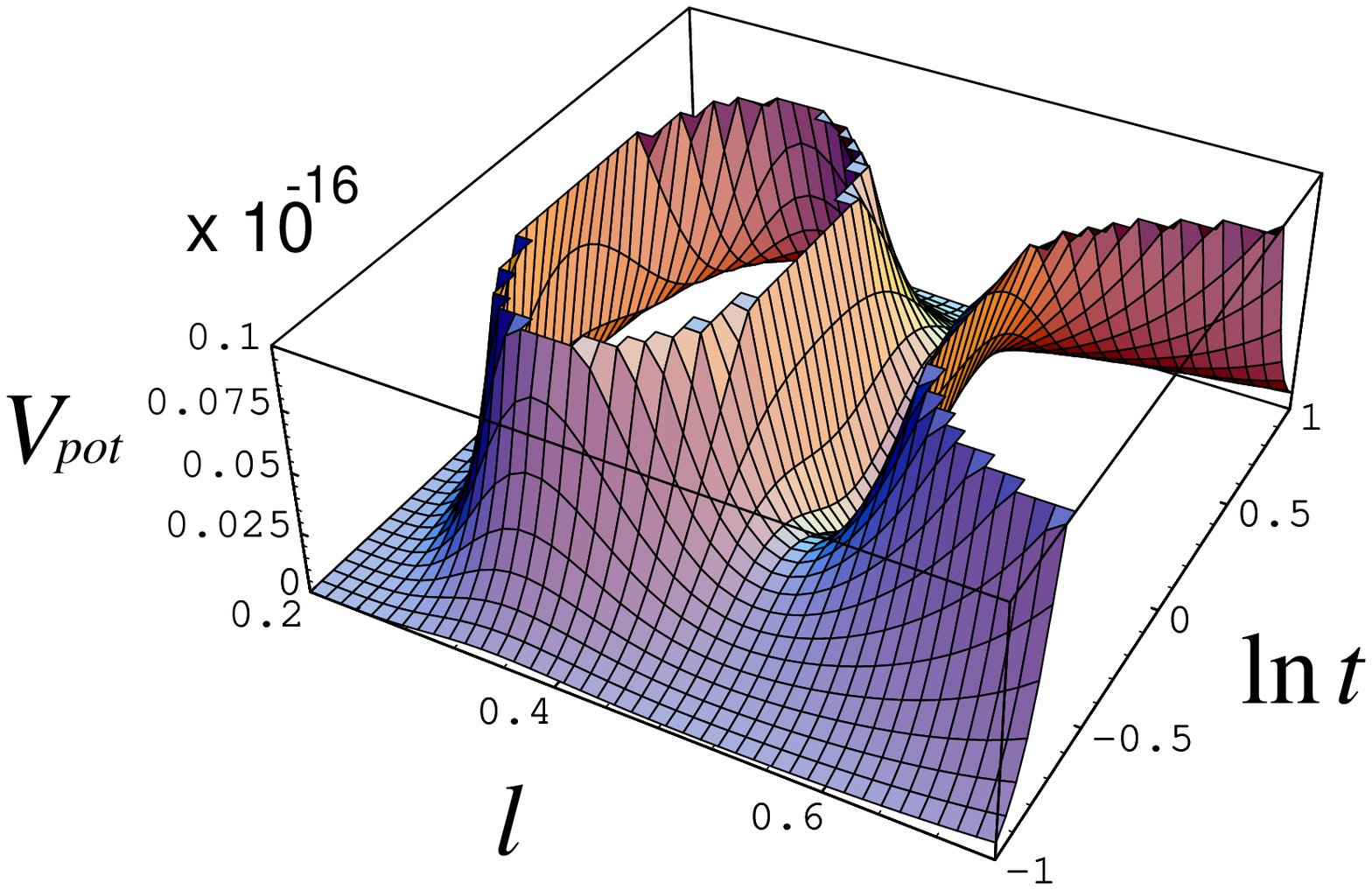}
\caption{The scalar potential $V_{pot}$ (in reduced Planck units)
            is plotted versus $\dilaton$ and $\ln t$.}
\end{figure}
Fig. 4.2 shows the scalar potential $V_{pot}$ for $\ell$ at the self-dual point
$t^I=1$, and Fig. 4.3 shows the scalar potential for $\ln t$ with $\ell$ fixed 
at its {\em vev}. 
\begin{figure}
\epsfxsize=12cm
\epsfysize=8cm
\epsfbox{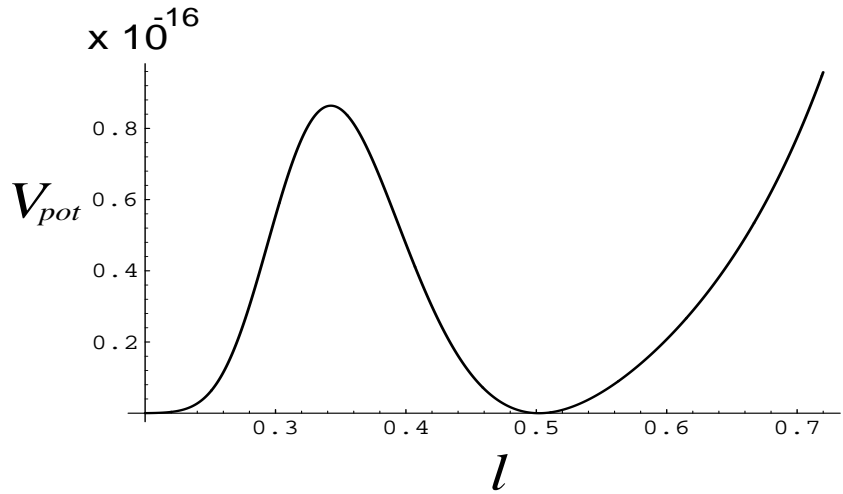}
\caption{The scalar potential $V_{pot}$ (in reduced Planck units)
   is plotted versus $\ell$ with $t^{I}=1$ (the self-dual point).}
\end{figure}
\begin{figure}
\epsfxsize=12cm
\epsfysize=8cm
\epsfbox{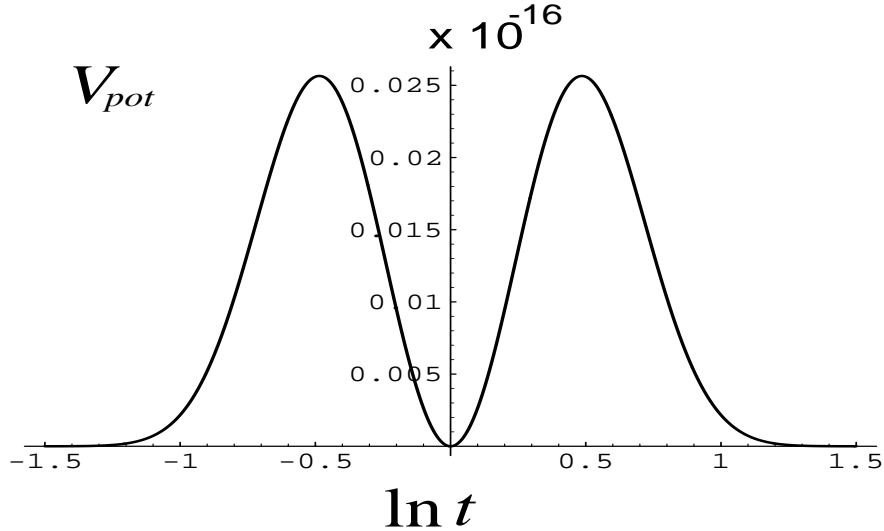}
\caption{The scalar potential $V_{pot}$ (in reduced Planck units)
   is plotted versus $\ln t$ with $\,\ell\,=\,\langle\,\ell\,\rangle$.}
\end{figure}
The qualitative features of the scalar potential are independent of the content
of the hidden sector. Fixing $A$ in each case by the condition 
$\langle\,V_{pot}\,\rangle=0$, we find for $\G_a = E_6$
\begin{equation}
A = \cases{7.324\cr7.359\cr7.381\cr}, \quad \langle\;\ell\;\rangle = 
\cases{0.502\cr0.501\cr0.500\cr}\approx g_s^2,\quad{\rm for}\quad 
N_{27} = \cases{1\cr2\cr3\cr}.
\end{equation}
For $\G_a = SU(3),\;N_3 = 1$, we find $A= 7.383,\;\langle\,\ell\,\rangle = 
0.500\approx g^2_s$. As will be discussed in Section 4.5, the scale of 
supersymmetry breaking in this case is far too small, and further decreases 
with increasing $N_3$.
\subsection{Two Gaugino Condensates}
We have 
\begin{eqnarray}
{\pp V_{pot}\over\pp \omega_1} &=& - {\pp V_{pot}\over\pp \omega_2} = 
- \rho_1\rho_2R_{12}\sin\omega_{12},\nonumber 
\end{eqnarray}
\begin{equation}
\sum_{abc}\beta_{ca}\rho_a\rho_bR_{ab}\cos\omega_{ab} =
\beta_{21}\(\rho_1^2R_{11} - \rho_2^2R_{22}\).
\end{equation}
Minimization with respect to $\omega_{1}$ requires either 
$\langle\,\sin\omega_{12}\,\rangle = 0$ or $\langle\, R_{12}\,\rangle = 0$. 
Identifying $b_1=b_+,\;b_2=b_-$, positivity of the scalar potential requires 
$R_{11}\ge 0$, which in turn implies $R_{12}>0,$ so the extrema in $\omega$ 
are at $\sin\omega_{12}=0$, with $\cos\omega_{12} = - 1\; (+1)$ corresponding 
to minima (maxima):
\begin{equation}
{\pp^2 V_{pot}\over\pp\omega_{12}^2} = 
- \rho_1\rho_2R_{12}\cos\omega_{12},\quad
m^2_{\omega_{12}} = \left<\,{3b^{2}_{+}\beta_{12}^{2}R_{12}\over 
2(1+b_+\ell)^2}\rho_1\rho_2 \,\right> . 
\end{equation}
There is also a small Im$t^I$-$\omega_{12}$ mixing. Note that while in contrast 
to the single condensate case, the dynamical axion is no longer massless, its 
mass is exponentially suppressed relative to the gravitino mass by a factor 
$\sim \langle\,\rho_2/\rho_1\,\rangle^{1/2}$. Therefore, in generic string 
models there is only one very light axion\footnote{As discussed in Section 
3.3.1, this statement is true in the context of both static and dynamical 
gaugino condensation, where the former is the effective description of the 
latter.} ({\it i.e.}, the model-independent axion). As will be discussed in 
Chapter 5, this very light axion has the right properties to be the QCD axion 
\cite{QCD}. 

For $\G = E_6\otimes SU(3)$, the potential is dominated by the $E_6$ gaugino 
condensate, and the results are the same as in (4.71). The only other gauge
groups in the restricted set considered here that are subgroups of $E_8$ are
$\G = [SU(3)]^{n},\; n\le 4$; these cannot generate sufficient supersymmetry 
breaking.
\section{Supersymmetry Breaking}
The pattern and scale of supersymmetry breaking are determined by the 
{\em vev}'s of the $F$ components of the chiral superfields. From the equations 
of motion for $\pi^\alpha$ and $\rho_a$ we obtain, at the self-dual point
$\langle\,F^I\,\rangle=0$:
\begin{eqnarray}
\langle\; F^\alpha \;\rangle &=& 
{(1+\ell\dg)\over4\ell^2 b_a}\pi^\alpha
\(\bar{u} + \ell\sum_b b_b \bar{u}_b\) 
\approx{3b_+^2\over4b_a}\pi^\alpha
\bar{u}_+\(1 + \ell b_+\)^{-1}, \quad b^\alpha_a\ne0,\nonumber \\
\langle\; F^a + \bar{F}^a \;\rangle &=& {1\over 4 \ell^2 b_a}
(1+\ell\dg)(1+\ell b_a)
\[ u_a \( \bar{u} + \ell \sum_b b_b \bar{u}_b \) + 
{\rm h.c.}\] \nonumber \\ &&
\approx {3 b^2_+ \over 4b_a}{1+\ell b_a \over 1+\ell b_+}
\( u_a \bar u_+ + \bar u_a u_+ \),  
\end{eqnarray}
where the approximations on the right hand sides are exact for a single 
gaugino condensate. The dominant contribution is from the gaugino condensate 
with the largest $\beta$-function coefficient:
\begin{equation}
\langle\; F^+ + \bar{F}^+ \;\rangle = 
 {3\rho^2_+b_+\over2}.
\end{equation}
It has been known for some time that, if the dominant supersymmetry breaking 
effects come from the dilaton rather than the moduli, the soft supersymmetry 
breaking parameters are naturally flavor blind, and non-universal squark and 
slepton masses that could induce unacceptably large flavor-changing neutral 
currents (FCNC) could be thereby avoided \cite{FCNC}. Therefore, the fact that 
the $F^I$ vanish in the vacuum is a desirable feature for phenomenology. And it
should be emphasized that this unique feature is just the natural consequence 
of modular invariance and a correct treatment of gaugino condensation in string
theory. In other words, a modular invariant treatment of gaugino condensation in
string theory naturally leads to the phenomenologically desirable
dilaton-dominated supersymmetry breaking scenario\footnote{As will be 
discussed in Section 4.7, this feature is absent in those works \cite{ppp} 
where modular invariance is incorporated but the constraint on gaugino 
condensation (4.4) has not been included.}, which is very impressive!
However, as we will see in Chapter 5, the dilaton-dominated supersymmetry 
breaking scenario is not always free from the FCNC problem, which means the 
the analysis of dilaton-dominated scenario in the past \cite{Brignole,FCNC} is
oversimplified. In fact, possible non-universal couplings of the matter 
superfields to the Green-Schwarz counterterm could induce non-universal squark 
and slepton masses. More discussion of this problem will be given in Chapter 5.

Another important parameter for soft supersymmetry breaking in the observable 
sector is the gravitino mass $m_{\tilde G}$. The derivation of the gravitino 
part of the Lagrangian again parallels the construction in Section 2.3.2. The 
gravitino mass $m_{\tilde G}$ is determined by the term:
\begin{eqnarray}
\Lag_{mass}(\psi) &=& 
-\frac{1}{8}\psi^m\sigma_{mn}\psi^n\sum_a\bar{u}_a
\Bigg\{{1+f\over\ell} + \,b'_a\ln(e^{2-K}\bar{u}_au_a) 
+ \sum_\alpha b^\alpha_a\ln(\pi^\alpha\bar{\pi}^\alpha) 
\nonumber \\ & & 
+ \sum_I\[bg^I - {b_a^I\over4\pi^2}\ln|\eta(t^I)|^2\]\Bigg\}
- e^{K/2}\bar{W}\psi^m\sigma_{mn}\psi^n + {\rm h.c.},
\end{eqnarray}
giving, when the equations of motion (4.26) are imposed,
\begin{equation}
m_{\tilde G} = {1\over3}\langle\,|M|\,\rangle = 
\frac{\, 1}{\, 4\,}\,\langle\,|\sum_ab'_au_a - 4e^{K/2}W|\,\rangle= 
\frac{\, 1}{\, 4\,}\,\langle\,|\sum_ab_au_a|\,\rangle\,\approx\,
{1\over 4}b_+\langle\,\rho_{+}\,\rangle. 
\end{equation}

The scale of supersymmetry breaking is governed by the {\em vev} (4.31) of 
the gaugino condensate with the largest $\beta$-function coefficient. This 
includes the usual suppression factor 
$\langle\,\rho_a\,\rangle\,\,\propto\,\, e^{-1/b_a g_s^2}$, where 
$g_s^2 = \langle\,2\ell/(1+f)\,\rangle$ is the effective squared coupling 
constant at the string scale. However, there are also other important 
parameters that determine the scale of the hierarchy between the supersymmetry 
breaking scale and the Planck scale. The dependence on the string moduli 
provides a second exponential suppression factor:
\begin{equation}
\langle\,\rho_a\,\rangle \;\;\propto\;\; 
\langle\;\prod_I |\eta(t^I)|^{2(b - b_a)/b_a}\;\rangle \;=\; 
|\eta(1)|^{6(b - b_a)/b_a} \;\approx\; e^{-\pi(b - b_a)/2b_a} .
\end{equation}
On the other hand, the numerical factor 
$\;\prod_\alpha|b^\alpha_a/4c_\alpha|^{-b^\alpha_a/b_a}\;$ generates an 
exponential enhancement if $c_\alpha\sim 1$. This is the largest numerical
uncertainty in our analysis.  {\it A priori}, $c_\alpha$ is related to the
Yukawa couplings of matter fields in the hidden sector. However, there is an
arbitrary normalization factor in the definition of $\Pi^\alpha$. If the
hidden-sector Yukawa couplings were known, it might be possible to estimate 
$c_\alpha$ by a matching condition for the {\em vev}'s of the second lines of 
(4.32) and (4.33). In our numerical analysis, we have set $c_\alpha = 1$.
Then, if the hidden gauge group with the largest $\beta$-function coefficient
is $\,\G_+= E_6\,$ with $\,3N_{27}\,$ matter chiral superfields in the 
fundamental representation, we obtain:
\begin{equation}
m_{\tilde G} = \cases{1.1\times 10^{-9}\cr3.3\times 10^{-11}
\cr1.65\times 10^{-15}\cr} \quad {\rm for} \quad N_{27} = 
\cases{1\cr2\cr3\cr},
\end{equation}
in reduced Planck units. For $\,\G_+= SU(3)\,$ with three matter chiral 
superfields in the fundamental representation, we obtain an unacceptably large 
gauge hierarchy: $m_{\tilde G} = 2.2\times 10^{-32};\;m_{\tilde G}$ decreases 
rapidly as $N_{3}$ increases, {\it i.e.} as the $\beta$-function coefficient 
decreases.
\section{Concluding Remarks}
In the class of models studied here, the introduction of a parameterization 
for stringy nonperturbative contributions to the K\"ahler potential for the 
dilaton generically allows a stable vacuum at a nontrivial, phenomenologically 
acceptable point in the dilaton/moduli space. In particular, when we impose 
the constraint that the cosmological constant vanishes, we find that in the 
linear multiplet formalism, the string moduli $t^I$ are stabilized at the
self-dual point, and their associated $F$ components vanish in the vacuum,
which results in a phenomenologically desirable dilaton-dominated
supersymmetry breaking scenario. This striking feature of string phenomenology
is in fact just the consequence of modular invariance and a correct treatment 
of gaugino condensation\footnote{As discussed in the appendix, an
inappropriate treatment of gaugino condensation and/or modular 
invariance is the reason why this unique feature of string phenomenology has 
been ignored in the past.}. Therefore, in this sense the experimental 
search for a dilaton-dominated supersymmetry breaking scenario can be regarded 
as an indirect test of the modular invariance of weakly-coupled heterotic 
string. 

A salient feature of our formalism is that there is little qualitative 
difference between a single condensate and a multi-condensate scenario. For 
several gaugino condensates with equal (or very similar) $\beta$-function
coefficients, the scalar potential reduces to that of the single gaugino 
condensate case, except that there may be flat directions. If 
$b_1=b_2=\cdots b_k$, then at the self-dual point $\rho_a/\rho_1=\zeta_a =$ 
constant and the potential vanishes identically in the direction 
$\sum_{a=1}^k\zeta_ae^{i\omega_a}=0,\;\rho_{a>k} = 0.$ This always has a 
solution if $\zeta_a = 1$, in which case the flat direction preserves 
supersymmetry and there is no barrier between this solution and the interesting,
supersymmetry breaking solution. For several gaugino condensates with different
$\beta$-function coefficients, the scalar potential is dominated by the gaugino
condensate(s) with the largest $\beta$-function coefficient, and the result is 
essentially the same as in the single gaugino condensate case, except that a 
very small mass is generated for the dynamical (model-independent) axion. In 
all cases, stringy nonperturbative corrections to the dilaton K\"ahler 
potential are required to stabilize the dilaton. This picture is very different
from previously studied ``racetrack'' models \cite{racetrack} where dilaton 
stabilization is achieved through cancellations among different gaugino 
condensates with similar $\beta$-function coefficients. The qualitative 
difference between an $E_8$ hidden sector and one with a product gauge group 
is the presence of hidden matter; in the $E_8$ case there is no hidden matter
and the scalar potential is independent of the moduli, which therefore remain 
undetermined in the classical vacuum of the effective condensate theory. More
phenomenological discussions of the model constructed in this chapter will be 
presented in Chapter 5.
\section{Appendix: Chiral Multiplet Formalism}
There has been interest in the question as to whether the linear and chiral
multiplet formalisms are equivalent at the quantum level. They are presumably
equivalent in the sense that technically we may always perform a duality 
transformation at the superfield level on the Lagrangian (4.5) so as to recast 
it entirely in terms of chiral supermultiplets. The resulting effective 
Lagrangian should be the chiral multiplet formalism with the gaugino 
condensates constrained by (2.12), and it is apt to be rather complicated
\cite{Burgess95,dilaton}.

The string phenomenology that we have constructed and studied so far is 
quite different from the ``conventional'' string phenomenology in several 
aspects. Besides the aforementioned linear--chiral duality question, the 
``conventional'' string phenomenology is different from ours in the sense 
that the constraint (2.12) on gaugino condensates has always been ignored, and
usually the treatment of modular invariance is incomplete or
incorrect in the ``conventional'' study of string phenomenology. Therefore, a 
more practical question that we address in this appendix is the extent to 
which our studies in Sections 4.1--4.6 can be reproduced if one takes as a 
starting point the usual chiral multiplet formalism for the dilaton with the 
gaugino condensates represented by {\em unconstrained} chiral superfields 
({\it i.e.}, the ``conventional'' approach), and modular invariance is ensured 
through the Green-Schwarz mechanism and string threshold corrections. In
particular, we would like to know how an inappropriate treatment of gaugino
condensation ({\it i.e.}, a treatment without the constraint (2.12) on gaugino 
condensates) might have affected our understanding of string phenomenology in
the past.

In the chiral multiplet formalism, the Green-Schwarz counterterm appears as a 
correction to the K\"ahler potential, which we take to be
\begin{equation}
K(S,T^I) = \ln(L) + {\tilde g}(L) 
+ \sum_Ig^I, \quad L^{-1} = S+\bar{S} - b\sum_Ig^I, 
\end{equation}
where ${\tilde g}$ is the correction from stringy nonperturbative effects in 
the chiral multiplet formalism\footnote{Notice that the vector superfield $L$ 
here is simply a convenient notation for $\,(S+\bar{S} - b\sum_Ig^I)^{-1}$. It
should not be confused with the $L$ used in the linear multiplet formalism.}.
Modular invariance of the Yang-Mills Lagrangian at the quantum level is assured 
by the transformation property of $S$ under (4.14):
\begin{equation}
S \to S + b\sum_IH^I,
\end{equation}
and modular covariance of the K\"ahler potential 
($K\,\to\,K+\sum_I(H^I +\bar{H}^I)$) requires that it depend on $S$ only 
through the vector superfield $L$ defined in (4.81). We introduce static 
gaugino and matter condensate superfields $U_a$ and $\Pi^\alpha$ as before,
but now the gaugino condensate chiral superfield 
\begin{equation}
U_a = e^{K/2}H_a^3 
\end{equation}
is not constrained by the constraint (2.12) or (4.42) because $H_a$ is taken 
to be an {\em unconstrained} chiral superfield in the treatment here. (This is
what has always been done in the conventional study of string phenomenology.) 
We construct the superpotential in analogy to (4.5), using the standard 
approach of Veneziano and Yankielowicz:
\begin{equation}
W_{tot} =  W_{cond} + W(\Pi), 
\end{equation}
where $W(\Pi)$ is the same as in (4.30), and
\begin{eqnarray}
W_{cond} &=& W_C + W_{VY} + W_{th}, \quad W_C = {1\over4}S\sum_aH_a^3,
\nonumber \\ W_{VY} &=& {1\over4}\sum_aH_a^3\(3b'_a\ln H_a + 
\sum_\alpha b^\alpha_a\ln\Pi^{\alpha}\), \nonumber \\ 
W_{th} &=& {1\over4}\sum_{a,I}{b^I_a\over8\pi^2}H_a^3\ln[\eta^2(T^I)],
\end{eqnarray}
where $W_C$ represents the classical contribution of gaugino condensation. 
$\,H_a^3$ transforms in the same way as $U_a$ under rigid chiral and conformal 
transformations, and the anomaly matching conditions give the same constraints 
on the coefficients $b$'s as in Section 4.2. Then it is straightforward to 
check that, under the modular transformation (4.14) with 
$H_a\to e^{-\sum_IH^I/3},$ we have $W_{cond}\to e^{-\sum_IH^I/3}W_{cond},$ as 
required by modular invariance of the Lagrangian. Summing the various 
contributions, the superpotential for $H_a$ can be written in the following 
form:
\begin{equation}
W_{cond} = {1\over4}\sum_a b'_aH_a^3\ln\lbr 
e^{S/b'_a}H^3_a\prod_\alpha
(\Pi^\alpha)^{b^\alpha_a/b'_a}
\prod_I[\eta(T^I)]^{-b^I_a/4\pi^2b'_a}\rbr. 
\end{equation}
The bosonic Lagrangian takes the standard form:
\begin{eqnarray}
\Lag_B &=& -{1\over2}{\cal R} - {1\over3}M\bar{M} + K_{i\m}\(F^i
\bar{F}^{\m} - \pp_\mu z^i\pp^\mu z^{\m}\) \nonumber \\ & & 
+ e^{K/2}\[F^i\(W_i + K_iW\) - \bar{M}W + {\rm h.c.}\],
\end{eqnarray}
where $Z^i = S,T^I,H_a,\Pi^\alpha,\;z^i=Z^I\lowest$. In our static model 
$K_{i\m},K_i = 0$ for $Z^i,Z^m = H_a,\Pi^\alpha$, and the equations of motion 
for $F^i$ give $W_i = 0$ for these fields. This determines the chiral 
superfields $H_a,\Pi^\alpha$ as holomorphic functions of $S,T^I$. Making the 
same restrictions on $W(\Pi)$ and the $b^\alpha_a$ as in Section 4.2, we obtain:
\begin{eqnarray}
H^3_a &=& e^{(2n+1)i\pi(b'_a - b_a)/b_a - b'_a/b_a}e^{-S/b_a}\prod_I
[\eta(T^I)]^{2(b - b_a)/b_a}\prod_\alpha\left|{b_a^\alpha/4c_\alpha}
\right|^{-b_a^\alpha/b_a},  \nonumber \\
\Pi^\alpha &=& -{b_a^\alpha\over4c_\alpha}H_a^3
\prod_I[\eta(T^I)]^{-2(q^\alpha_I
- 1)},\quad b_a^\alpha \ne 0. 
\end{eqnarray}
As in (4.31), the correct dependence of the gaugino condensates on the squared 
gauge coupling constant $\,\langle\; 2/{\rm Re}s \;\rangle,\; s = S\lowest$, is 
recovered. Note however that, in contrast to (4.31), the phases of gaugino 
condensate here are quantized once Im$s$ is fixed at its {\em vev}. Using these 
results gives
\begin{equation}
W_{tot} = W(S,T^I) = - {1\over4} \sum_a  b_aH_a^3.
\end{equation}
The scalar potential $V_{pot}$ is determined in the standard way after 
eliminating the remaining auxiliary fields through their equations of motion:
\begin{eqnarray}
M &=& - 3e^{K/2}W,\quad \bar{F}^{\m} = 
- e^{K/2}K^{i\m}\(W_i + K_iW\), 
\quad Z^i = S,T^I, \nonumber \\
V_{pot}(s,t^I,\bar{t}^I) &=& 
e^{K}\[K^{i\m}\(W_i + K_iW\)\(\bar{W}_{\m} + 
K_{\m}\bar{W}\) - 3|W|^2\].
\end{eqnarray}
The inverse K\"ahler metric for the K\"ahler potential (4.81) is: 
\begin{eqnarray}
K^{I\bar{J}} &=& {4({\rm Re}t^I)^2\over(1-bK_s)}\delta^{IJ}, \quad
K^{I\bar{s}} = - {2b{\rm Re}t^I\over(1-bK_s)}, \nonumber \\
K^{s\bar{s}} &=& 
{1 - bK_s + 3b^2K_{s\bar{s}}\over K_{s\bar{s}}(1-bK_s)},
\end{eqnarray}
and the scalar potential $V_{pot}$ reduces to 
\begin{eqnarray}
V_{pot} &=& 
{e^K\over1-bK_s}\bigg\{K_{s\bar{s}}^{-1}\(1-bK_s + 3b^2K_{s\bar{s}}\)
|W_s + K_sW|^2 + 4\sum_I\({\rm Re}t^I\)^2|W_I + K_IW|^2 
\nonumber \\ & &
- 2b\[\(\bar{W}_s + K_s\bar{W}\)\sum_I{\rm Re}t^I\(W_I + K_IW\) 
+ {\rm h.c.}\]
\bigg\} - 3e^K|W|^2.
\end{eqnarray}
We have
\begin{eqnarray}
- 2{\rm Re}t^I\(W_I + K_IW\) &=& -\sum_a{1\over4b_a}\[1-bK_s 
- {b-b_a\over b_a}{\rm Re}t^I\zeta(t^I)\]H^3_a, \nonumber \\
W_s + K_sW &=& \sum_a{1\over4b^2_a}\(1 - K_sb_a\)H^3_a ,
\end{eqnarray}
and the scalar potential can be written in the following form:
\begin{equation}
V_{pot} = {e^K\over16(1-bK_s)}\sum_{ab}|h_ah_b|^3\cos\omega_{ab}R_{ab}, 
\end{equation}
where here $\omega_a$ is the phase of $h^3_a= H^3_a\lowest, \;\omega_{ab}$ 
is defined as before, and 
\begin{eqnarray}
R_{ab} &=& b_ab_bf_{ab}(\ell) + (b-b_a)(b-b_b)\sum_I|1 + 
4{\rm Re}t^I\zeta(t^I)|^2, \quad \ell = L\lowest,\nonumber \\ 
f_{ab}(\ell) &=& 
(1 - bK_s)\[{(1- b_aK_s)(1- b_aK_s)\over b_ab_bK_{s\bar{s}}} - 3\]. 
\end{eqnarray}
In the absence of stringy nonperturbative effects, 
$K_s = -\ell,\;K_{s\bar{s}} = \ell^2,\;f_{ab} \to - 2b\ell$ as 
$\ell\to \infty$, and the scalar potential $V_{pot}$ is unstable in the strong
coupling direction, as expected. A positive definite scalar potential requires 
that $f_{++}(\ell)$ be positive semi-definite where, as before, $b_+$ is the 
largest $b_a$. Note that the perturbative expression for $f_{aa}(\ell)$ is 
negative for $b_a\ell>1.4$, while in the linear multiplet formalism the 
corresponding expression is negative only for $b_a\ell> 2.4$, so 
stringy nonperturbative effects are required to be more important in the 
{\em unconstrained} chiral multiplet formalism\footnote{{\em Unconstrained} 
chiral multiplet formalism means the chiral multiplet formalism without the
constraint (2.12) or (4.42).} here. If there is only one gaugino condensate, 
the self-dual point for the moduli is again a minimum, but 
$\langle\,F^I\,\rangle\ne 0$.  In the general case, the minimization equations 
for the moduli read:
\begin{eqnarray}
{\pp V_{pot}\over\pp t^I} &=& {e^K\over16(1-bK_s)}
\sum_{ab}|h_ah_b|^3\cos\omega_{ab}
\({2b\over n}\zeta(t^I)\sum_{c}\beta_{ca}R_{ab} 
+ {\pp\over\pp t^I}R_{ab}\) 
\nonumber \\ & & 
+ \(A + {2b\over n}\zeta(t^I)\sum_{a}{1\over b_a}\)V_{pot} , 
\end{eqnarray}
where $\beta_{ab}$ is defined as in (4.63). Again imposing 
$\langle\,V_{pot}\,\rangle= 0$, the minimum is shifted slightly away from the 
self-dual point if some $\beta_{ab}\ne 0$.

The effective Lagrangian constructed using the linear multiplet formalism -- 
like the string and field-theoretical loop-corrected Yang-Mills Lagrangian 
\cite{Gaillard92,KL} -- depends only on the variables $t^I$ and the modular 
invariant field $\ell$, so the Lagrangian is invariant under modular 
transformations on the $t^I$ alone. In contrast, the effective Lagrangian 
constructed using this {\em unconstrained} chiral multiplet formalism has an 
explicit $s$--dependence which accounts for the fact that the self-dual point 
is not the minimum. The {\em unconstrained} chiral multiplet construction 
forces a holomorphic coefficient for the interpolating superfield for the 
Yang-Mills composite superfield $\, U\simeq \WaWa$, and hence cannot faithfully 
reflect the non-holomorphic contribution from the Green-Schwarz counterterm. 
This is again related to the fact that the {\em unconstrained} chiral multiplet 
construction does not account for the constraint (2.12) or (4.42) which has to 
be satisfied by the gaugino condensate superfields. Our analysis in this 
appendix explicitly explains why in the past the study of string phenomenology 
using the {\em unconstrained} chiral multiplet formalism has not been able to 
predict moduli stabilization at the self-dual point and therefore a 
dilaton-dominated supersymmetry breaking scenario. This conclusion is also
consistent with previous works such as \cite{ppp} where the 
{\em unconstrained} chiral multiplet formalism was employed.
\newpage
\setcounter{chapter}{4} 
\chapter{Weakly-Coupled Heterotic String Phenomenology}
\hspace{0.8cm}
\setcounter{equation}{0} 
\setcounter{footnote}{0}
\newpage

\section{Introduction}
\hspace{0.8cm}
In Chapter 4, we have constructed string models which include supersymmetry 
broken at a realistic scale, a stabilized dilaton, moduli fields with couplings 
respecting modular invariance and a vanishing cosmological constant. We believe 
that it is sufficiently realistic to allow for a discussion of many
phenomenological issues associated with supersymmetry breaking, moduli 
physics and axion physics based on actual computations rather than educated 
guesses\footnote{As we shall see, several such educated guesses about string
phenomenology which have been regarded as standard turn out to be inappropriate 
according to our actual computations.}. Needless to say, we have no miraculous 
solution for either dilaton stabilization or the vanishing of the cosmological 
constant. Although these are incorporated in the model by fixing some 
parameters (only the second constraint requires fine tuning), the model is 
still predictive enough in many respects. In Sections 5.2 and 5.3, we comment 
on several problems associated with string moduli and axion. In particular,
these analyses are quite insensitive to the details of the string models, and
therefore the conclusions are fairly model-independent. In Section 5.4, we 
study the pattern of soft supersymmetry breaking parameters. As expected, the 
conclusions of this section are sensitive to the details of the specific 
string model under consideration. In Section 5.5, we comment on gauge coupling
unification in the presence of significant stringy non-perturbative effects.
In order to make the presentation transparent, in most sections we start with 
the known results and problems of string phenomenology studied in the
past\footnote{As discussed in the appendix of Chapter 4 and elsewhere, these
studies in the past are based on the {\em unconstrained} chiral multiplet
formalism.}. We then present the results obtained from the realistic model
constructed in Chapter 4. In particular, we emphasize how the standard lore
of string phenomenology is modified within our model, and how the problems
of string phenomenology could naturally be solved by these important
modifications\footnote{As we have seen and shall see, many so-called problems 
of weakly-coupled string phenomenology known in the past are not really 
problems of weakly-coupled string phenomenology itself. In fact, they are 
mostly due to our limited calculational power in string theory, little 
knowledge of its true vacuum structure, and an inappropriate treatment 
of gaugino condensation.}.
\section{Moduli Physics}
\subsection{Moduli Stabilization at the Self-Dual Point}
\hspace{0.8cm}
As discussed in Chapter 4, simply as a consequence of modular invariance and 
an appropriate treatment of gaugino condensation, the compactification moduli
$T^{I}$'s are stabilized at the self-dual point, 
$\langle\,t^{I}\,\rangle\,=\,1$. This means that the compactification scale 
$M_{comp}$ is actually close to the string scale $M_{s}$. This observation 
will be relevant to the discussion of the Newton's constant in Section 5.5. 
What's more interesting is that fact that, in the vacuum ({\it i.e.}, at the 
self-dual point), the $F$ components of $T_{I}$'s vanish. Therefore, although
$T_{I}$'s are stabilized by supersymmetry breaking effects, they do not
contribute to the breaking of supersymmetry. As emphasized before, this leads
a dilaton-dominated scenario of supersymmetry breaking. In the context of 
superstring, our study offers a rationale for the phenomenologically 
interesting dilaton-dominated scenario.
\subsection{Mass Hierarchy between Moduli and Gravitino}
\hspace{0.8cm}
At the perturbative level, the dilaton and moduli are are flat directions of 
the potential, and they are lifted only through non-perturbative effects. It is
often argued that the non-perturbative effects which break supersymmetry also
lift these flat directions. As we have learned from the standard lore of string
phenomenology, a naive oder-of-magnitude estimate concludes that string dilaton 
and moduli have masses of order (or no larger than) the gravitino mass 
\cite{Kaplan93,Carlos93}, where the natural scale of gravitino mass is about 
\mbox{1 TeV}. Obviously, these light dilaton and moduli fields with couplings 
suppressed by the Planck scale could lead to serious cosmological problems. A 
rough estimate for the decay rate $\Gamma$ of string dilaton or moduli is at 
most
\begin{equation} 
\Gamma\;\sim\;\frac{m^{3}}{8\pi {M'}_{P}^{2}},
\end{equation}
where $m$ is the mass of string dilaton or moduli, $M'_{P}=M_{P}/\sqrt{8\pi}$ 
is the reduced Planck scale and $M_{P}$ is the Planck scale. This slow decay 
rate is the source of cosmological problems. That is, relic dilaton and moduli 
produced in the very early universe survive to a dangerously late epoch. With 
the slow decay rate (5.1), they result in a low reheat temperature $T_{R}$ 
\cite{Kaplan93,Randall94}:
\begin{equation} 
T_{R}\;\sim\;5\left(\frac{m}{\mbox{TeV}}\right)^{3/2}\;\;\;\mbox{keV}.
\end{equation}
Such a low reheat temperature is inconsistent with successful nucleosynthesis
unless $\,m\,\geq\,\CO(3)\times 10^{4}$ GeV (if $\,T_{R}\,\geq\,\CO(1)$ MeV is
required.) According to the standard lore of string phenomenology, 
$\,m\,\geq\,\CO(3)\times 10^{4}$ GeV would imply an un-naturally large gravitino
mass, which is not acceptable. This is the so-called cosmological moduli problem
\cite{Kaplan93,Randall94,Berkooz94}, where the Polonyi problem is an earlier
version of this problem in the context of spontaneously broken supergravity 
\cite{Polonyi}. In order to solve the cosmological moduli problem, there have 
been attempts at a hierarchy between moduli and squark masses 
\cite{Berkooz94,Nir94}; however, none of them is realistic. There are also
possible cosmological solutions to the cosmological moduli problem, such as a
weak scale inflation \cite{Randall94}.

Now, let's leave the standard lore of string phenomenology and turn to the 
realistic model constructed in Chapter 4. One can easily extract from the 
scalar potential the masses of the dilaton and of the moduli, which are 
particularly relevant for cosmology. According to (4.70), one finds the mass 
of the moduli $m_{t^{I}}$ as follows:
\begin{equation} 
m_{t^{I}}\,\approx\,\left<\,
{1\over2}{(b-b_+)\over(1+b\ell)}\,\rho_+\,\right>.
\end{equation}
where $\rho_+$ is the hidden-sector gaugino condensate with the largest 
one-loop $\beta$-function coefficient $b_+$. As for the mass of the dilaton 
$m_{d}$, one finds:
\begin{equation} 
m_d \,\sim\, {1 \over b_+^2}\, m_{\tilde G}.
\end{equation}
According to (4.77), the gravitino mass is: 
$\,m_{\tilde G}\,\approx\,\frac{1}{4}b_{+}\langle\,\rho_{+}\,\rangle$. In
generic string models $b/b_{+}$ and $1/b_{+}^{2}$ are naturally large numbers, 
and therefore in contrast to the standard lore of string phenomenology our 
model has a natural hierarchy between the dilaton/moduli and squark/slepton 
masses. More precisely, in order to generate a realistic hierarchy of order 
$m_{\tilde G}\approx 10^{-15}M'_{P}\approx 10^3$ GeV, it is required that 
$b/b_+\approx 10$ for the string models under consideration. (Such an example 
has been presented in Section 4.5.) In this case, 
$m_{t^{I}}\approx 20 m_{\tilde G}\approx 20$ TeV and 
$m_d\sim 10^3m_{\tilde G}\approx 10^3$ TeV (where $m_{\tilde G}\approx 1$ TeV.) 
This natural hierarchy between the dilaton/moduli and squark/slepton masses 
could be sufficient to solve the cosmological moduli problem. This hierarchy
could also have other non-trivial cosmological implications. The implication of
such a hierarchy on the primordial black hole constraints has recently been
studied in \cite{Liddle}.

One may wonder why the mass of dilaton is particularly large in our model. In
fact, this specific feature has to do with the cancellation of the cosmological 
constant. In our model, it is implicitly assumed that the mechanism which breaks
supersymmetry is also responsible for the cancellation of the 
cosmological constant,
which is the minimal and most economical assumption\footnote{In our model, 
positivity of the scalar potential can always be imposed. One thus does not 
need to appeal to another source of supersymmetry breaking to cancel the 
cosmological constant.}. With this assumption, 
$\;\langle\,V_{pot}\,\rangle =0\;$ leads to 
$\;\langle\,1+\ell\dg\,\rangle\;\approx\;3b_{+}^{2}\langle\,\ell^{2}\,\rangle$.
According to (4.27), the kinetic term of dilaton contains the small factor 
$\langle\,1+\ell\dg\,\rangle$, which therefore leads to an enhancement of the 
mass of dilaton. On the other hand, there is so far very little insight about
how the cosmological constant problem should be solved. It is possible that
there are other sources which could contribute to the cancellation of 
cosmological constant. However, a detailed analysis of these more complicated 
scenarios is beyond the scope of our study here. We wish to emphasize that, 
even if $\;\langle\,1+\ell\dg\,\rangle\;$ might turn out to be, for example, an 
$\CO(1)$ number in some other more complicated solutions to the cosmological 
constant problem, the natural hierarchy between the dilaton/moduli and 
squark/slepton masses still exists as long as gaugino condensation is the major 
source of supersymmetry breaking; in this case we have 
$m_t\approx 20 m_{\tilde G}\approx 20$ TeV and 
$m_d\sim\left(1/b_{+}\right)m_{\tilde G}\sim 30m_{\tilde G}\approx 30$ TeV. 
\section{Axion Physics}
\hspace{0.8cm}
The invisible axion is an elegant solution to the strong CP problem. In string
theory, there seem to be many such axion candidates. However, as for the
weakly-coupled superstring, it has been argued that QCD cannot be the dominant
contribution to the potential of any string axion \cite{Banks96}, and therefore
none of the string axions is qualified for the QCD axion. For string axions 
associated with the compactification $T^{I}$ moduli, Peccei-Quinn symmetries 
are significantly broken by world-sheet instanton effects \cite{Banks96}. 
For the string model-independent axion, it has been argued (again using the 
{\em unconstrained} chiral multiplet formalism) that the model-independent 
axion cannot be the QCD axion due to stringy non-perturbative effects (of 
order $e^{-c/g_{s}}$ for the superpotential of dilaton) \cite{Banks94,Banks96}.
On the other hand, in the realistic model constructed in Chapter 4 where 
stringy non-perturbative effects are fully included using the linear multiplet 
formalism, the model-independent axion does have the right features to be the
QCD axion. The resolution for the stringy non-perturbative contribution, 
$e^{-c/g_{s}}$, to the superpotential of the dilaton is simple and impressive:
as argued in \cite{Banks94,Banks96} using the chiral multiplet formalism, it 
seems plausible that there should be significant $e^{-c\sqrt{S}}$ contributions
to the superpotential of dilaton, leading to the QCD axion 
problem raised by Banks and Dine \cite{Banks96}. On the other hand, in the 
linear multiplet formalism of string effective theory where the dilaton is
represented by a vector superfield $L$, it is simply impossible to write down 
any $L$-dependent contribution ({\it e.g.}, $e^{-c/\sqrt{L}}$) to the 
superpotential  -- a constraint coming from holomorphy. Therefore, in the 
linear multiplet formalism the QCD axion problem of Banks and Dine 
\cite{Banks96} is resolved in an elegant way, and one should re-examine the 
attractive possibility of the string model-independent axion being the QCD 
axion in this framework. 

For any of the string axions to solve the strong CP problem, there is also a
cosmological constraint. Cosmological considerations require the decay 
constant $F_{a}$ of the invisible axion to lie between $10^{10}$ GeV and 
$10^{12}$ GeV (the so-called axion window \cite{Abbott83,Raffelt90}). The upper
bound on the axion decay constant, $F_{a}\leq 10^{12}$ GeV, is due to the
requirement that the energy density of the coherent oscillations of the axion
be less than the critical density of the universe \cite{Abbott83}. However, in 
superstring theory the axion decay constant $F_{a}$ is naturally of order the 
Planck scale, and therefore the cosmological upper bound on $F_{a}$ is
seriously violated. Although it was shown by Choi and Kim \cite{Kim} that the
decay constant $F_{a}$ of the model-independent axion in the weakly-coupled 
heterotic string theory actually is $\,M'_{P}/16\pi^{2}\,\approx\,10^{16}$ GeV, 
this is still much larger than the cosmological upper bound. On the other hand, 
cosmological constraints could be quite scheme-dependent; for example, it has 
been pointed out that the entropy production due to the decays of massive 
particles dilutes the axion density and therefore raise the upper bound on 
$F_{a}$ \cite{Dine83}. Based on the above idea Kawasaki, Moroi and Yanagida 
\cite{Moroi} have proposed a refined scenario where the Polonyi fields of 
supergravity models are natural candidates for entropy production. The new 
cosmological upper bound on $F_{a}$ in this scheme is:
\begin{equation} 
F_{a}\,\leq\,5\times 10^{15} 
\left(\frac{m_{\phi}}{10\;\mbox{TeV}}\right)^{-3/4}\;\;\;\mbox{GeV},
\end{equation}
where $m_{\phi}$ is the mass of the Polonyi field. In order to keep successful 
primordial nucleosynthesis in this scheme, $m_{\phi}$ should be larger than 
about 10 TeV. With $m_{\phi}\approx 10$ TeV, $F_{a}\leq 5\times 10^{15}$ GeV 
and therefore the string model-independent axion is almost consistent with this 
new upper bound. However, $m_{\phi}\geq 10$ TeV seems un-natural according to 
the standard lore of string phenomenology where one expects 
$m_{\phi}\approx m_{\tilde G} \approx 1$ TeV. On the contrary, the 
cosmological scenario of Kawasaki {\it et al} naturally occurs in our model 
constructed in Chapter 4. As discussed in Section 5.2, in our model there is a 
natural hierarchy between the moduli and gravitino masses 
($m_{t^{I}}\approx 20 m_{\tilde G}\approx 20$ TeV), and therefore the decays of
moduli serve the purpose of raising the cosmological upper bound on $F_{a}$ to 
a value consistent with the $F_{a}$ of string model-independent axion. This
natural hierarchy is indeed a desirable feature of our model since it not only
could solve the cosmological moduli problem but also keeps the energy density 
of the oscillations of string model-independent axion from overclosing the
universe.

One particularly interesting aspect of our model constructed using the linear
multiplet formalism of gaugino condensation in Chapter 4 is axion physics. 
Pseudoscalar fields are the phases $\omega_a$ of the condensates and the 
so-called model-independent axion which is dual to the fundamental 
antisymmetric tensor field. The latter couples in a universal way to the 
$F^{a\mu\nu}{\tilde F}^{a}_{\mu\nu}$ term of each gauge subgroup. If again we 
look at the dynamical model with one $E_8$ gaugino condensate in Chapter 3, we 
find that out of the two possible pseudoscalar the condensate phase is very 
heavy whereas the string model-independent axion remains massless. This is 
obviously the supersymmetric counterpart of what happens with the scalars. If 
we allow for more than one gaugino condensate, the model-independent axion 
acquires a very small mass\footnote{Higher-dimension operators might give 
extra contributions to the mass of this axion. However, these contributions can
be argued to be negligible using discrete R symmetry \cite{Banks94}.}
(typically exponentially suppressed relative to the 
gravitino mass by a factor of order $\langle\,\rho_2 /\rho_1\,\rangle^{1/2}$ 
in the two-condensate case according to (4.73)). Furthermore, as we have seen 
in Section 5.2, the axions associated with the $T^{I}$ moduli get masses of 
order $20\,m_{\tilde G}$. Therefore, we are always left with only one very 
light axion, the model-independent axion, and it has the right properties to 
be the QCD axion. Remember that there are two kinds of non-perturbative effects 
in our model ({\it i.e.}, the field-theoretical non-perturbative effects of 
hidden-sector gaugino condensation constrained by (2.12) and stringy 
non-perturbative effects), and they are best described using the linear 
multiplet formalism. In contrast to the argument against the string 
model-independent axion as the QCD axion \cite{Banks96} in the presence of 
generic stringy non-perturbative effects using the {\em unconstrained} chiral 
multiplet formalism, in our model the model-independent axion can indeed be 
the QCD axion. As explained before, the reason why the model-independent axion 
has the desirable features in the linear multiplet formalism are a correct 
treatment of gaugino condensation and the fact that such stringy 
non-perturbative effects of dilaton are actually forbidden in the 
superpotential due to holomorphy. As for the decay constant $F_a$ of the 
model-independent axion in our model, there is an additional reduction factor 
of $\,\left\langle\,2\ell^{2}(1+\ell\dg)\,\right\rangle^{1/2}$ compared 
to the result obtained by Choi and Kim \cite{Kim}. As discussed in Section 5.2,
this reduction factor comes from the fact that the kinetic term of dilaton in 
(4.27) contains the small factor 
$\;\langle\,1+\ell\dg\,\rangle\;\approx\;3b_{+}^{2}\langle\,\ell^{2}\,\rangle$
when $\,\langle\,V_{pot}\,\rangle =0\,$ is imposed. More precisely, this 
reduction factor is about $\,\left\langle\,2\ell^{2}(1+\ell\dg)\,
\right\rangle^{1/2}\,\approx\,\left\langle\,\sqrt{6}b_{+}\ell^2\,
\right\rangle\,\approx\,1/50\,$ if the gravitino mass is about 1 TeV. Besides 
the fact that the cosmological scenario of Kawasaki {\it et al} naturally 
occurs in our model, this reduction in the model-independent axion's decay
constant is certainly desirable from the viewpoint of the cosmological upper
bound on $F_{a}$. Indeed, with this reduction factor the axion decay constant 
in our model is $\,F_{a}\,\approx\,2\times 10^{14}$ GeV, which is truly 
consistent with the upper bound on $F_{a}$ ($\,\approx\,5\times 10^{15}$ GeV)
imposed by the scenario of Kawasaki {\it et al}.
\section{Soft Supersymmetry Breaking Parameters}
\hspace{0.8cm}
In contrast to the studies of moduli and axion, the analysis of soft
supersymmetry breaking parameters is much more sensitive to the very details 
of a string model. Unfortunately, our current knowledge of string models is 
still limited. Although in the following we will try to discuss soft
supersymmetry breaking parameters in a model-independent way whenever it is
possible, yet it should be kept in mind that our analysis cannot cover all the 
interesting possibilities and therefore should not be regarded as final. 

It is straightforward to compute the soft supersymmetry breaking terms, that 
are generated at the condensation scale 
$\mu_{cond}\,\approx\,\langle\,\rho_{+}\,\rangle^{1/3}$, 
for our model constructed in Chapter 2. The gaugino masses $m_{\lambda_b}$ are:
\begin{equation} 
m_{\lambda_b} = -\left<\,{g^2_b(\mu_{cond})\over8\ell^2}\(1+\ell\dg\)\sum_a 
\(1 + b_a\ell\)\bar{u}_a\,\right> \,\approx\,
-{3\over8}{g^2_b(\mu_{cond}) b^2_+\over1 + b_+\langle\,\ell\,\rangle }
\langle\,\bar{u}_+\,\rangle . 
\end{equation}
Notice that the expression of gaugino masses contains the small factor 
$\langle\,1+\ell\dg\,\rangle$ discussed at the end of Section 5.2, and 
therefore gaugino masses are suppressed by $b^{2}_{+}$ after 
$\,\langle\,V_{pot}\,\rangle =0\,$ is imposed. Therefore, it is possible
that this suppression of gaugino masses could be relieved in models with a 
more complicated mechanism of cosmological constant cancellation. 

The soft terms in the scalar potential are sensitive to the -- as yet 
unknown -- details of matter-dependent contributions to the Green-Schwarz
counterterm and string threshold corrections. We neglect the 
former\footnote{If string threshold corrections are determined by a holomorphic 
function, they cannot contribute to the scalar masses.}, and write the 
Green-Schwarz counterterm as follows:
\begin{equation}
V_{GS} = b\sum_Ig^I + \sum_Ap_Ae^{\sum_Iq^A_Ig^I}|\Phi^A|^2 
+ \CO(|\Phi^A|^4), 
\end{equation}
where the $\Phi^A$ are gauge nonsinglet chiral superfields, the $q^I_A$ are
their modular weights, and the full K\"ahler potential reads
\begin{equation}
K = k(V) + \sum_Ig^I + \sum_Ae^{\sum_Iq^A_Ig^I}|\Phi^A|^2 + \CO(|\Phi^A|^4).
\end{equation}
Under these assumptions, the scalar masses and cubic ``$A$ terms'' are given,
respectively, by the following:
\begin{eqnarray}
m_A^2 &=& {1\over16}\left<\,\left|\sum_a u_a{(p_A - b_a)\over
(1 + p_A\ell)}\right|^2\,\right>\,\approx\,{1\over16}\left<\,{(p_A - b_+)^2
\over(1 + p_A\ell)^2} \rho_+^2\,\right>, \nonumber \\ 
V_A(\phi) &=& { 1\over 4}e^{K/2}\sum_{a,A}\bar{u}_a\phi^AW_A(\phi)\[
\,{p_A-b_a\over1+p_A\ell}  + b_a -\(1+\ell\dg\)
{1+b_a\ell\over3\ell} \,\] + {\rm h.c.}\nonumber \\
&\approx& { 1\over 4}e^{K/2}\bar{u}_+\[\,\sum_A{p_A-b_+\over1+p_A\ell}
\phi^AW_A(\phi) + {3b_+\over1+b_+\ell}W(\phi) \,\] + {\rm h.c.}, 
\end{eqnarray}
where $\phi = \Phi\lowest$ and $W(\Phi)$ is the cubic superpotential for chiral
matter superfields. Note that the squared scalar masses are always positive. 
As concluded in Section 4.6, we find in our model that moduli $t^{I}$ are 
stabilized at the self-dual point and their associated 
$\langle\,F^{I}\,\rangle$ vanish in the vacuum, which results in a
dilaton-dominated supersymmetry breaking scenario. According to (5.9), both the 
scalar masses and $A$ terms are indeed independent of their modular weights by 
virtue of the fact that $\langle\,F^{I}\,\rangle =0$. For the FCNC constraints,
this feature of dilaton-dominated scenario is a potential advantage over a
moduli-dominant scenario. In the past, it was generally believed that a
dilaton-dominated scenario results in universal soft supersymmetry breaking
parameters due to the universality of dilaton couplings \cite{FCNC}. However,
here we wish to stress that the above statement did not take into account the 
matter-dependent contributions to the Green-Schwarz counterterm, and therefore
a dilaton-dominated scenario does not guarantee universal soft supersymmetry 
breaking parameters. It is clear from the computations of our dilaton-dominated 
scenario in (5.9) that soft supersymmetry breaking parameters are universal 
-- and unwanted flavor-changing neutral currents are thereby suppressed -- if 
the matter couplings ($p_{A}$) to the Green-Schwarz counterterm are also 
universal. Unfortunately, so far there is little knowledge of $p_{A}$'s;
therefore, the best we can do right now is to study the consequences of several 
seemingly reasonable choices of $p_{A}$'s. One possibility is that $p_{A}$'s 
are universal; thus we have universal soft supersymmetry breaking parameters 
and in this case $A$ terms in (5.9) reduce to
\begin{equation}
V_A(\phi) \approx {3\over4}e^{K/2}\bar{u}_+{p_A\(1+2b_+\ell\)
-b^{2}_{+}\ell\over(1+p_A\ell)(1+b_+\ell)}W(\phi) \,+\, {\rm h.c.} 
\equiv Ae^{K/2}W(\phi) \,+\, {\rm h.c.}. 
\end{equation}
For example, if the Green-Schwarz counterterm is simply independent of the 
matter fields $\Phi^A$ ({\it i.e.}, $p_A = 0$), we have 
$m_A = m_{\tilde G},\;A \approx 2m_{\lambda}$. As for choices of non-universal
$p_{A}$'s, a possibility is that the Green-Schwarz counterterm depends only on
the radii $R_I$ of the three compact tori that determine the untwisted-sector
part of the K\"ahler potential (5.8): 
$$ K = k(V) - \sum_I\ln(2R_I^2) + \CO(|\Phi^A_{\rm twisted}|^2),$$ 
where $2R^2_I = T^I + \bar{T}^I - \sum_A|\Phi^A_I|^2$ in string units. In this 
case, $p_A = b$ for the untwisted chiral superfields $\Phi^A_I$, and $p_A = 0$
for the twisted chiral superfields $\Phi^A_{\rm twisted}$. The untwisted 
scalars have masses comparable to the moduli masses:
$\,m_{\rm untwisted} = m_t/2 \approx A/3$. Finally, we note that if 
$b_+ \approx b/10 \approx 1/30$, gaugino masses are suppressed relative to the 
gravitino mass at the condensation scale $\mu_{cond}\sim 10^{-4}M'_{P}$: 
$m_\lambda \sim m_{\rm twisted}/40$. If there is a sector with $p_A =
b$ and a Yukawa coupling of order one involving $SU(3)$ (anti-) triplets 
({\it e.g.}, $\bar{D}DN$, where $N$ is a standard model singlet), its two-loop
contribution to gaugino masses \cite{twoloop} can be more important than the 
standard one-loop contribution, generating a physical mass for gluinos that is 
well within experimental bounds for $m_{\tilde G}\approx 1$ TeV. Such a 
coupling could also generate a {\em vev} for $N$, thus breaking possible 
additional $U(1)$'s at a scale $\sim 10$ TeV. The phenomenologically required 
$\mu$ term of the MSSM may also be generated by the {\em vev} of a Standard
Model gauge singlet or by one of the other mechanisms that have been proposed 
in the literature \cite{muterm}.

In contrast to the case of universal $p_{A}$'s, for the case of non-universal 
$p_{A}$'s one has to worry about the FCNC problem. Scenarios in which the 
sparticles of the first two generations have masses as high as $20$ TeV have 
in fact been proposed \cite{cohen} to solve the FCNC problem. However, it has
recently been pointed out that such scenarios may suffer from a negative 
scalar top mass squared driven by two-loop renormalization group evolution
\cite{Hitoshi}\footnote{We thank Hitoshi Murayama for pointing out this 
problem to us.}. Clearly, a better understanding of the matter dependence of 
the Green-Schwarz counterterm is required to make precise predictions for soft 
supersymmetry breaking. Nevertheless our model suggests soft supersymmetry 
breaking patterns that may differ significantly from those generally assumed in
the context of the MSSM.  Phenomenological constraints such as current
limits on sparticle masses, gauge coupling unification and a charge and
color invariant vacuum can be used to restrict the allowed values of $p_A$'s 
as well as the low-energy spectrum of the string effective field theory. 
To conclude, we would like to stress that the model presented above is 
certainly not final and some of the results obtained, especially on the
low-energy sector of the theory, may receive modifications. Possible sources of
modification are the presence of an anomalous $U(1)$ symmetry \cite{messenger} 
or a constant term in the superpotential that breaks modular invariance
\cite{wr,horava}. 
\section{Gauge Coupling Unification and the Newton's Constant}
\hspace{0.8cm}
String non-perturbative corrections necessary to stabilize the dilaton could 
make significant corrections to the unification of gauge couplings.
The functions $f(\ell)$ and $g(\ell)$ introduced above and the threshold 
corrections whose form is dictated by $T$ duality invariance contribute as
follows to the value of couplings at unification:
\begin{eqnarray}
g_a^{-2} (M_s) &=& g_s^{-2} + {C_a \over 8 \pi^2} \ln (\lambda e)
-{1 \over 16 \pi^2} \sum_I b_a^I \ln (t^I + \bar t^I)|\eta^2(t^I)|^2,\\
g_s^{-2} &=& {1+f \over 2 \ell}, \; \; \; \; M_s^2 = \lambda g_s^2 M_{P}^2,
\end{eqnarray}
with
\begin{equation}
\lambda = {1\over 2} e^{g-1} (1+f)
\end{equation}
Let us note however that this parameter is worth $1/(2e)\approx 0.18$ in the
perturbative case and $e^{-1.65}\approx 0.19$ in the one gaugino condensate 
model. 

Let us take this opportunity to clarify a confusing statement in the 
literature about gauge coupling unification in weakly-coupled superstring.
It is often stated that one can determine from the low-energy values of gauge 
couplings the precise value of the gauge coupling unification scale, 
$M_{GUT}$, to be the $M_{GUT}^{(MSSM)}\,=\, 3 \times 10^{16}$ GeV based on the 
MSSM. We think that this 
is a misleading statement since most string models constructed so far that 
hold a claim for being realistic include new forms of matter which perturb the 
evolution of the gauge couplings at some intermediate threshold \cite{unif}.
In fact, as for the string models considered in this study, the unification
scale $M_{GUT}$ should naturally be the string scale $M_{s}$ 
\cite{Gaillard92}. Furthermore, the compactification scale $M_{comp}$ is also 
close to the string scale because the compactification moduli are stabilized 
at the self-dual point. Therefore, one naturally expects 
\begin{equation}
M_{GUT}\;\sim\;M_{s}\;\sim\;M_{comp}.
\end{equation}

Finally, let's make a short remark on the Newton's constant $G_{N}$. For 
the weakly-coupled heterotic string, it has been shown by E. Witten 
\cite{Witten96} that there exists a lower bound on the Newton's constant
as follows.
\begin{equation}
G_{N}\;\;\geq\;\;\frac{\alpha_{GUT}^{4/3}}{M_{comp}^{2}}.
\end{equation}
If one simply takes $M_{comp}$ to be $M_{GUT}^{(MSSM)}$, the resulting lower
bound on the Newton's constant is indeed too large. On the other hand, in our
study the compactification moduli are actually stabilized at the self-dual 
point, $\langle\,t^{I}\,\rangle\,=\,1$. Therefore, the compactification scale 
is quite close to the string scale. According to (5.14), one should take 
$M_{comp}$ to be of order $M_{s}$, and the resulting lower bound on the 
Newton's constant is of order $\alpha_{GUT}^{4/3}/M_{s}^{2}$. This lower bound
on $G_{N}$ is certainly small enough. 
\section{Concluding Remarks}
\hspace{0.8cm}
As discussed in Chapter 1, the weakly-coupled heterotic string theory is 
known to 
have problems with dilaton/moduli stabilization, supersymmetry breaking, 
gauge coupling unification, QCD axion, as well as cosmological problems 
involving dilaton/moduli and axion. In the literature some of these problems 
are often treated as evidence against the weakly-coupled heterotic string
theory. However, it is actually hard to say whether these problems are 
inherent to the weakly-coupled heterotic string theory or they simply reflect 
our ignorance of important string
dynamics. Furthermore, some of these problems will probably re-appear even in
the study of the strong-coupling limit of the heterotic string theory. In this 
work we study these problems by adopting the point of view that they arise 
mostly due to our
limited calculational power, little knowledge of of the full vacuum structure,
and an inappropriate treatment of gaugino condensation. Indeed, after a careful
review one finds that the phenomenological studies of the weakly-coupled 
heterotic string theory in the literature contain several essential flaws. It 
is therefore of utmost importance to correct these flaws and then re-examine 
the problems of weakly-coupled heterotic string theory. In conclusion, 
three essential changes to the standard lore of string phenomenology have to 
be made. The first essential change is about the effective field theory of 
the weakly-coupled heterotic string. It is emphasized that the linear multiplet 
formalism rather than the chiral multiplet formalism is the 
appropriate\footnote{We would like emphasize that, in consideration of the
chiral-linear duality shown in \cite{Burgess95}, in principle the linear
multiplet formalism should be equivalent to the {\em constrained} chiral
multiplet formalism. However, as discussed in \cite{dilaton}, the chiral-linear
duality is apt to be very complicated, especially when the full quantum
corrections are included; therefore, there should exist a formalism where the
physics allows a simpler description. It is in this sense that the linear
multiplet formalism is the appropriate formalism.}
framework for the effective field theory of the weakly-coupled heterotic string.
The second essential change is the inclusion of possible stringy
non-perturbative effects in addition to the usual field-theoretical
non-perturbative effects produced by gaugino condensation. The third essential
change is an improved treatment of gaugino condensation by including the
constraint (2.12). As discussed in Chapter 2, the last two changes are most
naturally implemented using the linear multiplet formalism. Finally, notice that
full modular invariance is always maintained in our construction. This is
important because modular invariance is supposed to be an exact quantum 
symmetry of closed string theory \cite{ns}. 

   In Chapters 2--4, the linear multiplet formalism with the aforementioned
features is constructed for an E$_8$ model as well as a generic orbifold model.
It is particularly transparent in this framework to realize how the dilaton can
be stabilized by stringy non-perturbative contributions to the K\"ahler
potential.\footnote{Of course, still we don't know how to calculate these
stringy non-perturbative effects. However, the point is that these effects
are at least under good control here.} Furthermore, supersymmetry can be broken
at a realistic scale once the dilaton is stabilized. As for the moduli, they are
always stabilized at their self-dual points where the moduli actually do not
contribute to supersymmetry breaking -- a beautiful consequence of modular
invariance and an appropriate treatment of gaugino condensation. 
Phenomenologically,
we always have a dilaton-dominated scenario of supersymmetry breaking. The fact
that the compactification moduli are stabilized at the self-dual point also
leads to a small enough lower bound on the Newton's constant \cite{Witten96}. 
As for the masses of 
moduli, in contrast to the standard lore of string phenomenology a careful 
analysis reveals that there is a natural hierarchy between moduli and 
gravitino masses. It is not difficult to see how this hierarchy arises: in a
generic orbifold model with realistic supersymmetry breaking scale, there is
already a natural hierarchy between the E$_{8}$ $\beta$-function coefficient 
$b$ (associated with the Green-Schwarz counterterm) and the $b_{a}$ of the 
largest hidden gauge subgroup ($b/b_+\approx 10$). Such a hierarchy between
moduli and gravitino masses has important cosmological consequences. As
discussed in Chapter 5, it not only could solve the cosmological moduli problem
but also keeps the energy density of the oscillations of the string
model-independent axion from overclosing the universe. As for the strong CP
problem, there is always only one very light axion (the model-independent 
axion) in our model, and it does have the right features to be the QCD axion
in contrast to the conclusion of Banks and Dine \cite{Banks96}. The difference
between our result and that of Banks and Dine has to do with our improved
treatment of gaugino condensation and a holomorphy argument associated
with the linear multiplet which is unique to the linear multiplet formalism. In
conclusion, it is fair to say that these problems of the weakly-coupled 
heterotic string theory can be solved or are much less severe.

   As expected, the origin of the cosmological constant remains a mystery here
although it is indeed under better control and the cosmological constant can be
fine tuned to zero in our treatment. Again, a final resolution of this problem
might have to wait for a complete understanding of superstring dynamics. The 
other unsettled issue in this work is the soft supersymmetry breaking pattern. 
Although our model always predicts a dilaton-dominated scenario of supersymmetry
breaking, yet in contrast to the standard lore of string phenomenology we
point out that whether a dilaton-dominated scenario predicts universal soft
supersymmetry breaking parameters actually depends on whether the matter
couplings to the Green-Schwarz counterterm are universal. To settle this
issue, a better understanding of the matter dependence of the Green-Schwarz
counterterm for generic string models is certainly required; it deserves 
further studies and could lead to a rich phenomenology. Another potential
problem of this work is that the gaugino masses might be too small. Whether
this is a serious problem or not can be very model-dependent, especially in the
context of superstrings where one generically encounters scenarios beyond the
MSSM. In conclusion, we emphasize that this work is certainly not final, and it
is very important to understand more about the non-perturbative aspects of
superstrings, realistic string model building and the phenomenology. 
After a careful re-examination of the aforementioned problems of the 
weakly-coupled heterotic string theory, it is also hoped that those 
misunderstandings of the current status of weakly-coupled heterotic string
theory in the literature are clarified by this work.

\end{document}